\documentclass[10pt]{article}
\usepackage{amsmath}
\usepackage{graphicx}
\usepackage{color}
\usepackage[table,xcdraw]{xcolor}
\usepackage{orcidlink}
\usepackage{hyperref}
\hypersetup{colorlinks=true, linkcolor=blue, citecolor=blue, urlcolor=blue}
\usepackage{float}
\usepackage{multirow}
\usepackage{array}
\usepackage{longtable}
\usepackage{booktabs}
\usepackage{colortbl}
\usepackage{subcaption}
\usepackage{geometry}
\RequirePackage[numbers,sort&compress]{natbib}
\begin{document}
\baselineskip=14pt

\begin{center}
\LARGE{Dynamics of test particles, QPOs and thermodynamics of charged Euler-Heisenberg AdS black holes with a cloud of strings and dark matter}
\end{center}

\vspace{0.3cm}
\begin{center}
{\bf Faizuddin Ahmed\orcidlink{0000-0003-2196-9622}}\footnote{\bf faizuddinahmed15@gmail.com}\\
{\it Department of Physics, Royal Global University, Guwahati, 781035, Assam, India}\\

{\bf Ahmad Al-Badawi\orcidlink{0000-0002-3127-3453}}\footnote{\bf ahmadbadawi@ahu.edu.jo}\\
{\it Department of Physics, Al-Hussein Bin Talal University 71111, Ma’an, Jordan}\\

{\bf \.{I}zzet Sakall{\i}\orcidlink{0000-0001-7827-9476}}\footnote{\bf izzet.sakalli@emu.edu.tr (Corresp. author)}\\
{\it Physics Department, Eastern Mediterranean University, Famagusta 99628, North Cyprus via Mersin 10, Turkey}
\end{center}

\vspace{0.3cm}

\begin{abstract}
We investigate the geodesic motion of test particles around a charged Euler-Heisenberg Anti-de Sitter black hole (CEH-AdS-BH) with a cloud of strings (CoS) and surrounded by perfect fluid dark matter (PFDM). Our analysis examines how the interplay between Euler-Heisenberg nonlinear electrodynamic corrections, string cloud, and PFDM distributions modifies fundamental BH properties. We study dynamics of both massless photons and massive neutral test particles, deriving the effective potentials and trajectory equations that govern particle motion in this exotic spacetime. The photon sphere analysis shows systematic increases in shadow radii with string cloud parameter variations, while innermost stable circular orbit (ISCO) calculations demonstrate competing effects between string-induced gravitational weakening and electromagnetic charge contributions. We examine scalar and electromagnetic field perturbations using the Klein-Gordon and Maxwell equations, establishing the perturbative potentials that characterize wave propagation in the modified background. The scalar perturbative potential exhibits characteristic barrier structures influenced by all exotic matter components, while electromagnetic perturbations show fundamental differences due to their vector nature. Our thermodynamic investigation extends the conventional framework by incorporating additional intensive variables corresponding to cosmic string and perfect fluid dark matter contributions. The specific heat capacity analysis reveals divergence points that demarcate thermodynamically stable and unstable regions, while the Gibbs free energy exhibits modified profiles that extend Anti-de Sitter black hole thermodynamics into new parameter regimes.
\end{abstract}

{\bf Keywords}: Euler-Heisenberg black holes; Topological defects; Dark matter; Gravitational thermodynamics

\section{Introduction}\label{sec1}

The theoretical investigation of black holes (BHs) has evolved dramatically since Einstein's original formulation of general relativity, with modern research encompassing diverse extensions \cite{isz00} that incorporate quantum field theory, nonlinear electrodynamics, and exotic matter distributions \cite{isz01,isz02,Balakin:2021jby,Bittencourt:2023ikm,Afonso:2018mxn,Kruglov:2023xxm,Gullu:2020ant}. Among these extensions, the study of charged BHs in Anti-de Sitter (AdS) spacetimes has gained particular prominence due to their relevance in holographic duality and their rich thermodynamic properties that mirror phase transitions in ordinary matter systems \cite{isz03,isz04,Chen:2016gzz,Cinti:2025kqh,Ahmed:2023snm}.

The Euler-Heisenberg (EH) theory emerges as a natural extension of Maxwell's linear electrodynamics when quantum corrections become significant in strong electromagnetic field regimes \cite{isz05,isz06}. This nonlinear electrodynamic (NLED) theory, first derived by Heisenberg and Euler in 1936, accounts for photon-photon scattering and vacuum polarization effects that become measurable when electromagnetic field strengths approach the critical field $E_c = m_e^2 c^3/(e\hbar) \approx 1.3 \times 10^{18}$ V/m \cite{isz07}. In the context of BH physics, EH corrections modify the standard Reissner-Nordström solution, leading to charged Euler-Heisenberg AdS black holes (CEH-AdS-BH) that exhibit altered horizon structures, modified thermodynamic properties, and distinct observational signatures \cite{isz08,isz09}.

The incorporation of topological defects, particularly cosmic strings (CoS), into BH spacetimes introduces additional layers of complexity that reflect the rich structure of the early universe. CoS are one-dimensional topological defects predicted to form during symmetry-breaking phase transitions in the primordial cosmos, characterized by their linear energy density and the angular deficit they introduce into spacetime geometry \cite{isz10,isz11}. When CoS intersect BH horizons or extend through BH interiors, they fundamentally alter the causal structure and create distinctive modifications to particle trajectories, thermodynamic properties, and gravitational wave emission patterns \cite{isz12,isz13}. The parameter $\gamma$ quantifies the strength of CoS effects, with values approaching unity leading to increasingly dramatic departures from standard BH behavior.

Perfect Fluid Dark Matter (PFDM) has emerged as an important component in modern astrophysical and cosmological models, motivated by compelling observational evidence suggesting that non-baryonic dark matter constitutes nearly 27\% of the total energy density of the universe \cite{Planck2018}. In contrast to the conventional cold dark matter paradigm, which models dark matter as pressureless dust, PFDM includes a non-vanishing pressure component that can significantly affect the space-time geometry, particularly in the vicinity of compact astrophysical objects \cite{Li2014, Xu2018}. The influence of PFDM is often encoded in a coupling parameter $\beta$, which quantifies the interaction strength of the dark fluid with the gravitational field. This leads to characteristic logarithmic corrections in the metric function, arising from the assumed equation of state of the dark fluid. Such corrections have been shown to alter geodesic motion, gravitational lensing, and black hole shadow profiles \cite{Li2021, Hou2023}.

The theoretical motivation for studying CEH-AdS-BH solutions with CoS and PFDM stems from multiple converging factors. From a fundamental physics perspective, these systems provide testing grounds for quantum field theory in curved spacetime, where EH corrections become comparable to geometric curvature effects in the strong-field regime near BH horizons \cite{isz18,isz19}. The presence of CoS introduces topological complexity that may be relevant for understanding the universe's large-scale structure and the formation mechanisms of supermassive BHs observed in galactic centers \cite{isz20,isz21}. Additionally, PFDM effects offer phenomenological approaches to modeling dark matter influences on BH dynamics, potentially bridging the gap between theoretical predictions and observational constraints from gravitational wave detections and direct imaging campaigns \cite{isz22,isz23}.

Recent observational breakthroughs have transformed BH physics from a purely theoretical discipline into an empirical science with unprecedented precision. The Event Horizon Telescope's direct imaging of supermassive BH shadows in M87* and Sagittarius A* has provided the first visual confirmation of BH existence while establishing new constraints on alternative gravity theories and exotic matter models \cite{isz24,isz25}. Simultaneously, gravitational wave detections by LIGO-Virgo collaborations have opened new windows into BH merger dynamics, revealing populations of stellar-mass BHs with properties that challenge conventional formation scenarios \cite{isz26,isz27}. These observational advances create urgent needs for theoretical frameworks capable of predicting distinctive signatures that could distinguish exotic BH models from their classical counterparts. In this regard, CEH-AdS-BH solutions with CoS and PFDM encompasses a rich phenomenological landscape where multiple competing effects determine observable properties. The EH parameter $\alpha$ governs the strength of quantum electrodynamic corrections, becoming increasingly important as electric charge approaches extremal values. The CoS parameter $\gamma$ introduces angular deficit effects that modify global spacetime topology, while the PFDM coupling $\beta$ contributes additional gravitational attractions with characteristic logarithmic dependences. The AdS curvature radius $\ell_p$ sets the scale for cosmological constant effects, creating confining potentials that fundamentally change particle dynamics compared to asymptotically flat scenarios.

Our investigation aims to establish a theoretical framework for CEH-AdS-BH spacetimes incorporating both CoS and PFDM, addressing several fundamental questions that bridge theoretical consistency with observational viability. We seek to determine how the interplay between EH corrections, topological defects, and exotic matter distributions affects the basic properties of BH solutions, including horizon structures, causal relationships, and stability criteria. The geodesic analysis addresses how test particle trajectories, both massive and massless, respond to the modified gravitational potentials created by these exotic matter components, with particular attention to observationally relevant quantities such as photon sphere radii, shadow characteristics, and innermost stable circular orbit (ISCO) locations. The perturbation analysis investigates field propagation characteristics in these modified spacetimes, examining how scalar and electromagnetic perturbations respond to the complex potential structures arising from multiple exotic matter contributions. This analysis provides crucial information about stability properties, quasi-normal mode spectra, and greybody radiation characteristics that could serve as observational signatures distinguishing these exotic BH models from conventional solutions. The thermodynamic investigation explores phase transition behaviors, stability criteria, and critical phenomena in extended phase spaces where CoS and PFDM parameters serve as additional thermodynamic variables with corresponding intensive conjugates.

The paper is organized as follows: Section~\ref{sec2} establishes the fundamental spacetime structure of CEH-AdS-BH with CoS surrounded by PFDM, including detailed horizon analysis and geometric properties. Section~\ref{sec3} investigates geodesic dynamics of test particles, encompassing photon trajectories, shadow formation, massive particle orbits, ISCO configurations, and fundamental frequency analysis. Section~\ref{sec4} examines scalar field perturbations and their propagation characteristics in the modified spacetime background. Section~\ref{sec5} analyzes electromagnetic perturbations and their distinctive features compared to scalar field dynamics. Section~\ref{sec6} presents the complete thermodynamic framework, including phase transitions, stability analysis, and critical phenomena in extended phase spaces. Section~\ref{sec7} summarizes our findings and discusses future research directions emerging from this study.

 \section{CEH-AdS-BH with CoS Surrounded by PFDM: Spacetime Structure and Horizon Analysis} \label{sec2}

The theoretical investigation of BHs in modified gravitational theories has gained substantial momentum in recent years, particularly with the advent of sophisticated observational techniques such as the Event Horizon Telescope. Among the most intriguing extensions to classical general relativity are those incorporating NLED theories, exotic matter distributions, and modified spacetime geometries. In this context, the EH theory represents a paradigmatic example of NLED that emerges naturally from QED considerations at strong electromagnetic field regimes.

Building upon the foundational work in Ref. \cite{sec2is01}, where the authors derived exact solutions for EH BHs surrounded by PFDM and investigated their optical and thermodynamic properties, subsequent research by \cite{sec2is02} extended these investigations to charged AdS configurations. The present work advances this research frontier by incorporating an additional exotic matter component: CoS, which introduce topological defects that can significantly alter the spacetime geometry and associated physical phenomena.

The motivation for studying such complex BH solutions stems from both theoretical completeness and potential astrophysical relevance. CoS are predicted to form during phase transitions in the early universe \cite{sec2is03}, while PFDM represents a phenomenological approach to modeling dark matter effects around compact objects \cite{sec2is04}. The combination of these exotic matter fields with EH NLED creates a rich parameter space for exploring modified BH physics and their observational signatures \cite{sec2is05}.

The line element describing our CEH-AdS-BH with CoS surrounded by PFDM adopts the standard spherically symmetric, static form :
\begin{equation}
    ds^2=-h(r)\,dt^2+\frac{dr^2}{h(r)}+r^2\,(d\theta^2+\sin^2 \theta\,d\phi^2),\label{bb1}
\end{equation}
where the metric function $h(r)$ reads \cite{sec2is02}:
\begin{equation}
    h(r)=1-\gamma-\frac{2\,M}{r}+\frac{r^2}{\ell^2_p}+\frac{Q^2}{r^2}-\frac{\alpha\,Q^4}{20\,r^6}+\frac{\beta}{r}\,\ln\frac{r}{|\beta|}.\label{bb2}
\end{equation}
Each term in this metric function carries profound physical significance. The parameter $M$ represents the BH mass, while $0 \leq \gamma <1$ characterizes the CoS parameter, quantifying the deficit angle introduced by the cosmic string. The electric charge $Q$ couples to the EH NLED through the parameter $\alpha$, which measures the strength of QED corrections beyond the linear Maxwell theory. The PFDM contribution enters through $\beta>0$, the coupling parameter that governs the interaction between the BH and the surrounding dark matter distribution. Finally, $\ell_p$ denotes the AdS curvature radius, related to the cosmological constant via $\frac{1}{\ell^2_p}=-\frac{\Lambda}{3}$ \cite{sec2is06}.

The remarkable versatility of this solution becomes apparent when examining various limiting cases. Setting $Q=0=\beta$ recovers the AdS Letelier BH solution \cite{isz13}, which represents the canonical CoS BH in AdS spacetime. Alternatively, the limit $\gamma=0=\beta$ yields the EH-AdS BH solution \cite{sec2is08,sec2is09,sec2is10}, isolating the effects of NLED in AdS geometry. The case $Q=0=\gamma$ reduces to the Schwarzschild-AdS BH surrounded by PFDM, while the most restrictive limit $\gamma=0=\alpha=\beta$ reproduces the classical Reissner-Nordström AdS solution.

Table \ref{ads_eh_horizons_no_naked} offers a concise analysis of horizon structures, highlighting the phenomenology resulting from various parameter interactions. This exploration across the parameter space $\gamma \in \{0.0, 0.1, 0.2, 0.577216\}$, $Q \in \{0.0, 0.5, 1.0\}$, $\alpha \in \{0.0, 0.1\}$, and $\beta \in \{0.0, 0.1\}$ demonstrates several key features. The empty bracket [] for $\gamma = 0.577216$, $Q = 1.0$, $\alpha = 0$, $\beta = 0.1$ indicates a naked singularity configuration (no horizons exist). For vanishing parameters, the Schwarzschild-AdS case yields the expected single horizon at $r_h = 1.0$. The introduction of PFDM ($\beta > 0$) typically shifts the horizon inward, reflecting the additional gravitational attraction. Most intriguingly, certain parameter combinations exhibit double horizons, characteristic of non-extremal BH configurations, while others display single horizons corresponding to extremal states.

\begin{figure}[ht!]
    \centering
    \includegraphics[width=0.7\linewidth]{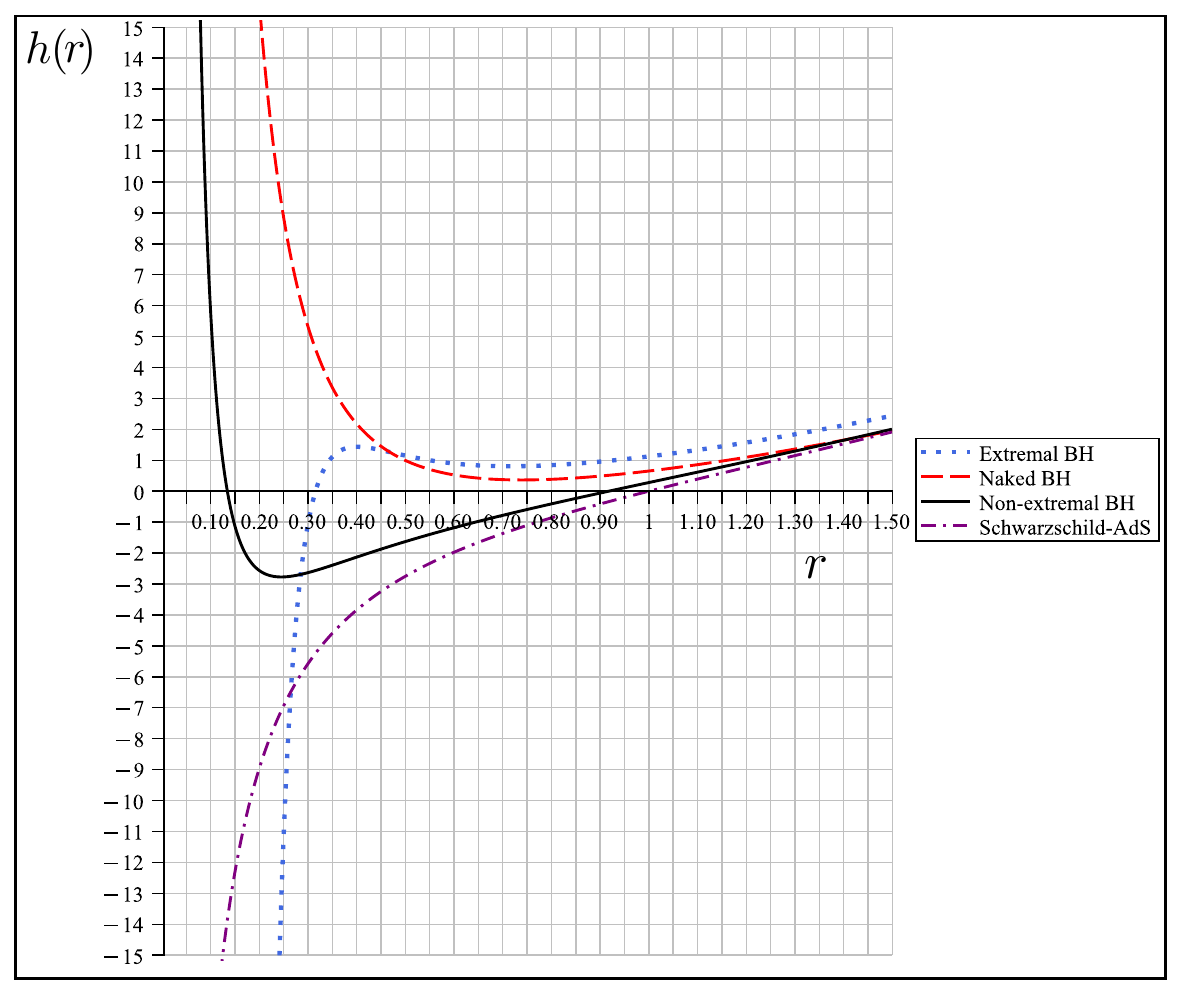}
    \caption{\footnotesize Behavior of the metric function $h(r)$ as a function of $r$, plotted for fixed $M = 1$ and $\ell_p = 1$. The figure shows: (i) Extremal BH with $\gamma = 0.1$, $Q = 1.0$, $\alpha = 0.1$, $\beta = 0.1$ (blue, dotted); (ii) Naked singularity with $\gamma = 0.577216$, $Q = 1$, $\alpha = 0$, $\beta = 0.1$ (red, dashed); (iii) Non-extremal BH with $\gamma = 0.2$, $Q = 0.5$, $\alpha = 0.0$, $\beta = 0.1$ (black, solid); and (iv) Schwarzschild-AdS with $\gamma = 0.0$, $Q = 0.0$, $\alpha = 0.0$, $\beta = 0.0$ (purple, dash-dot). The spacetime is static, spherically symmetric, and asymptotically AdS, demonstrating various horizon configurations.}
    \label{fig:metric_function}
\end{figure}

The metric function behavior, illustrated in Fig. \ref{fig:metric_function}, provides crucial insights into the spacetime causal structure. The extremal BH configuration (blue dotted line) shows $h(r)$ touching zero at a single point, indicating the merger of inner and outer horizons. The naked singularity case (red dashed line) demonstrates the absence of horizons, revealing the central singularity to external observers. The non-extremal BH (black solid line) exhibits the characteristic double-zero structure, while the reference Schwarzschild-AdS case (purple dash-dot line) displays the standard single horizon behavior.

\begin{figure}[ht!]
    \centering
    \includegraphics[width=0.4\linewidth]{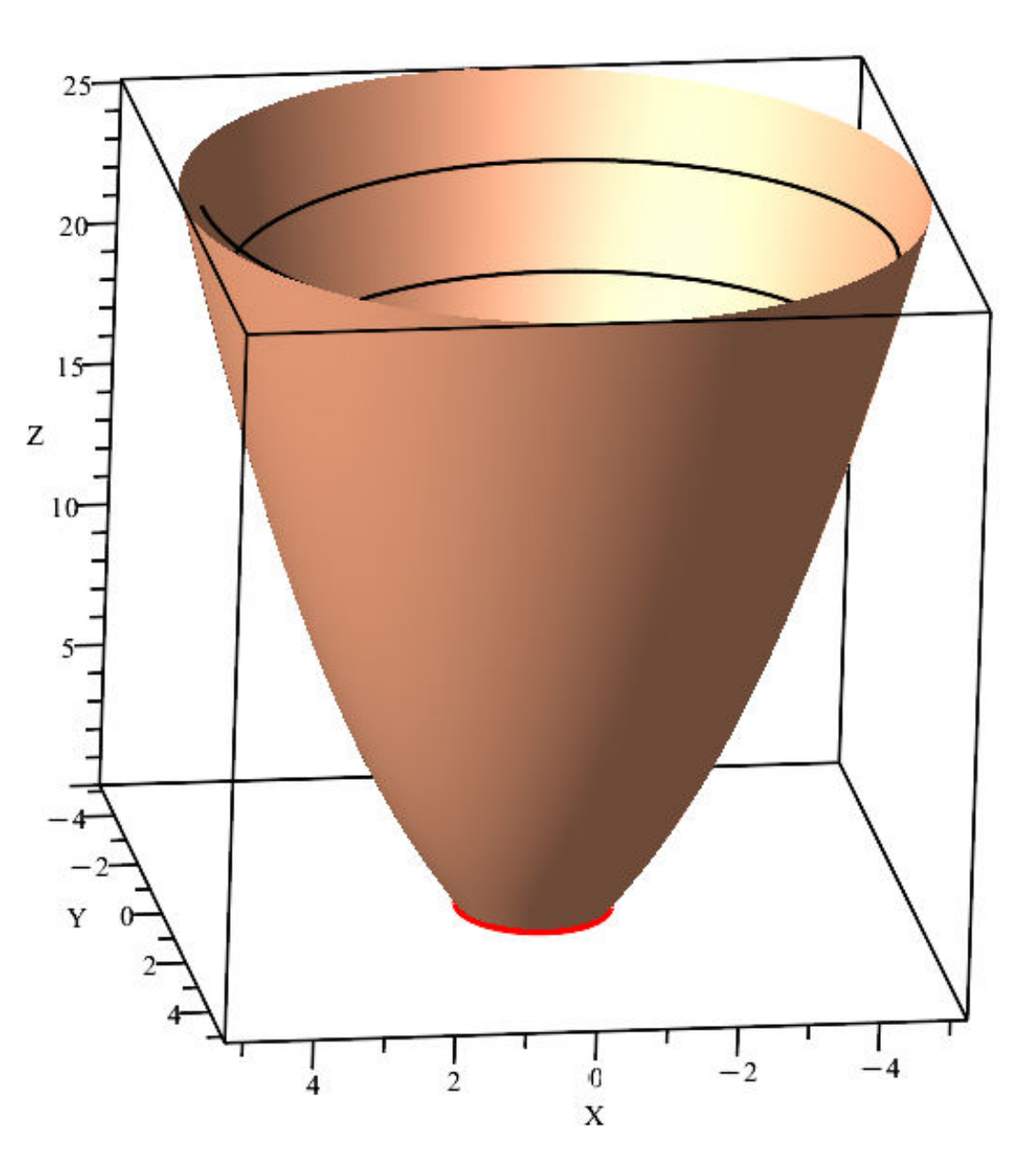}
    \caption{\footnotesize 3D embedding diagram of the metric function $h(r)$ of CEH-AdS-BH with CoS. The physical parameters are chosen as $M = 1$, $\gamma = 0.1$, $Q = 0.5$, $\alpha = 0.1$, $\beta = 0.1$, and $\ell_p = 1$.}
    \label{fig:embedding_horizons}
\end{figure}

To complement the analytical investigation, Fig. \ref{fig:embedding_horizons} presents a three-dimensional embedding diagram that visualizes the spacetime geometry. This representation, constructed for the specific parameter set $M = 1$, $\gamma = 0.1$, $Q = 0.5$, $\alpha = 0.1$, $\beta = 0.1$, and $\ell_p = 1$, illustrates the warped spacetime structure from the event horizon ($r_{+}=1.028886510$) extending to $r = 5$. The embedded surface reveals the characteristic throat structure, with the red ring marking the event horizon boundary and the black trajectory representing the path of an infalling object.

The critical value $\gamma = 0.577216$ deserves special attention, as it represents a threshold beyond which certain parameter combinations fail to produce viable BH solutions, instead yielding naked singularities. This behavior underscores the delicate balance required to maintain physically reasonable BH configurations in the presence of multiple exotic matter components.

\setlength{\tabcolsep}{12pt}
\begin{longtable}{|c|c|c|c|c|}
\hline
\rowcolor{gray!50}
\textbf{$\gamma$} & \textbf{$Q$} & \textbf{$\alpha$} & \textbf{$\beta$} & \textbf{Horizon(s)} \\
\hline
\endfirsthead
\hline
\rowcolor{gray!50}
\textbf{$\gamma$} & \textbf{$Q$} & \textbf{$\alpha$} & \textbf{$\beta$} & \textbf{Horizon(s)} \\
\hline
\endhead
0.0 & 0.0 & 0.0 & 0.0 & [1.0000000] \\
\hline
0.0 & 0.0 & 0.0 & 0.1 & [0.9414211] \\
\hline
0.0 & 0.0 & 0.1 & 0.1 & [0.9414211] \\
\hline
0.0 & 0.5 & 0.0 & 0.1 & [0.1366430,\ 0.8593557] \\
\hline
0.0 & 0.5 & 0.1 & 0.1 & [0.8595781] \\
\hline
0.0 & 1.0 & 0.1 & 0.1 & [0.3136560] \\
\hline
0.1 & 0.0 & 0.0 & 0.1 & [0.9666234] \\
\hline
0.1 & 0.0 & 0.1 & 0.1 & [0.9666234] \\
\hline
0.1 & 0.5 & 0.0 & 0.1 & [0.1354863,\ 0.8882160] \\
\hline
0.1 & 0.5 & 0.1 & 0.1 & [0.8884003] \\
\hline
0.1 & 1.0 & 0.1 & 0.1 & [0.3154782] \\
\hline
0.2 & 0.0 & 0.0 & 0.1 & [0.9921955] \\
\hline
0.2 & 0.0 & 0.1 & 0.1 & [0.9921955] \\
\hline
0.2 & 0.5 & 0.0 & 0.1 & [0.1343699,\ 0.9173454] \\
\hline
0.2 & 0.5 & 0.1 & 0.1 & [0.9174986] \\
\hline
0.2 & 1.0 & 0.1 & 0.1 & [0.3174102] \\
\hline
0.577216 & 0.0 & 0.0 & 0.1 & [1.0912882] \\
\hline
0.577216 & 0.0 & 0.1 & 0.1 & [1.0912882] \\
\hline
0.577216 & 1.0 & 0.0 & 0.1 & [] \\
\hline
0.577216 & 0.5 & 0.0 & 0.1 & [0.13047954,\ 1.0288081] \\
\hline
0.577216 & 0.5 & 0.1 & 0.1 & [1.0288864] \\
\hline
0.577216 & 1.0 & 0.1 & 0.1 & [0.32598503] \\
\hline
\caption{\footnotesize Horizons obtained for various values of the model parameters $\gamma \in \{0.0, 0.1, 0.2, 0.577216\}$, $Q \in \{0.0, 0.5, 1.0\}$, $\alpha \in \{0.0, 0.1\}$, and $\beta \in \{0.0, 0.1\}$, with fixed $M = 1$ and $\ell_p = 1$, excluding naked singularity cases (no horizons). The spacetime is static, spherically symmetric, and asymptotically AdS, exhibiting configurations such as extremal BHs (single horizon) or non-extremal BHs (double horizons). The Schwarzschild-AdS case, corresponding to $\gamma = 0$, $Q = 0$, $\alpha = 0$, $\beta = 0$, yields a single horizon.}
\label{ads_eh_horizons_no_naked}
\end{longtable}

\section{Dynamics of Test Particles in CEH-AdS-BH Spacetime with CoS and PFDM} \label{sec3}

The study of particle motion in modified gravitational backgrounds represents a cornerstone of modern theoretical astrophysics, providing crucial insights into the nature of spacetime curvature and its observational manifestations. In CEH-AdS-BH spacetimes, the geodesic dynamics of both massless and massive test particles undergo significant modifications due to the complex interplay between EH NLED corrections, CoS topological defects, and PFDM distributions \cite{sec3is01,sec3is02}. These exotic matter components fundamentally alter the effective gravitational potential experienced by particles, leading to distinctive signatures in photon trajectories, particle orbits, and associated phenomena such as BH shadows and accretion disk dynamics.

The theoretical framework governing particle motion in such complex gravitational environments extends far beyond the classical general relativistic treatment. The presence of CoS introduces angular deficit effects that modify the global topology of spacetime \cite{isz12}, while PFDM contributes additional gravitational attractions that can significantly alter orbital stability criteria. Simultaneously, EH NLED corrections become particularly pronounced in the strong-field regime near the BH event horizon, where QED effects begin to dominate over linear Maxwell electrodynamics \cite{isz05,isz07}.

Our systematic investigation aims to elucidate how these multiple exotic components collectively influence geodesic motion, with particular emphasis on identifying potentially observable signatures that could distinguish such modified BH solutions from their conventional counterparts. This analysis becomes increasingly relevant in light of recent advances in BH imaging techniques and gravitational wave detection capabilities, which offer unprecedented opportunities to probe the strong-field regime of gravity \cite{isz24}.

\subsection{Photon Dynamics and Null Geodesics in Modified Spacetime}

Photons, being massless particles, follow null geodesics whose structure provides direct information about the causal properties of spacetime geometry. In our CEH-AdS-BH background, photon trajectories exhibit rich phenomenology arising from the competition between various gravitational and electromagnetic effects. The AdS boundary conditions introduce a confining potential that fundamentally alters photon propagation compared to asymptotically flat scenarios \cite{isz03}, while the EH corrections modify the electromagnetic field distribution in ways that can significantly impact light deflection angles and capture cross-sections.

For the spherically symmetric, static spacetime described by Eq. (\ref{bb1}), we restrict our analysis to the equatorial plane $\theta=\pi/2$ without loss of generality. The Lagrangian density governing photon motion takes the standard form:
\begin{equation}
   \mathcal{L}=\frac{1}{2}\,g_{\mu\nu}\,\frac{dx^{\mu}}{d\lambda}\,\frac{dx^{\nu}}{d\lambda},\label{cc1}
\end{equation}
where $\lambda$ represents an affine parameter along the geodesic, and $g_{\mu\nu}$ denotes the metric tensor components.

The null geodesic condition, derived from our metric (\ref{bb1}), yields:
\begin{equation}
   -h(r)\,\dot{t}^2+\frac{\dot{r}^2}{h(r)}+r^2\,\dot{\phi}^2=0.\label{cc2}
\end{equation}

The spherical symmetry and static nature of our spacetime ensure the existence of two Killing vector fields: the temporal $\xi_{(t)} \equiv \partial_{(t)}$ and rotational $\xi_{(\phi)} \equiv \partial_{(\phi)}$. These symmetries give rise to conserved quantities - the energy $\mathrm{E}$ and angular momentum $\mathrm{L}$ - which govern the geodesic motion according to:
\begin{eqnarray}
   &&\dot{t}=\frac{\mathrm{E}}{h(r)},\label{cc3}\\
   &&\dot{\phi}=\frac{\mathrm{L}}{r^2},\label{cc4}\\
   &&\dot{r}=\sqrt{\mathrm{E}^2-V_\text{eff}(r)},\label{cc5}
\end{eqnarray}
where the effective potential for null geodesics is given by:
\begin{equation}
   V_\text{eff} (r)=\frac{\mathrm{L}^2}{r^2}\,h(r)=\frac{\mathrm{L}^2}{r^2}\,\left[1-\gamma-\frac{2\,M}{r}+\frac{r^2}{\ell^2_p}+\frac{Q^2}{r^2}-\frac{\alpha\,Q^4}{20\,r^6}+\frac{\beta}{r}\,\ln\left(\frac{r}{\beta}\right)\right].\label{cc6}
\end{equation}

\begin{figure}[ht!]
   \centering
   \includegraphics[width=0.32\linewidth]{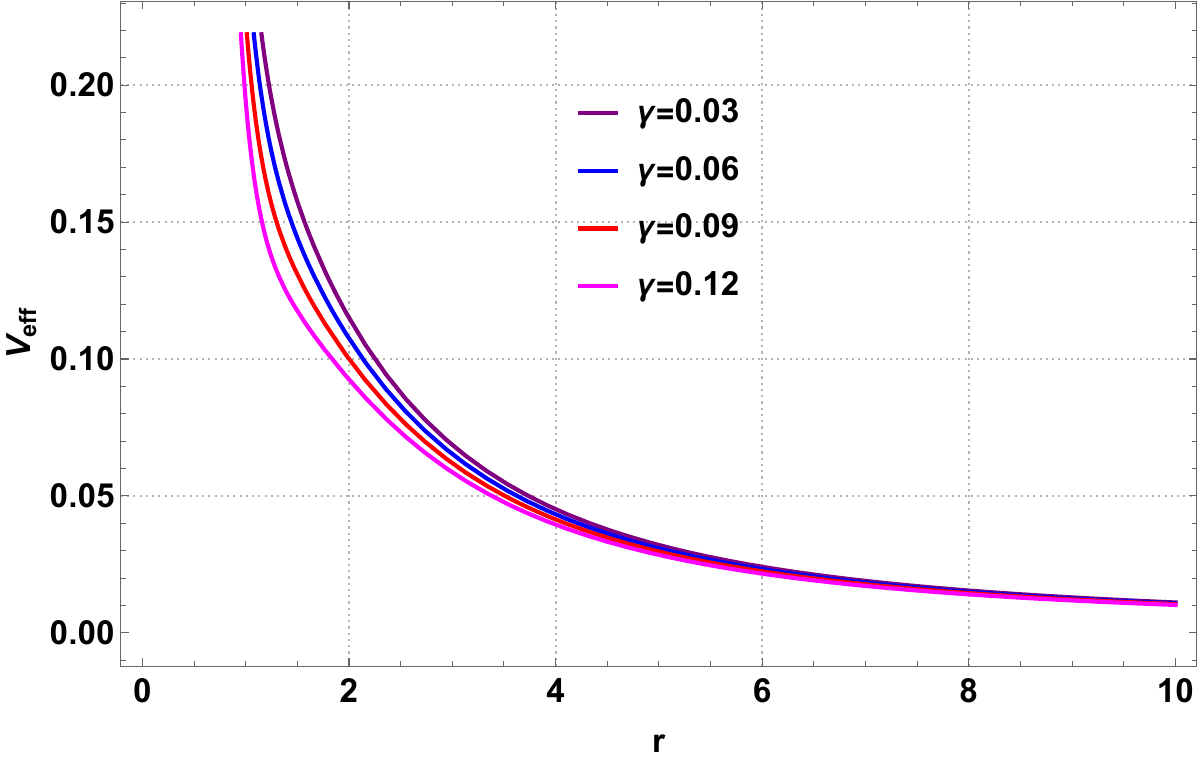}\quad
   \includegraphics[width=0.32\linewidth]{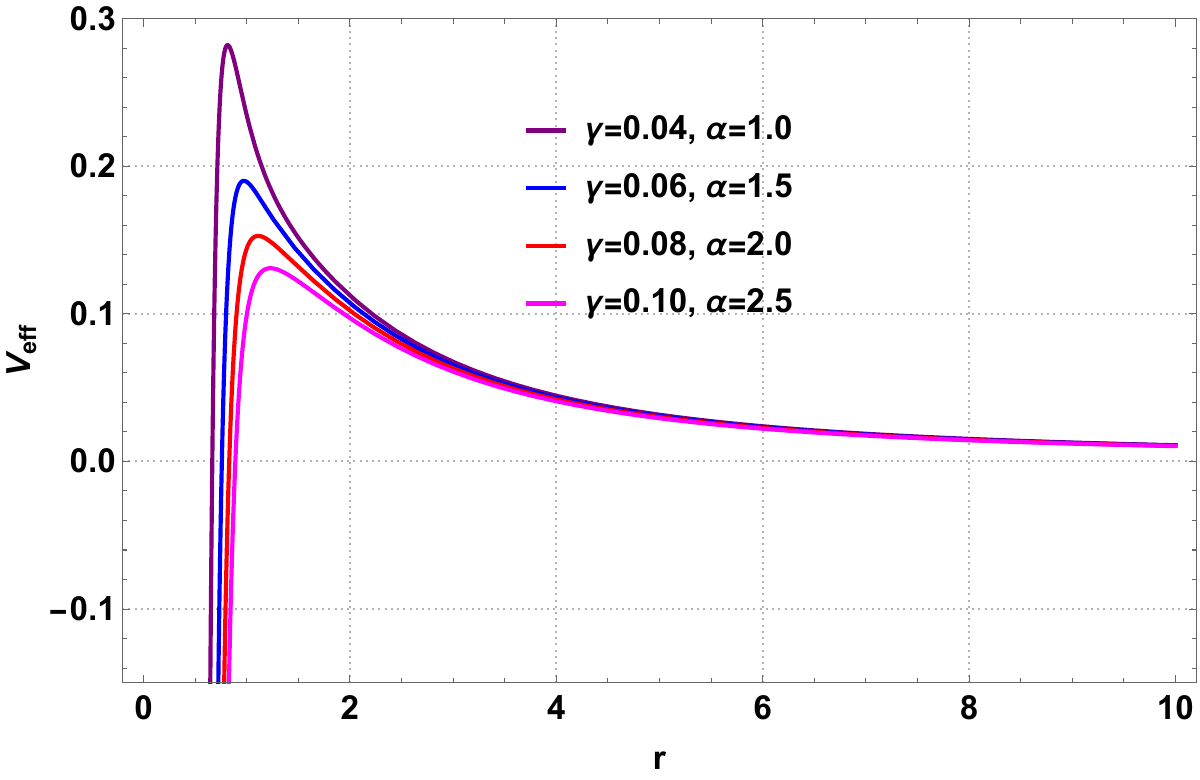}\quad
   \includegraphics[width=0.32\linewidth]{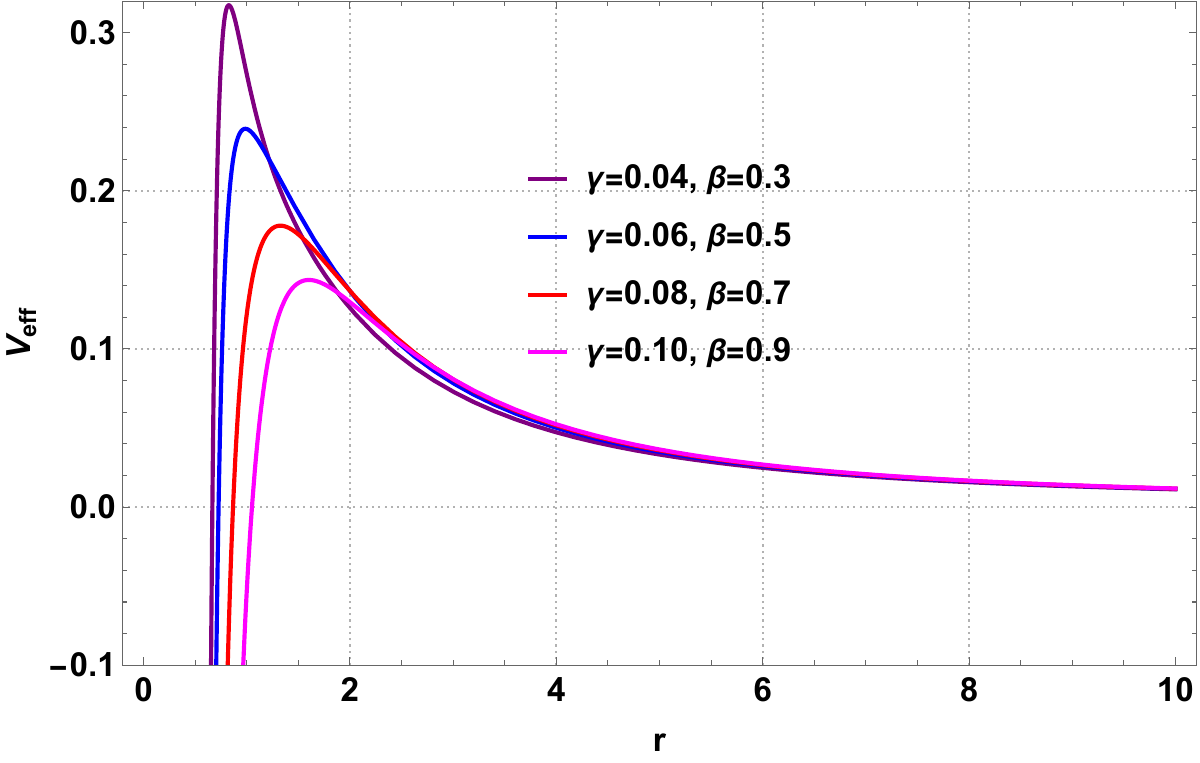}\\
   (i) $\alpha=0.2=\beta$ \hspace{4cm} (ii) $\beta=0.1$ \hspace{4cm} (iii) $\alpha=1$
   \caption{\footnotesize Radial profile of the effective potential $V_\text{eff}(r)$ for different values of CoS parameter $\gamma$, combination of CoS and CEH-AdS-BH parameters, \(\gamma,\alpha\), and combination of CoS and coupling constant of DM, \(\gamma,\beta\). Here, we set $M=1,\,\ell_p=20,\,\mathrm{L}=1,\,Q=1$.}
   \label{fig:effective-potential}
\end{figure}

The effective potential structure, illustrated in Fig. \ref{fig:effective-potential}, reveals the complex interplay between the various physical parameters. The CoS parameter $\gamma$ introduces a uniform shift in the potential, effectively modifying the global gravitational strength. The EH parameter $\alpha$ contributes high-order corrections that become significant at small radii, while the PFDM coupling $\beta$ introduces logarithmic modifications that can substantially alter the potential shape in intermediate regions.

From expression (\ref{cc6}), one can observe that the effective potential governing the dynamics of photon particles is influenced by geometric and physical parameters. These include string parameter $\gamma$, the electric charge $Q$, the BH mass $M$, CEH-AdS-BH parameter $\alpha$, the coupling parameter $\beta$ of PFDM, and the AdS radius $\ell_p$. Moreover, the conserved angular momentum $\mathrm{L}$ modifies this potential.

The orbital dynamics can be characterized through the equation of motion:
\begin{equation}
   \left(\frac{dr}{d\phi}\right)^2=\frac{\dot{r}^2}{\dot{\phi}^2}=r^4\,\left[\frac{1}{\mathrm{b^2}}-\frac{1}{r^2}\,\left\{1-\gamma-\frac{2\,M}{r}+\frac{r^2}{\ell^2_p}+\frac{Q^2}{r^2}-\frac{\alpha\,Q^4}{20\,r^6}+\frac{\beta}{r}\,\ln\left(\frac{r}{\beta}\right)\right\}\right],\label{cc7}
\end{equation}
where $\mathrm{b}$ represents the impact parameter. Through the transformation $u=1/r$, this reduces to:
\begin{equation}
   \left(\frac{du}{d\phi}\right)^2=\mathcal{K}(u),\label{cc8}
\end{equation}
with:
\begin{equation}
   \mathcal{K}(u)=\frac{1}{\mathrm{b^2}}+\frac{1}{\ell^2_p}-u^2\,\left\{1-\gamma-2\,M\,u+Q^2\,u^2-\frac{\alpha\,Q^4\,u^6}{20}-\beta\,u\,\ln(\beta\,u)\right\}.\label{cc9}
\end{equation}

Differentiating with respect to $\phi$ yields the highly nonlinear trajectory equation:
\begin{equation}
   \frac{d^2u}{d\phi^2}+(1-\gamma)\,u=\left\{\frac{\beta}{2}+3\,M+\frac{3\,\beta}{2}\,\ln(\beta\,u)\right\}\,u^2-2\,Q^2\,u^3+\frac{\alpha\,Q^4}{5}\,u^7.\label{cc10}
\end{equation}

Equation~(\ref{cc10}) is a highly nonlinear second-order differential equation that describes photon trajectories in the BH background. It is important to note that obtaining an exact analytical solution to this equation is, in general, not feasible. We observe that the photon trajectory is influenced by geometric and physical parameters. These include string parameter $\gamma$, the electric charge $Q$, the BH mass $M$, CEH-AdS-BH parameter $\alpha$, and the coupling parameter $\beta$ of DM.

In the PFDM-absent limit ($\beta=0$), the equation simplifies to:
\begin{equation}
   \frac{d^2u}{d\phi^2}+(1-\gamma)\,u=3\,M\,u^2-2\,Q^2\,u^3+\frac{\alpha\,Q^4}{5}\,u^7,\label{cc11}
\end{equation}
while for uncharged configurations ($Q=0$), it reduces to:
\begin{equation}
   \frac{d^2u}{d\phi^2}+(1-\gamma)\,u=\left[\frac{\beta}{2}+3\,M+\frac{3\,\beta}{2}\,\ln(\beta\,u)\right]\,u^2.\label{cc12}
\end{equation}

The effective radial force experienced by photons, derived from the effective potential gradient, takes the form:
\begin{equation}
   \mathrm{F}_\text{eff}=\frac{\mathrm{L}^2}{r^3}\,\left[1-\gamma-\frac{(3\,M+\beta/2)}{r}+\frac{2\,Q^2}{r^2}-\frac{\alpha\,Q^4}{5\,r^6}+\frac{3\,\beta}{2\,r}\,\ln\frac{r}{|\beta|}\right].\label{cc13}
\end{equation}
From expression (\ref{cc13}), one can observe that the effective radial force experienced by photon particles is altered by the geometric and physical parameters. These include string parameter $\gamma$, the electric charge $Q$, the BH mass $M$, CEH-AdS-BH parameter $\alpha$, and the coupling parameter $\beta$ of DM.

\begin{figure}[ht!]
   \centering
   \includegraphics[width=0.32\linewidth]{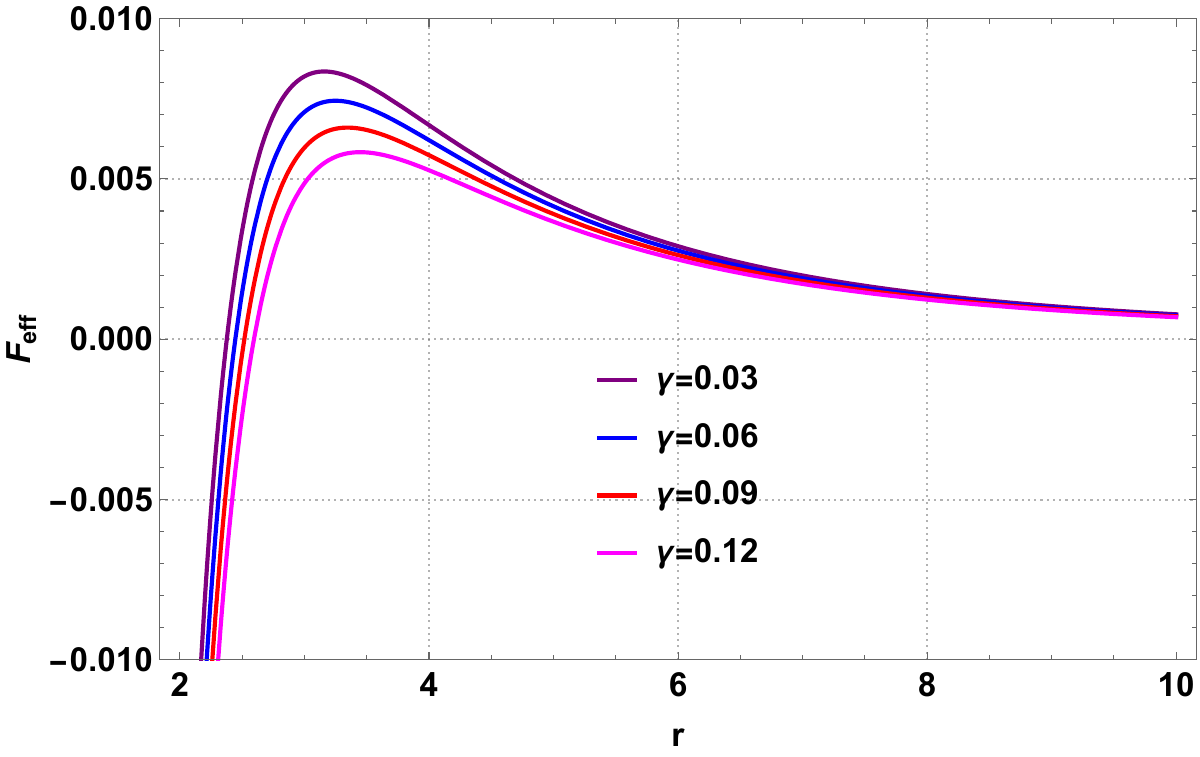}\quad
   \includegraphics[width=0.32\linewidth]{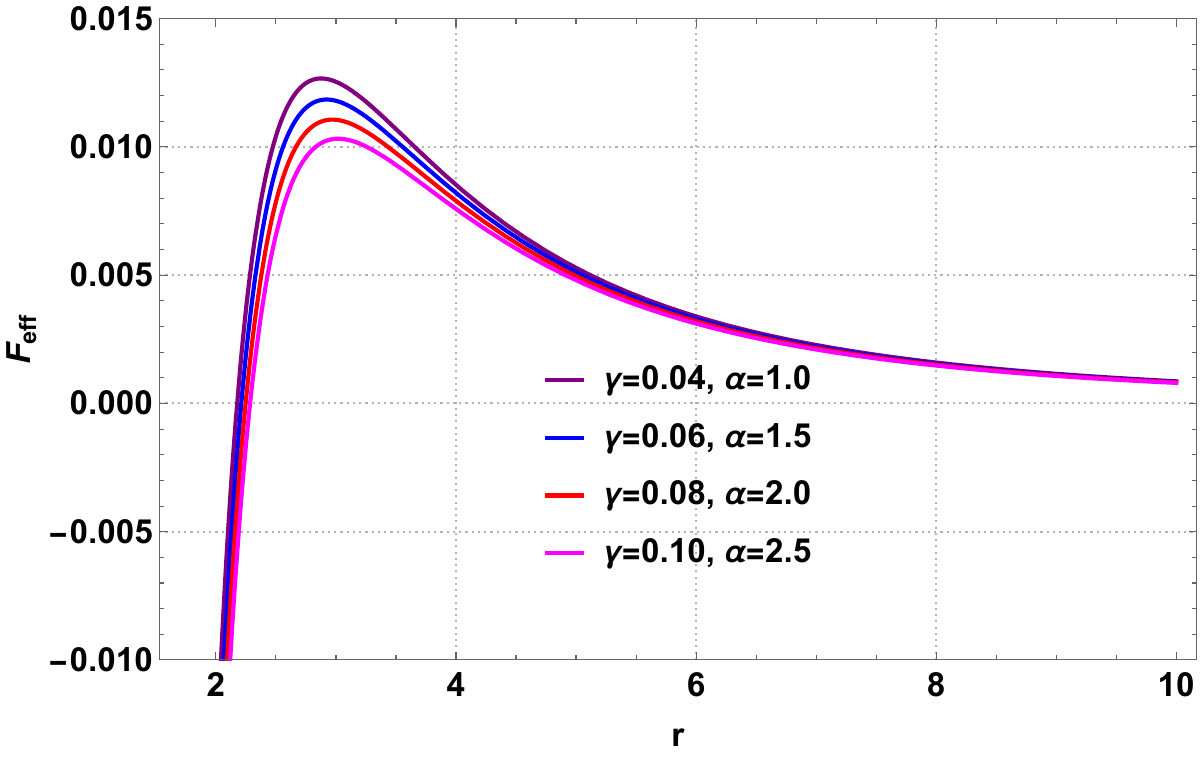}\quad
   \includegraphics[width=0.32\linewidth]{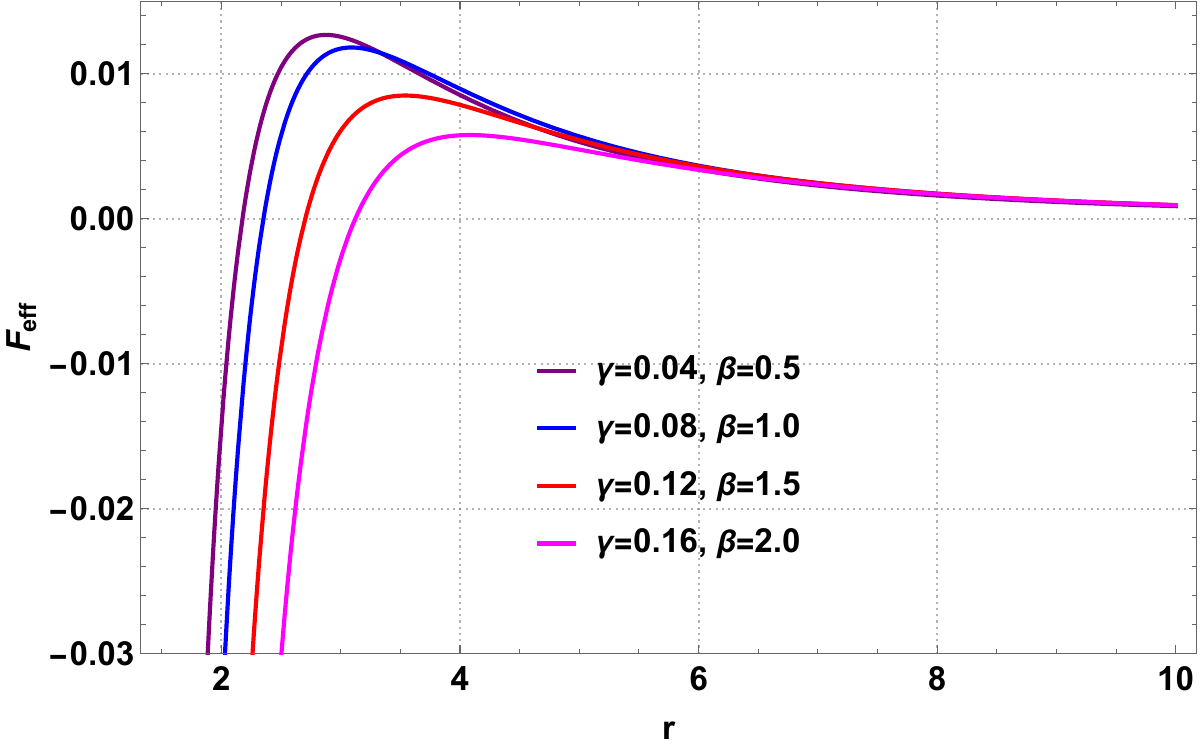}\\
   (i) $\alpha=0.2=\beta$ \hspace{4cm} (ii) $\beta=0.5$ \hspace{4cm} (iii) $\alpha=1$
   \caption{\footnotesize Radial profile of the effective force $F_\text{eff}(r)$ for different values of CoS parameter $\gamma$, combination of CoS and CEH-AdS-BH parameters, \(\gamma,\alpha\), and combination of CoS and coupling constant of DM, \(\gamma,\beta\). Here, we set $M=1,\,\ell_p=20,\,\mathrm{L}=1,\,Q=0.25$.}
   \label{fig:effective-force}
\end{figure}

Figure \ref{fig:effective-force} demonstrates how this force varies with different parameter combinations, revealing the intricate balance between attractive and repulsive contributions from various exotic matter components.

Next, we focus into the circular orbits and study photon sphere size. For circular null orbits of $r=r_\text{ph}$, the conditions $\dot{r}=0$ and $\ddot{r}=0$ must satisfied. Using Eq. (\ref{cc5}), these conditions simplify as follows:
\begin{equation}
   \mathrm{E}^2=V_\text{eff}\Big{|}_{r=r_\text{ph}}.\label{cc14}
\end{equation}
And
\begin{equation}
   \frac{dV_\text{eff}}{dr}\Big{|}_{r=r_\text{ph}}=0.\label{cc15}
\end{equation}
The first condition gives us the critical impact parameter for photon particles at radius $r=r_\text{ph}$ and is given by
\begin{equation}
   \mathrm{b}_c=\frac{r}{\sqrt{1-\gamma-\frac{2\,M}{r}+\frac{r^2}{\ell^2_p}+\frac{Q^2}{r^2}-\frac{\alpha\,Q^4}{20\,r^6}+\frac{\beta}{r}\,\ln\frac{r}{|\beta|}}}\Big{|}_{r=r_\text{ph}}.\label{cc16}
\end{equation}
The critical impact parameter for photon particles is modified by the string parameter $\gamma$, the electric charge $Q$, the BH mass $M$, CEH-AdS-BH parameter $\alpha$, the coupling parameter $\beta$ of DM, and the AdS radius $\ell_p$. Thus, the size of BH shadow size also influence by these parameters and hence modified the result in comparison to the standard charged BH solution.

The relation (\ref{cc15}) using the effective potential $V_\text{eff}(r)$ given in Eq.(\ref{cc6}) determines the radius of the photon sphere satisfying the following equation:
\begin{eqnarray}
   1-\gamma-\frac{(6\,M+\beta)}{2\,r}+\frac{2\,Q^2}{r^2}-\frac{\alpha\,Q^4}{5\,r^6}+\frac{3\,\beta}{2\,r}\,\ln\frac{r}{|\beta|}=0.\label{cc17}
\end{eqnarray}

From Eq.~(\ref{cc17}), we observe that the radius of photon sphere is altered by the geometric and physical parameters. These include string parameter $\gamma$, the electric charge $Q$, the BH mass $M$, CEH-AdS-BH parameter $\alpha$, and the coupling parameter $\beta$ of DM.

Noting that an analytical solution of the above equation (\ref{cc17}) is quite a challenging task. However, one can determine numerical values of photon sphere radius $r=r_\text{ph}$ by choosing suitable values of the geometric and physical parameters $\gamma, M, \alpha, \beta$ and $Q$.

\begin{table}[ht!]
\centering
\begin{tabular}{|c|c|c|}
\hline
\boldmath$\gamma$ & \boldmath$r_\text{ph}$ for $\alpha=0.25$ & \boldmath$r_\text{ph}$ for $\alpha=0.50$ \\ 
\hline
0.02 & 2.28611672 & 2.28611951 \\
\hline
0.04 & 2.32771744 & 2.32772004 \\
\hline
0.06 & 2.37094416 & 2.37094659 \\
\hline
0.08 & 2.41589666 & 2.41589891 \\
\hline
0.10 & 2.46268314 & 2.46268523 \\
\hline
0.12 & 2.51142120 & 2.51142313 \\
\hline
0.14 & 2.56223883 & 2.56224062 \\
\hline
0.16 & 2.61527564 & 2.61527730\\
\hline
0.18 & 2.67068421 & 2.67068574\\
\hline
0.20 & 2.72863163 & 2.72863303\\
\hline
\end{tabular}
\caption{\footnotesize Numerical values of the photon sphere radius $r_\text{ph}$ for various values of $\gamma$ with $\alpha=0.25$ and $\alpha=0.50$. Constants used: $M=1$, $\beta=0.25$, $Q=0.25$.}
\label{tab:1}
\end{table}

\begin{table}[ht!]
\centering
\begin{tabular}{|c|c|c|}
\hline
\boldmath$\gamma$ & \boldmath$r_\text{ph}$ for $\alpha=0.25$ & \boldmath$r_\text{ph}$ for $\alpha=0.50$ \\ 
\hline
0.02 & 2.12954433 & 2.12961254\\
\hline
0.04 & 2.17103929 & 2.17110238 \\
\hline
0.06 & 2.21415132 & 2.21420960 \\
\hline
0.08 & 2.25897978 & 2.25903354 \\
\hline
0.10 & 2.30563245 & 2.30568196 \\
\hline
0.12 & 2.35422644 & 2.35427196 \\
\hline
0.14 & 2.40488923 & 2.40493102 \\
\hline
0.16 & 2.45775989 & 2.45779818 \\
\hline
0.18 & 2.51299039 & 2.51302542\\
\hline
0.20 & 2.57074718 & 2.57077916\\
\hline
\end{tabular}
\caption{\footnotesize Numerical values of the photon sphere radius $r_\text{ph}$ for various values of $\gamma$ with $\alpha=0.25$ and $\alpha=0.50$. Constants used: $M=1$, $\beta=0.25$, $Q=0.50$.}
\label{tab:2}
\end{table}

\begin{table}[ht!]
\centering
\begin{tabular}{|c|c|c|c|c|c|}
\hline
\multirow{2}{*}{\boldmath$\gamma$} & \multicolumn{2}{c|}{\(\alpha=0.2\)} & \multicolumn{2}{c|}{\(\alpha=0.4\)} \\ \cline{2-5}
& \boldmath$r_{\text{ph}}$ & \boldmath$R_s$ & \boldmath$r_{\text{ph}}$ & \boldmath$R_s$ \\ \hline
0.02 & 2.371782673 & 3.91793 & 2.371783453 & 3.91793 \\
0.04 & 2.416058701 & 4.02731 & 2.416059427 & 4.02731 \\
0.06 & 2.462098429 & 4.14184 & 2.462099104 & 4.14184 \\
0.08 & 2.510011621 & 4.26185 & 2.510012247 & 4.26185 \\
0.10 & 2.559917442 & 4.38770 & 2.559918021 & 4.38770 \\
0.12 & 2.611945498 & 4.51979 & 2.611946034 & 4.51979 \\
0.14 & 2.666237013 & 4.65853 & 2.666237507 & 4.65853 \\
0.16 & 2.722946165 & 4.80438 & 2.722946621 & 4.80438 \\
0.18 & 2.782241627 & 4.95783 & 2.782242045 & 4.95783 \\
0.20 & 2.844308312 & 5.11942 & 2.844308696 & 5.11942 \\
\hline
\end{tabular}
\caption{\footnotesize Numerical values of the photon sphere radius $r_\text{ph}$ and shadow radius $R_s$ for various values of $\gamma$ with $\alpha=0.2$ and $\alpha=0.4$. Constants used: $M=1$, $\beta=0.2$, $Q=0.2$.}
\label{tab:3}
\end{table}

Finally, we determine the BH shadow size and examine how various geometrical parameters influence the radius of the shadow. The BH shadow radius refers to the apparent angular size of the dark region that a BH casts against a background of light, caused by the extreme gravitational lensing near the BH \cite{Sucu:2025fwa,Ahmed:2025vww,Jusufi:2018kmk,Ahmed:2025did}. This dark region arises due to the deflection of light rays around the BH, and is bounded by the so-called photon sphere-a spherical surface where photons can travel in unstable circular orbits under the influence of the BH's gravitational field.

\begin{figure}[ht!]
   \centering
   \includegraphics[width=0.32\linewidth]{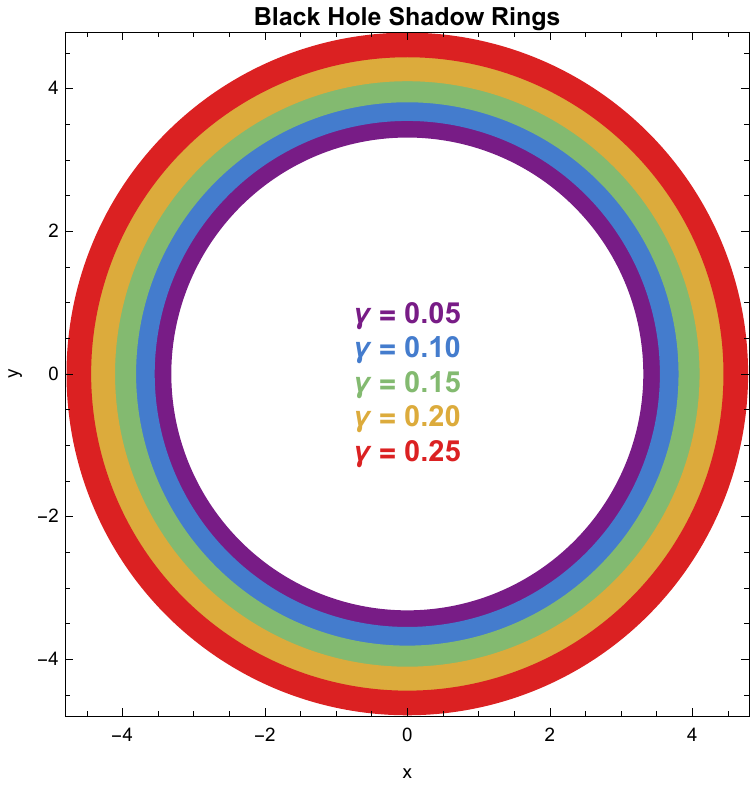}\quad
   \includegraphics[width=0.32\linewidth]{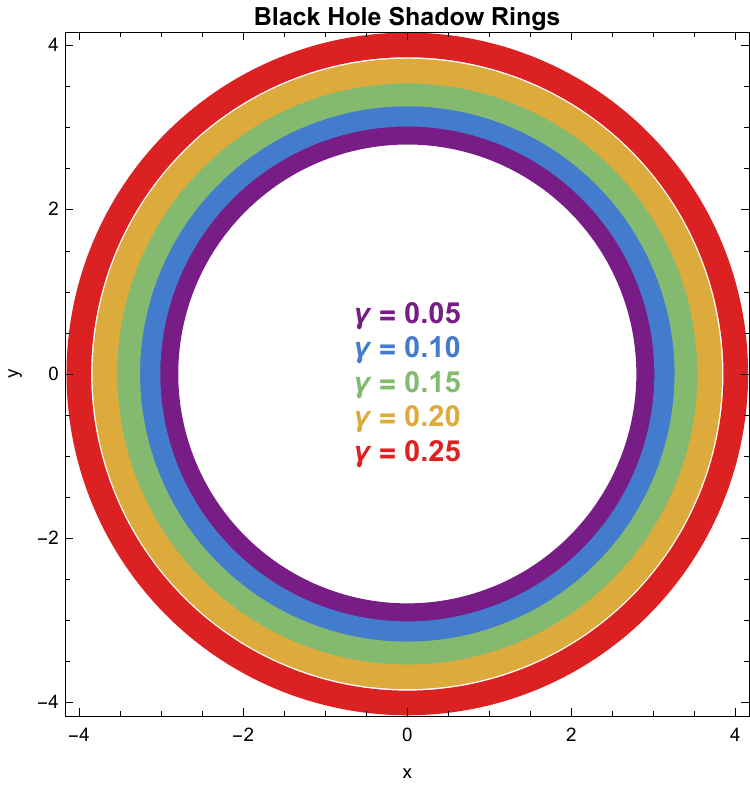}\quad
   \includegraphics[width=0.32\linewidth]{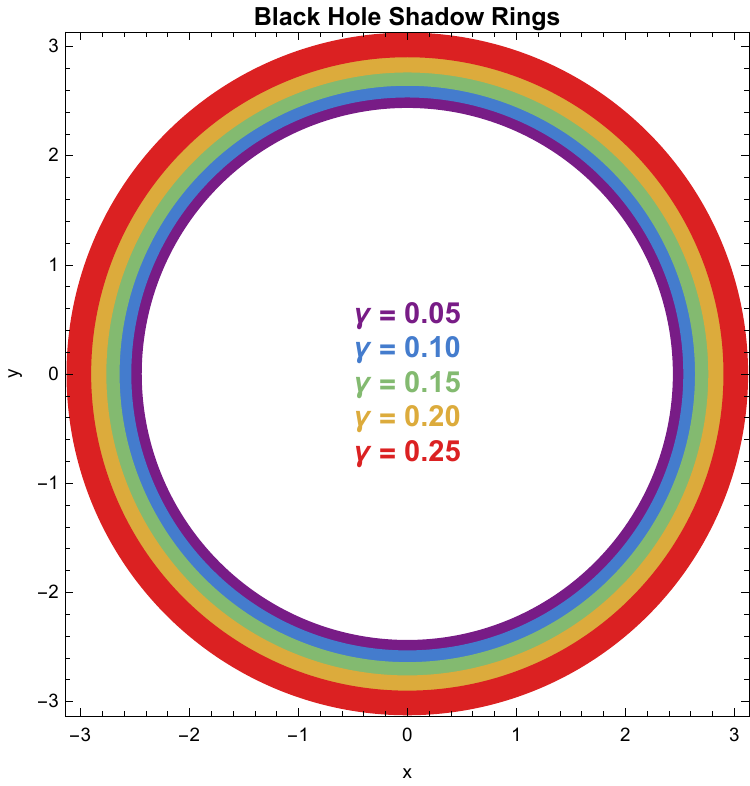}\\
   (a) $Q=0.25$ \hspace{4cm} (b) $Q=0.75$ \hspace{4cm} (c) $Q=1.25$
   \caption{\footnotesize Effects of electric charge $Q$ and CoS parameter $\gamma$ on BH shadow size. Here, $M=1,\,\ell_p=25,\,\alpha=0.2,\,\beta=0.5$}
   \label{fig:shadow}
\end{figure}

Photons slightly perturbed inward from this radius spiral into the event horizon, while those perturbed outward escape to infinity. The boundary between these two behaviors, when projected onto the observer's sky, forms the black hole shadow.

The radius of the shadow corresponds to the critical impact parameter $\mathrm{b}_c$, which is the maximum impact parameter for photons that are captured by the BH rather than escaping. This critical value is directly linked to the photon sphere radius $r=r_\text{ph}$. For non-rotating BH solution, this size can be determined using the following formula:
\begin{equation}
   R_s=\mathrm{b}_c=\left(\frac{\mathrm{L}}{\mathrm{E}}\right)\Big{|}_{r=r_\text{ph}}.\label{cc18}
\end{equation}

Using Eq. (\ref{cc16}), we find the following expression for shadow radius
\begin{equation}
   R_s=\frac{r_\text{ph}}{\sqrt{1-\gamma-\frac{2\,M}{r_\text{ph}}+\frac{r_\text{ph}^2}{\ell^2_p}+\frac{Q^2}{r_\text{ph}^2}-\frac{\alpha\,Q^4}{20\,r_\text{ph}^6}+\frac{\beta}{r_\text{ph}}\,\ln\frac{r_\text{ph}}{|\beta|}}}.\label{cc19}
\end{equation}
Correspondingly, the angular radius $\vartheta_s$ of the shadow observed at a distance $D$ from the BH is given by:
\begin{equation}
   \vartheta_\text{s}=\frac{R_s}{D}.\label{cc20}
\end{equation}

From Eq.~(\ref{cc19}), we observe that the size of the shadow cast by the BH is altered by the string parameter $\gamma$, electric charge $Q$, BH mass $M$, CEH-AdS-BH parameter $\alpha$, and the coupling parameter $\beta$ of the DM. In Table \ref{tab:3}, we present numerical values of the shadow size by varying string parameter $\gamma$, the CEH-AdS-BH parameter $\alpha$, while keeping other parameters fixed. Understanding these dependencies is essential for interpreting observations such as those made by the Event Horizon Telescope (EHT), which aim to constrain BH properties through direct imaging of their shadows.

In Fig. \ref{fig:shadow}, we present the shadow rings cast by the black hole (BH) for varying values of the electric charge $Q$ and CoS parameter $\gamma$, while keeping all other parameters fixed. Our observations indicate that, for a fixed value of electric charge, the size of the shadow rings increases with the CoS parameter $\gamma$. This occurs because the cosmic string parameter $\gamma$ effectively reduces the gravitational binding, leading to larger photon sphere radii and consequently larger shadow sizes. Conversely, for a given value of $\gamma$, the shadow ring size decreases as the electric charge $Q$ increases due to enhanced electromagnetic repulsion effects.

\subsection{Motion of neutral particle}

Massive neutral particles follow time-like geodesics influenced by an effective potential shaped by gravitational attraction, the curvature effects associated with BH charge, and the asymptotic structure imposed by the AdS boundary conditions. This potential governs the existence and stability of circular orbits, including both stable and unstable configurations. Of particular interest is the innermost stable circular orbit (ISCO), whose radius depends sensitively on the BH's parameters. The inclusion of electric charge and the negative cosmological constant in the AdS background can significantly shift the ISCO radius compared to the uncharged or asymptotically flat scenarios.

Furthermore, when the BH solution is coupled with a cloud of strings and surrounded by perfect fluid dark matter, the resulting modifications to the space-time geometry introduce additional contributions to the effective potential. These contributions further alter the dynamics of test particles, specifically impacting the ISCO radius and the overall orbital structure. Understanding these modifications is crucial for accurately modeling particle motion and extracting potential astrophysical signatures in such extended gravitational backgrounds.

The motion of a neutral test particle in the curved spherically symmetric space-time can be analyzed using the Hamiltonian approach, as discussed in Refs. \cite{sec3is09,sec3is10,sec3is13,sec3is14,sec3is17,sec3is18,sec3is20} given by
\begin{equation}
   \mathrm{H}=\frac{1}{2}\,g^{\mu\nu}\,p_{\mu}\,p_{\nu}+\frac{1}{2}\,m^2,\label{ss1}
\end{equation}
where $m$ is the mass of neutral particles, $p^{\mu}=m\,u^{\mu}$ is the four-momentum, $u^{\mu}=dx^{\mu}/d\tau$ is the four-velocity equation, and $\tau$ is the appropriate time of the neutral particle. Also, the Hamilton equations of motion are given: 
\begin{equation}
   \frac{dx^{\mu}}{d\lambda}\equiv m\,u^{\mu}=\frac{dH}{dp_{\mu}},\label{ss2}
\end{equation}
and
\begin{equation}
   \frac{dp_{\mu}}{d\lambda}=-\frac{\partial H}{\partial x^{\mu}},\label{ss3}
\end{equation}
where the affine parameter is given by $\lambda=\tau/m$.

As stated earlier, the chosen BH solution is static and spherically symmetric. Consequently, the metric tensor $g_{\mu\nu}$ is independent of the cyclic coordinates, namely the temporal coordinate $t$ and the azimuthal coordinate $\phi$. Due to this symmetry, the corresponding components of the particle's four-momentum, specifically $p_t$ and $p_{\phi}$, are conserved along its geodesics. These conserved quantities are given by
\begin{eqnarray}
   &&\frac{p_t}{m}=-h(r)\,\dot{t}=-\mathcal{E},\label{ss4}\\
   &&\frac{p_{\phi}}{m}=r^2\,\sin^2 \theta\,\dot{\phi}=\mathcal{L}_0,\label{ss5} 
\end{eqnarray}
where $\mathcal{E}=\mathrm{E}/\mu$ and $\mathcal{L}_0=\mathrm{L}/\mu$, respectively are the specific energy and angular momentum per unit mass of the neutral test particles. Moreover, the conjugate momentum associated with $\theta$ coordinate is given by
\begin{equation}
   \frac{p_{\theta}}{m}=r^2\,\dot{\theta}.\label{ss6}
\end{equation}
The components of the four-velocity $u^{\mu}$ in the time \(u^t\), azimuthal \(u^{\phi} \) and angular \(u^{\theta}\) directions obey the governing equations given in the following forms:
\begin{align}
   \dot{t}&=\frac{\mathcal{E}}{h(r)},\label{ss6a}\\
   \dot{\phi}&=\frac{\mathcal{L}_0}{r^2\,\sin^2 \theta},\label{ss6bb}\\
   \dot{\theta}&=\frac{p_{\theta}}{m\,r^2}.\label{ss6cc}
\end{align}

Moreover, massive test particles follow the time-like trajectories and thereby using the normalization condition $g_{\mu\nu}\,u^{\mu}\,u^{\nu}=-1$ and our metric (\ref{bb1}), we find
\begin{equation}
   -h(r)\,\dot{t}^2+\frac{1}{h(r)}\,\dot{r}^2+r^2\,\sin^2\theta\,\dot{\phi}^2+r^2\,\dot{\theta}^2=-1.\label{ss7}
\end{equation}
Substituting $\dot{t}$, $\dot{\theta}$ and $\dot{\phi}$, we finally arrive
\begin{equation}
   -\frac{\mathcal{E}^2}{h(r)}+\frac{\dot{r}^2}{h(r)}+\frac{\mathcal{L}^2_0}{r^2\,\sin^2 \theta}+\frac{p^2_{\theta}}{m^2\,r^2}=-1.\label{ss8}
\end{equation}
The above equation can be re-written as,
\begin{equation}
   \dot{r}^2+\left(\frac{\mathcal{L}^2_0}{r^2\,\sin^2 \theta}+1\right)\,h(r)+\frac{p^2_{\theta}}{m^2\,r^2}\,h(r)=\mathcal{E}^2.\label{ss9}
\end{equation}

With these, the Hamiltonian Eq. (\ref{ss1}) can be expressed as follows:
\begin{equation}
  H=\frac{1}{2}\,h(r)\,p^2_r+\frac{1}{2}\,\frac{p^2_{\theta}}{r^2}+\frac{1}{2}\,\frac{m^2}{h(r)}\,\left[U_\text{eff}(r,\theta)-\mathcal{E}^2\right],\label{ss9aa}
\end{equation}
where \(U_\text{eff}(r,\theta)\) is the effective potential and takes the following form:
\begin{equation}
   U_\text{eff}(r,\theta)=\left(\frac{\mathcal{L}^2_0}{r^2\,\sin^2 \theta}+1\right)\,h(r).\label{ss10a}
\end{equation}

In the equatorial plane defined by $\theta=\pi/2$ and $p_{\theta}=0$, the effective potential of time-like geodesic reduces as
\begin{equation}
   U_\text{eff}(r)=\left(\frac{\mathcal{L}^2_0}{r^2}+1\right)\,h(r).\label{ss10}
\end{equation}

\begin{figure}[ht!]
   \centering
   \includegraphics[width=0.32\linewidth]{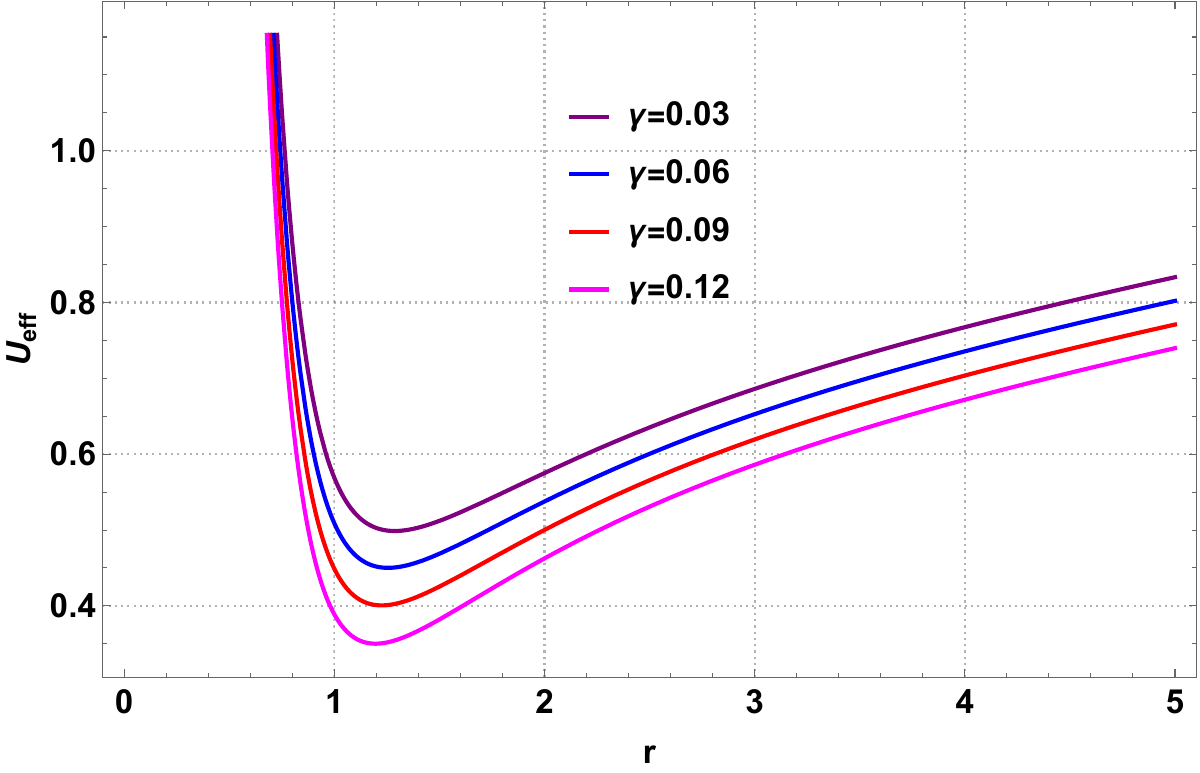}\quad
   \includegraphics[width=0.32\linewidth]{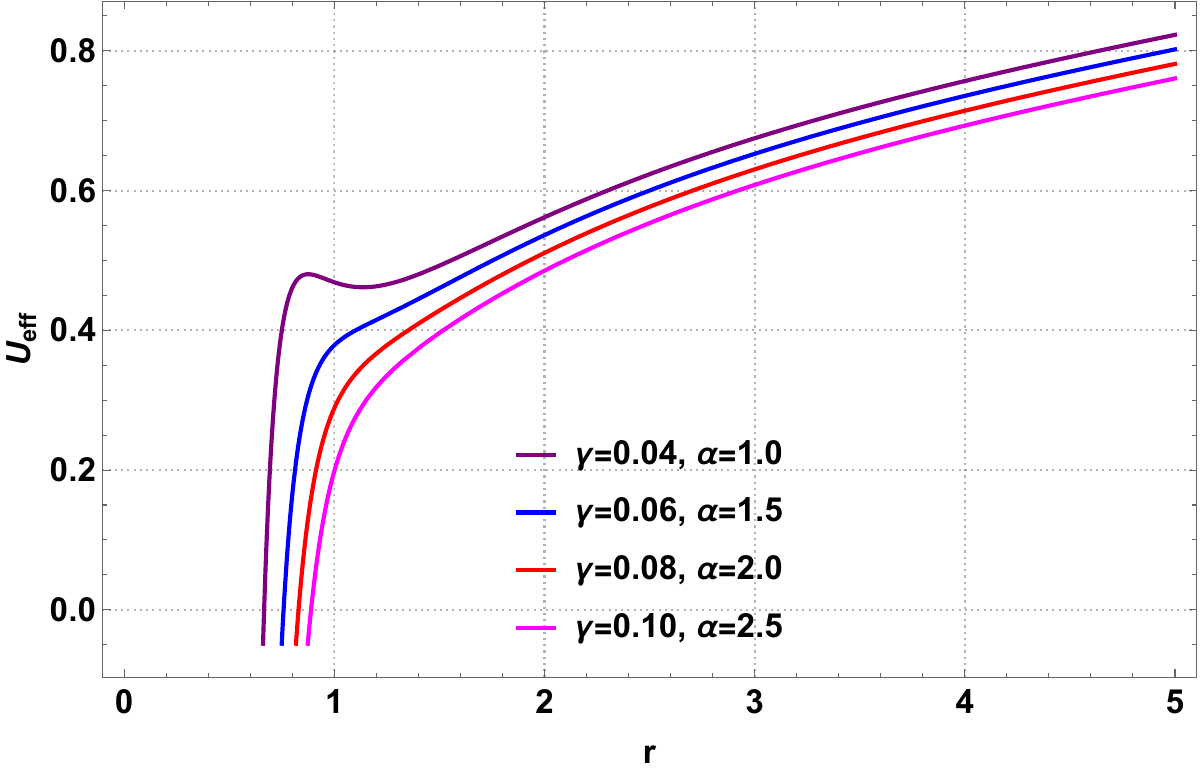}\quad
   \includegraphics[width=0.32\linewidth]{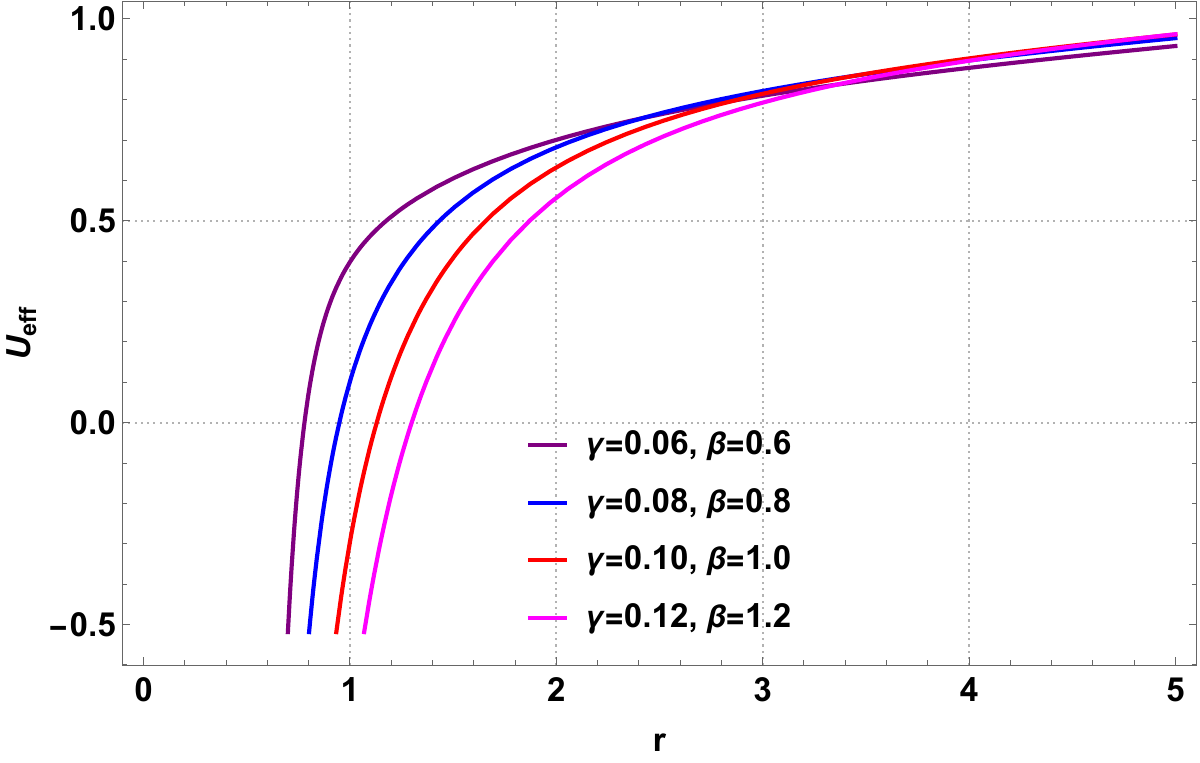}\\
   (i) $\alpha=0.2=\beta$ \hspace{4cm} (ii) $\beta=0.2$ \hspace{4cm} (iii) $\alpha=1$
   \caption{\footnotesize Radial profile of the effective potential $U_\text{eff}(r)$ of neutral test particles in the equatorial plane $\theta=\pi/2$ for different values of CoS parameter $\gamma$, combination of CoS and CEH-AdS-BH parameters, \((\gamma,\alpha)\), and combination of CoS and coupling constant of DM, (\(\gamma,\beta\)). Here, we set $M=1,\,\ell_p=20,\,\mathrm{L}=1,\,Q=1$.}
   \label{fig:effective-potential-neutral}
\end{figure}

From expression (\ref{ss10}), we observe that the effective potential governing the dynamics of neutral test particles is altered by the geometric and physical parameters. These include string parameter $\gamma$, the electric charge $Q$, the BH mass $M$, CEH-AdS-BH parameter $\alpha$, and the coupling parameter $\beta$ of DM. More the conserved angular momentum $\mathcal{L}_0$ and mass $m$ of the test particles also modifies this potential.

Next, we determine the effective radial force experiences by the neutral test particles in the gravitational field of a BH, indicating whether the particle is attracted toward or repelled from the BH. We compute the effective radial force using the effective potential given in Eq.~(\ref{ss10}), which is given by (considering equatorial plane $\theta=\pi/2$):
\begin{eqnarray}
   \mathcal{F}_\text{eff}&=&-\frac{1}{2}\,\frac{\partial U_\text{eff}}{\partial r}\nonumber\\
   &=&\frac{\mathcal{L}^2_0}{r^3} \Bigg[1 - \gamma - \frac{(3M+\beta/2)}{r}+ \frac{2\,Q^2}{r^2} - \frac{\alpha\, Q^4}{5\, r^6} + \frac{3\,\beta}{2\,r} \ln\left( \frac{r}{\beta} \right)\Bigg] -\frac{1}{r}\,\left( \frac{r^2}{\ell_p^2} + \frac{M+\beta/2}{r} - \frac{Q^2}{r^2} + \frac{3\,\alpha\, Q^4}{20\, r^6} - \frac{\beta}{2\,r} \ln\left( \frac{r}{\beta} \right)\right).\label{ss17}
\end{eqnarray}

\begin{figure}[ht!]
   \centering
   \includegraphics[width=0.32\linewidth]{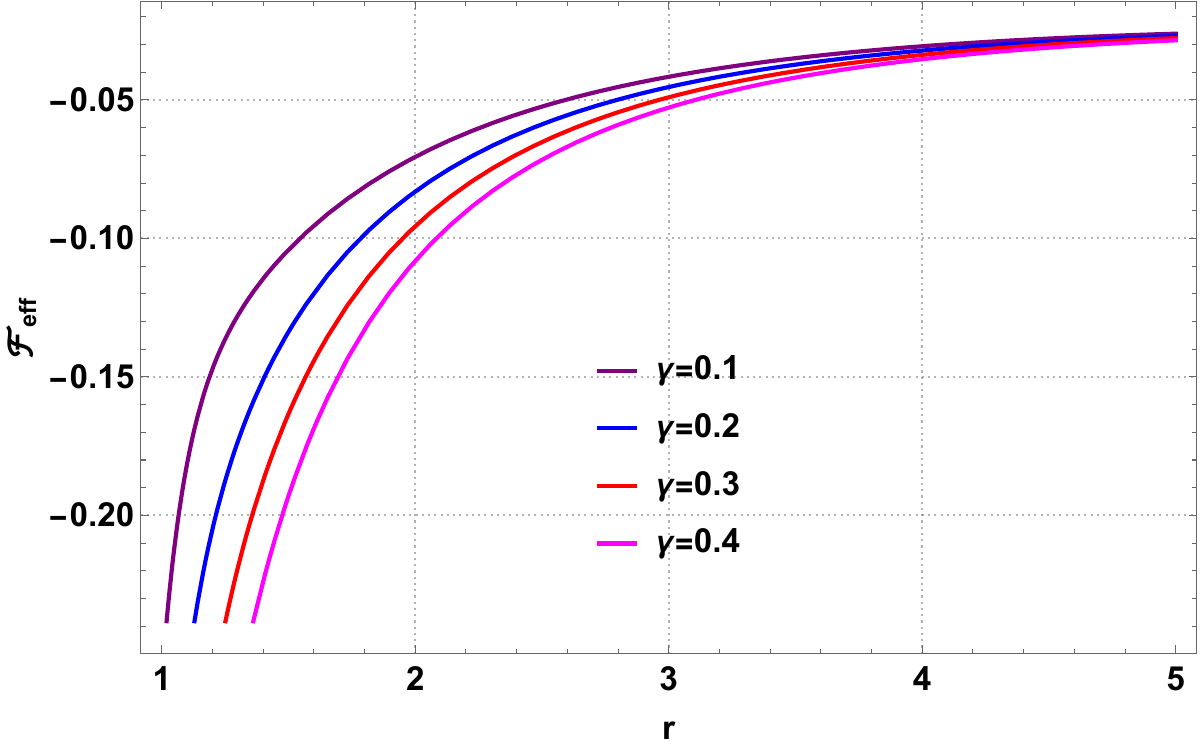}\quad
   \includegraphics[width=0.32\linewidth]{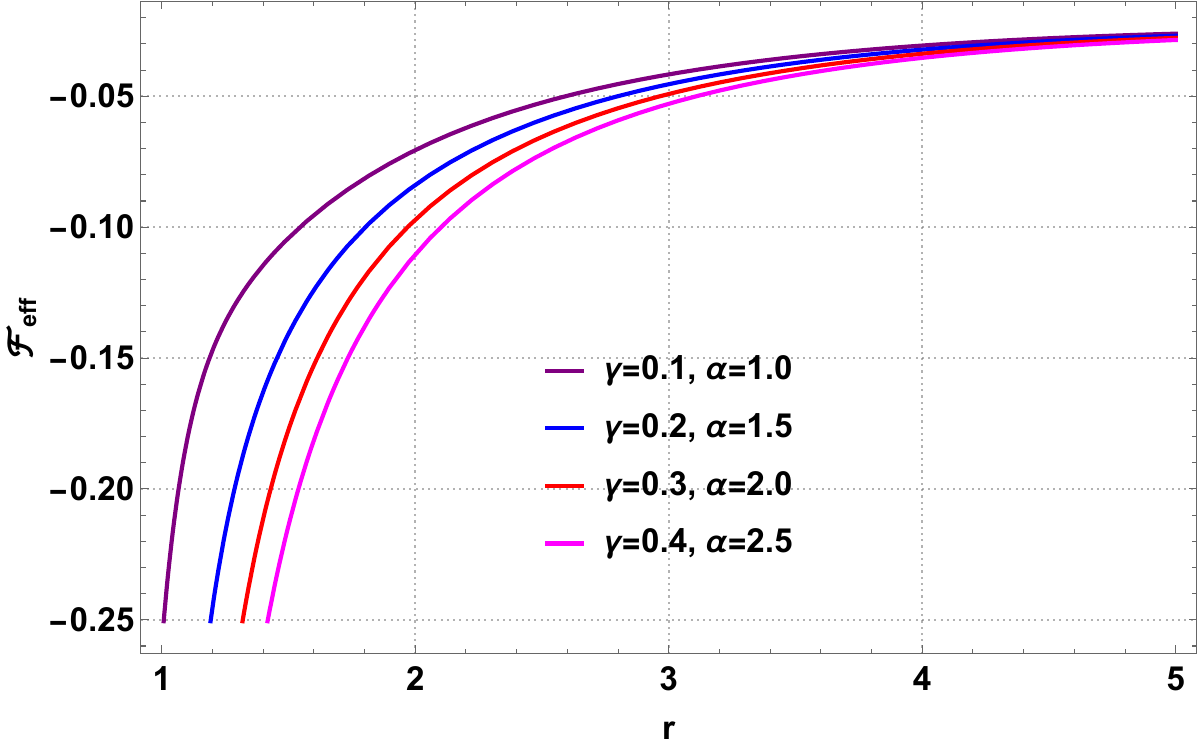}\quad
   \includegraphics[width=0.32\linewidth]{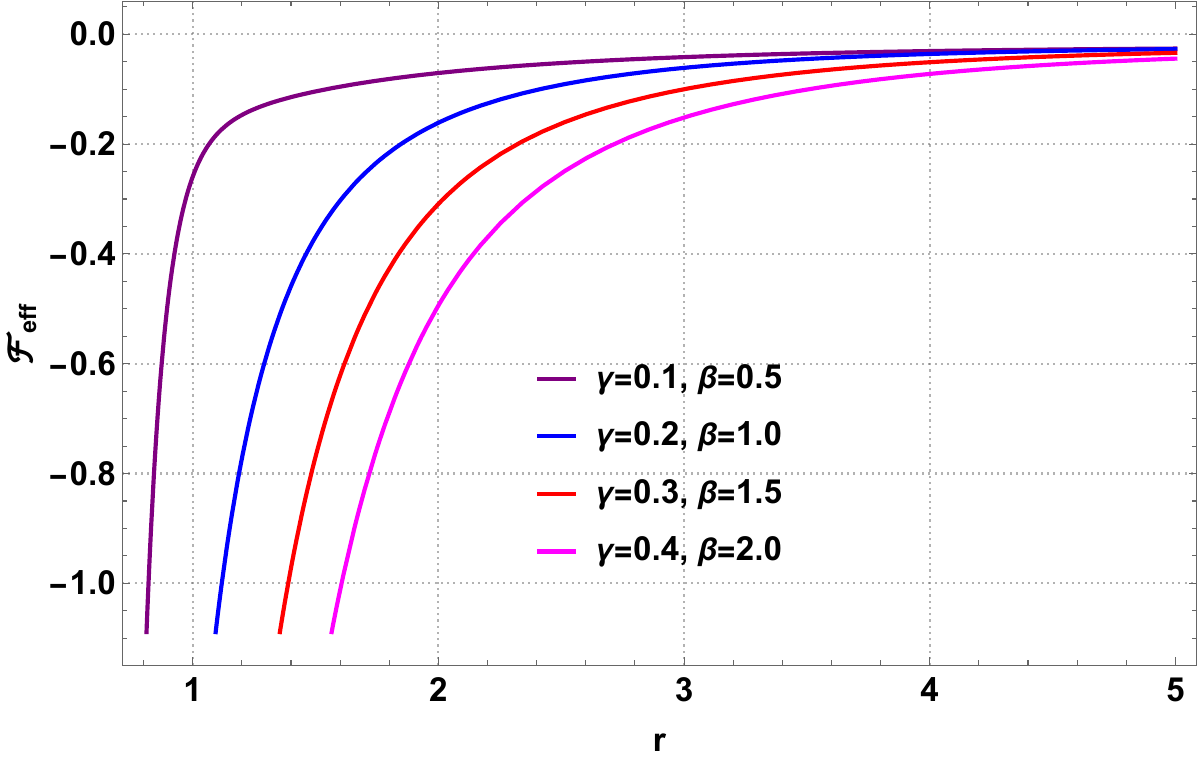}\\
   (i) $\alpha=1,\,\beta=0.5$ \hspace{4cm} (ii) $\beta=0.5$ \hspace{4cm} (iii) $\alpha=1$
   \caption{\footnotesize Radial profile of the effective force $\mathcal{F}_\text{eff}(r)$ experienced by neutral test particles for different values of CoS parameter $\gamma$, combination of CoS and CEH-AdS-BH parameters, \(\gamma,\alpha\), and combination of CoS and coupling constant of DM, \(\gamma,\beta\). Here, we set $M=1,\,\ell_p=20,\,\mathrm{L}=1,\,Q=0.25$.}
   \label{fig:effective-force-neutral}
\end{figure}

From expression (\ref{ss17}), one can observe that the effective radial force experienced by neutral test particles is altered by the geometric and physical parameters. These include string parameter $\gamma$, the electric charge $Q$, the BH mass $M$, CEH-AdS-BH parameter $\alpha$, and the coupling parameter $\beta$ of DM.

We now study motion of neutral particle in circular orbits in the equatorial plane, $\theta=\pi/2$ for which the conditions $\dot{r}=0$ and $\ddot{r}=0$ must be satisfied. These conditions implies the following two relations:
\begin{eqnarray}
   &&\mathcal{E}^2=U_\text{eff}(r),\label{ss11}\\
   &&U'_\text{eff}(r)=0.\label{ss12}
\end{eqnarray}

Simplifying these conditions leads to the determination of key physical quantities that characterize the motion of time-like neutral particles orbiting the BH. Specifically, these quantities are the specific angular momentum and specific energy of the particles. The specific angular momentum refers to the angular momentum per unit mass of the particle, while the specific energy represents the total energy per unit mass, both measured relative to an observer at infinity. These parameters are fundamental in describing stable and unstable orbits around the BH and are expressed mathematically by the following relations:
\begin{eqnarray}
   &&\mathcal{L}_0=\sqrt{\frac{r^3\,h'(r)}{2\,h-r\,h'}}=r\,\sqrt{\frac{\frac{r^2}{\ell_p^2} + \frac{M+\beta/2}{r} - \frac{Q^2}{r^2} + \frac{3\,\alpha\, Q^4}{20\, r^6} - \frac{\beta}{2\,r} \ln\left( \frac{r}{\beta} \right)}{1 - \gamma - \frac{(3M+\beta/2)}{r}+ \frac{2\,Q^2}{r^2} - \frac{\alpha\, Q^4}{5\, r^6} + \frac{3\,\beta}{2\,r} \ln\left( \frac{r}{\beta} \right)}},\label{ss13}\\
   &&\mathcal{E}_0=\pm\,\sqrt{\frac{2\,h^2(r)}{2\,h(r)-r\,h'(r)}}=\pm\,\frac{\left(1-\gamma-\frac{2\,M}{r}+\frac{r^2}{\ell^2_p}+\frac{Q^2}{r^2}-\frac{\alpha\,Q^4}{20\,r^6}+\frac{\beta}{r}\,\mbox{ln}\left(\frac{r}{\beta}\right)\right)}{\sqrt{1 - \gamma - \frac{(3M+\beta/2)}{r}+ \frac{2\,Q^2}{r^2} - \frac{\alpha\, Q^4}{5\, r^6} + \frac{3\,\beta}{2\,r} \ln\left( \frac{r}{\beta} \right)}}.\label{ss14}
\end{eqnarray}

\begin{figure}[ht!]
    \centering
    \includegraphics[width=0.32\linewidth]{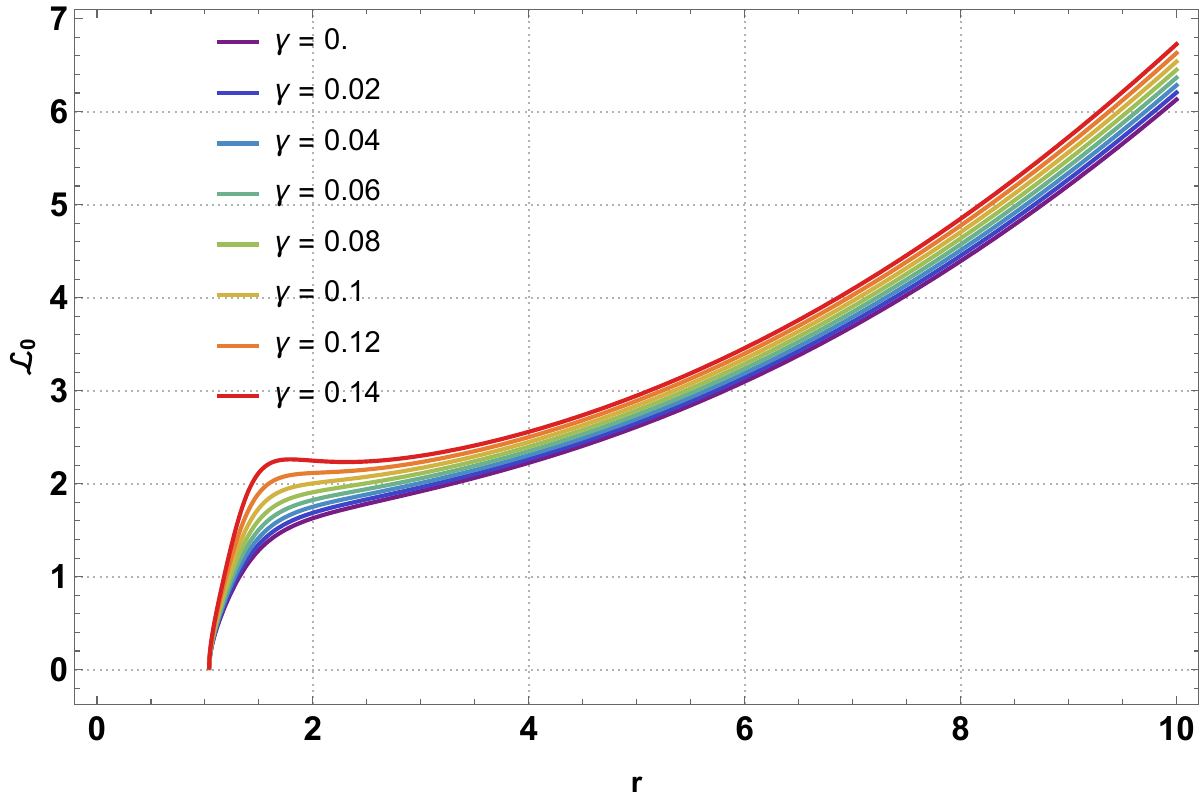}\quad
    \includegraphics[width=0.32\linewidth]{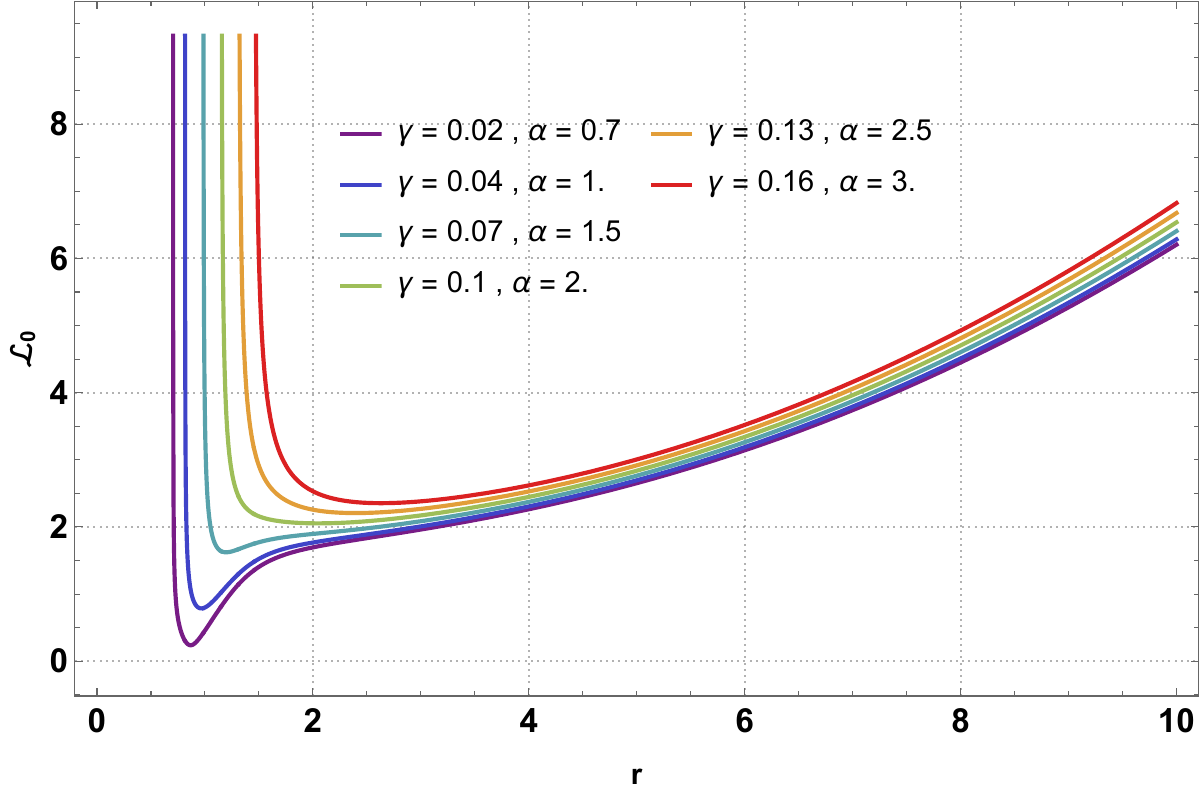}\quad
    \includegraphics[width=0.32\linewidth]{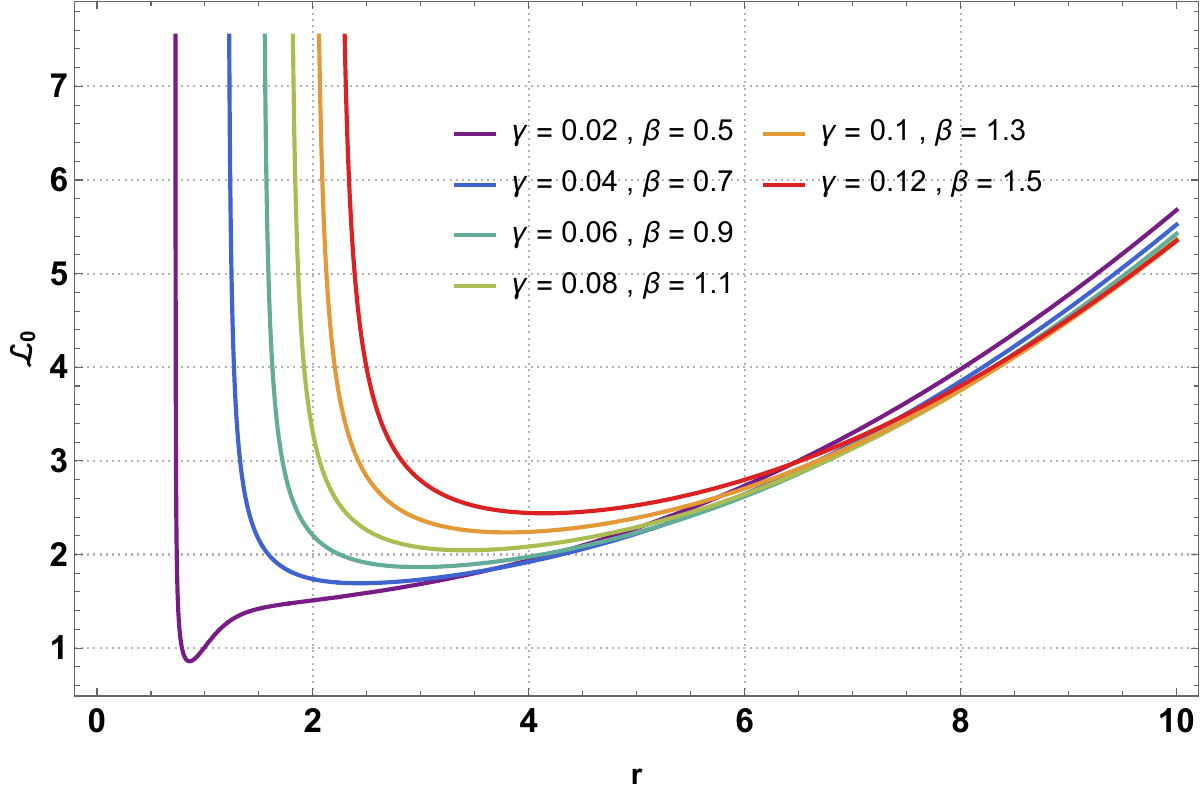}\\
    (i) $\alpha=0.2,\,\beta=0.2$ \hspace{4cm} (ii) $\beta=0.2$ \hspace{4cm} (iii) $\alpha=0.5$ 
    \caption{\footnotesize behavior of the specific angular momentum \(\mathcal{L}_0\) for different values of $\gamma,\,\alpha$ and $\beta$. Here $M=1\,\ell_p=20,\,Q=1$ }
    \label{fig:angular}
    \hfill\\
    \includegraphics[width=0.32\linewidth]{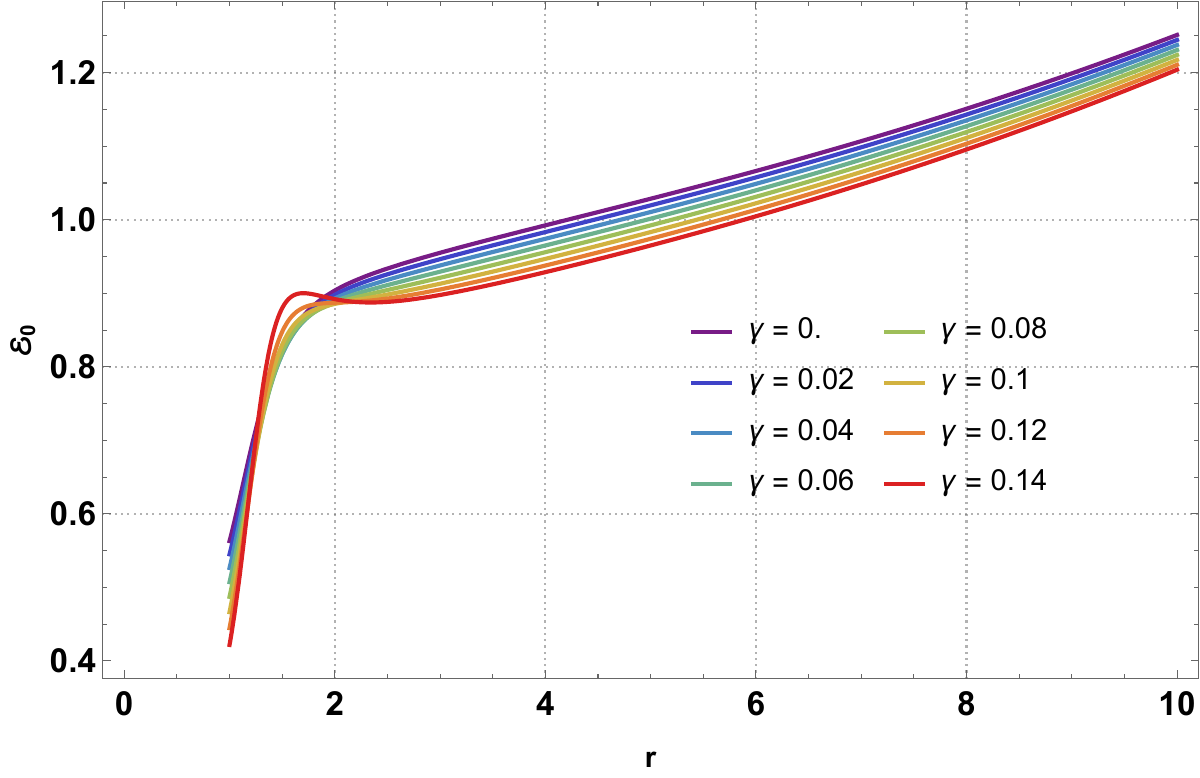}\quad
    \includegraphics[width=0.32\linewidth]{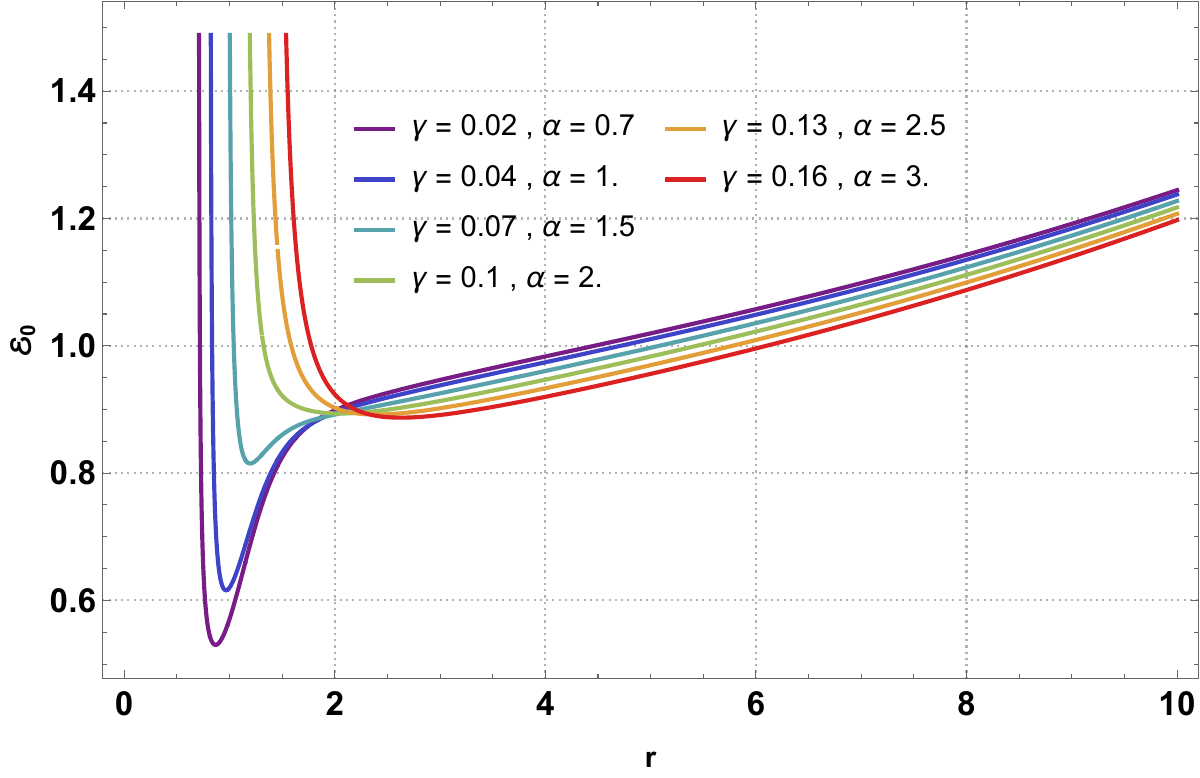}\quad
    \includegraphics[width=0.32\linewidth]{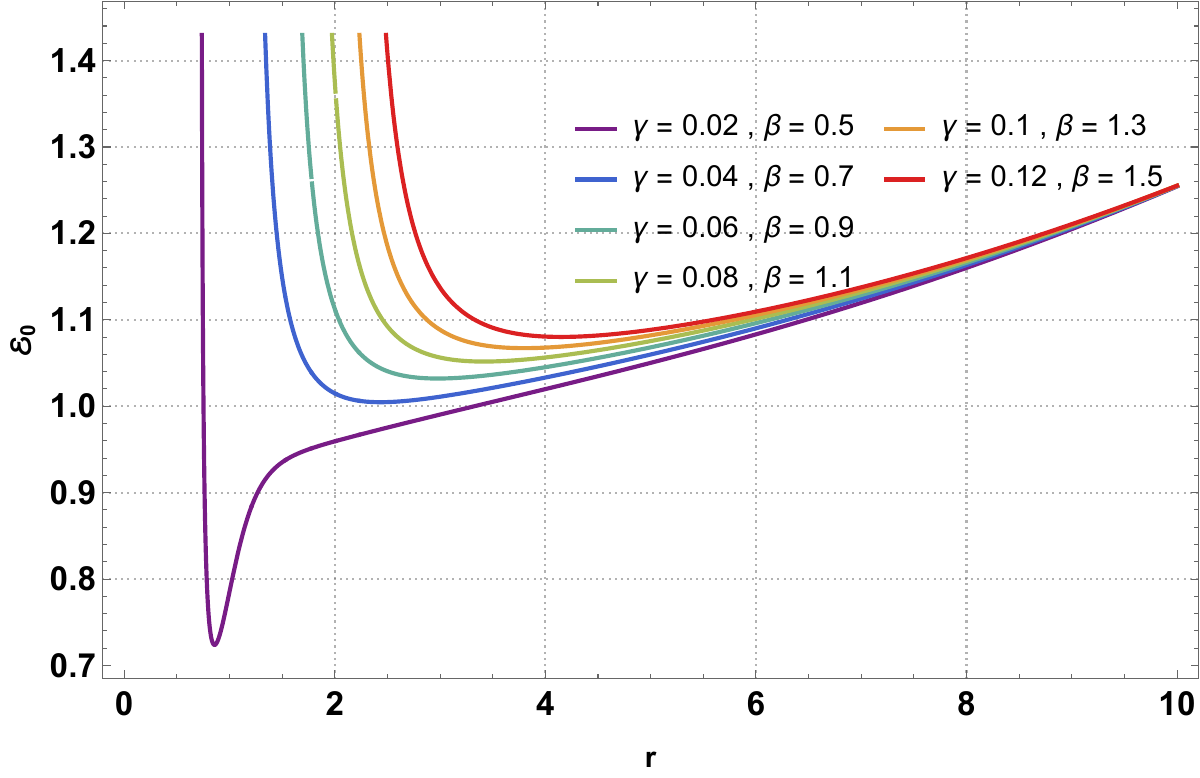}\\
    (i) $\alpha=0.2,\,\beta=0.2$ \hspace{4cm} (ii) $\beta=0.2$ \hspace{4cm} (iii) $\alpha=0.5$ 
    \caption{\footnotesize behavior of the specific energy \(\mathcal{E}_0\) for different values of $\gamma,\,\alpha$ and $\beta$. Here $M=1\,\ell_p=20,\,Q=1$ }
    \label{fig:energy}
\end{figure}

From expressions (\ref{ss13}) and (\ref{ss14}), we observe that the specific angular momentum and specific energy of the test particles is altered by the geometric and physical parameters. These include string parameter $\gamma$, the electric charge $Q$, the BH mass $M$, CEH-AdS-BH parameter $\alpha$, and the coupling parameter $\beta$ of DM. Moreover, the specific angular momentum is altered by the angular coordinate $\theta$.

We focus on an important physical properties of neutral particle. This is called the innermost stable circular orbit (ISCO). The stability of circular orbits is critical to studying test particle dynamics around BHs. Circular orbits are key indicators of the gravitational potential's nature and provide insights into the interplay between the CEH-AdS-BH deformation and the surrounding PFDM distribution including the cloud of strings. 

For a circular orbit to be stable, small perturbations in the radial direction should not lead to unbounded oscillations or escape trajectories \cite{isz01,sec3is11,sec3is12,sec3is15,sec3is16,sec3is19}. This can be determined by applying the following conditions to the effective potential \( U_\text{eff}(r) \):
\begin{align}
U_\text{eff}&=\mathcal{E}^2, \\
\frac{dU_\text{eff}}{dr} &= 0 \quad \text{(circular orbit condition)}, \\
\frac{d^2U_\text{eff}}{dr^2} &= 0 \quad \text{(marginal stability condition)}.\label{ss15}
\end{align}

Using the effective potential expression given in Eq. (\ref{ss10}), one can determine the position of ISCO. The expression of the last equation is as follows:
\begin{equation}
   r\,h(r)\,h''(r)-2\,r\,h'^2(r)+3\,h(r)\,h'(r)=0.\label{ss16}
\end{equation}
Substituting the metric function $h(r)$ and after simplification, one can find a polynomial equation of $r_\text{ISCO}$ as $f(r_\text{ISCO})=0$. Noted that analogue to the photon sphere radius, it is quite challenging task to determine the radius of ISCO. However, numerical values can be obtained by choosing suitable values of parameter $\gamma, M, \beta$ and $Q$. 

\begin{table}[h!]
\centering
\begin{tabular}{|c|c|c|}
\hline
\boldmath$\gamma$ & \boldmath$r_\text{ISCO}$ for $\alpha = 0.25$ & \boldmath$r_\text{ISCO}$ for $\alpha = 0.50$ \\
\hline
0.02 & 3.73053  & 3.73057  \\
\hline
0.04 & 3.77469  & 3.77473  \\
\hline
0.06 & 3.81997  & 3.82000  \\
\hline
0.08 & 3.86642  & 3.86645  \\
\hline
0.10 & 3.91411  & 3.91414  \\
\hline
0.12 & 3.96311  & 3.96314  \\
\hline
0.14 & 4.01349  & 4.01351  \\
\hline
0.16 & 4.06533  & 4.06536  \\
\hline
0.18 & 4.11873  & 4.11875  \\
\hline
0.20 & 4.17378  & 4.17380  \\
\hline
\end{tabular}
\caption{\footnotesize Numerical result for ISCO radius $r_\text{ISCO}$ for different values CoS parameter $\gamma$ with $\alpha=0.25$ and $\alpha=0.5$. Here, $M=1,\,\ell_p=25,\,\beta=0.5,\,Q=0.50$.}
\label{tab:4}
\end{table}

\begin{table}[ht!]
\centering
\begin{tabular}{|c|c|c|}
\hline
\boldmath$\gamma$ & \boldmath$r_\text{ISCO}$ for $\alpha = 0.25$ & \boldmath$r_\text{ISCO}$ for $\alpha = 0.50$ \\
\hline
0.02 & 3.24775 & 3.24824 \\
\hline
0.04 & 3.29697 & 3.29743 \\
\hline
0.06 & 3.34727 & 3.34769 \\
\hline
0.08 & 3.39870 & 3.39908 \\
\hline
0.10 & 3.45131 & 3.45166 \\
\hline
0.12 & 3.50517 & 3.50550 \\
\hline
0.14 & 3.56036 & 3.56066 \\
\hline
0.16 & 3.61694 & 3.61722 \\
\hline
0.18 & 3.67501 & 3.67526 \\
\hline
0.20 & 3.73466 & 3.73489 \\
\hline
\end{tabular}
\caption{\footnotesize Numerical result for ISCO radius $r_\text{ISCO}$ for different values CoS parameter $\gamma$ with $\alpha=0.25$ and $\alpha=0.5$. Here, $M=1,\,\ell_p=25,\,\beta=0.5,\,Q=0.75$.}
\label{tab:5}
\end{table}

In Tables~\ref{tab:4} and \ref{tab:5}, we present the numerical values of the ISCO radius corresponding to various values of CoS parameter \( \gamma \), for two choices of the CEH-AdS-BH deformation parameter, namely \( \alpha = 0.25 \) and \( \alpha = 0.50 \). The analysis is performed for two distinct values of the BH's electric charge: \( Q = 0.50 \) and \( Q = 0.75 \). Our results show a clear trend: for fixed values of \( \alpha \) and \( Q \), the ISCO radius increases with increasing \( \gamma \). This indicates that the presence of higher values of (\( \gamma \)) tends to push the ISCO outward, suggesting a weakening of the gravitational binding at smaller radii. Conversely, for a fixed \( \gamma \) and \( \alpha \), an increase in the electric charge \( Q \) leads to a decrease in the ISCO radius. This behavior can be attributed to the enhanced electrostatic repulsion due to the charge, which effectively counters the gravitational attraction and allows stable circular orbits to exist closer to the BH. This behavior can be attributed to the fact that CoS introduces angular deficit effects that effectively weaken the local gravitational field strength, requiring larger radii $r$ for stable circular orbits. The electric charge $Q$ creates electromagnetic repulsion that partially counteracts gravitational attraction, allowing stable orbits to exist closer to the BH.
 These results highlight the competing effects of CoS parameter \( \gamma \) and the electric charge \( Q \) on the structure of particle orbits in the modified BH geometry. 

\subsection{Fundamental Frequencies}

The motion of test particles in the space-time around BHs can be decomposed into radial, azimuthal, and vertical oscillations. The frequencies associated with these oscillations, known as fundamental frequencies, are essential to understanding the dynamics of accretion disks and the origin of QPOs. In the presence of QED gravity, CoS and PFDM, the fundamental frequencies are modified due to changes in the spacetime geometry. The three fundamental frequencies are:

\begin{center}
   {\bf I.\, Keplerian Frequency}
\end{center}

The Keplerian frequency, $\nu_{K,i}$, is a fundamental frequency associated with the orbital motion of a test particle around a central massive object. It describes the azimuthal angular velocity of the particle along a stable circular orbit and is crucial to understanding the dynamics of the accretion disk and the QPO phenomenon. The Keplerian frequency is given by:
\begin{equation}
   \nu _{K,r,\theta} = \frac{1}{2\,\pi }\frac{c^3}{G\,M}\, \Omega _{r,\theta, \phi }\ , [{\textrm{Hz}}]\ . \label{pp1}
\end{equation}
Azimuthal frequency \(\Omega_{\phi}\), which is the frequency of Keplerian orbits of test particles in the azimuthal direction, is given by:
\begin{equation}
   \nu_K=\Omega_{\phi}=\frac{d\phi}{dt}=\frac{\dot{\phi}}{\dot{t}}=\frac{\omega_{\phi}}{\dot{t}}=\frac{h'(r)}{2\,r}\Big{|}_{r=r_c}=\frac{1}{r^2}\,\left[\frac{r^2}{\ell_p^2} + \frac{M+\beta/2}{r} - \frac{Q^2}{r^2} + \frac{3\,\alpha\, Q^4}{20\, r^6} - \frac{\beta}{2\,r} \ln\left( \frac{r}{\beta} \right) \right].\label{pp2}
\end{equation}

\begin{center}
   {\bf II.\, Radial and Vertical Frequencies}
\end{center}

The radial and vertical angular frequencies \(\nu_r=\Omega_r\) and \(\nu_{\theta}=\Omega_{\theta}\) are the frequency of oscillations of the neutral test particles in the radial direction along the stable orbits, which can be determined from the second derivatives of the effective potential by $r$ and $\theta$ coordinates, respectively are:
\begin{align} 
\Omega_r^2=-\frac{1}{2\, g_{rr}\,(u^t)^2}\, \frac{\partial^2 U_{\text {eff}}}{\partial r^2} \bigg{|}_{r = r_c}.\label{pp3} 
\end{align}
And
\begin{align} 
\Omega^2_{\theta}=- \frac{1}{2\, g_{\theta \theta }\, (u^t)^2} \frac{\partial ^2 U_{\text {eff}}}{\partial \theta ^2} \bigg{|}_{r = r_c}\ . \label{pp4}
\end{align}
Noted that these are the frequencies of neutral test particles as measured by a static distance observer. According to a local observer, the harmonic oscillatory motion frequencies are provided by: 
\begin{align}
   \omega_r^2&=-\frac{1}{2\, g_{rr}}\, \frac{\partial^2 U_{\text {eff}}}{\partial r^2} \bigg{|}_{r = r_c},\label{pp5} \\
   \omega^2_{\theta}&=-\frac{1}{2\, g_{\theta \theta }} \frac{\partial ^2 U_{\text {eff}}}{\partial \theta ^2} \bigg{|}_{r = r_c}\ . \label{pp6}
\end{align}
Here $\Omega^2_i=\frac{\omega^2_i}{(u^t)^2}$ with $u^t=\dot{t}=\sqrt{\frac{2}{2\,h(r)-r\,h'(r)}}$ is the relation between these two types of frequencies as measured by a local and a static distant observer. 

In our case, we find these as follows:
\begin{align}
   \omega_r^2&=-\frac{1}{2}\,\frac{\left(1-\gamma-\frac{2\,M}{r}+\frac{r^2}{\ell^2_p}+\frac{Q^2}{r^2}-\frac{\alpha\,Q^4}{20\,r^6}+\frac{\beta}{r}\,\mbox{ln}\left(\frac{r}{\beta}\right)\right)}{\left(1 - \gamma - \frac{(3M+\beta/2)}{r}+ \frac{2\,Q^2}{r^2} - \frac{\alpha\, Q^4}{5\, r^6} + \frac{3\,\beta}{2\,r} \ln\left( \frac{r}{\beta} \right)\right)}\times\nonumber\\
   &\Bigg[\left(1-\gamma-\frac{2\,M}{r}+\frac{r^2}{\ell^2_p}+\frac{Q^2}{r^2}-\frac{\alpha\,Q^4}{20\,r^6}+\frac{\beta}{r}\,\mbox{ln}\left(\frac{r}{\beta}\right)\right)\,\left(\frac{2}{\ell_p^2} - \frac{4M}{r^3} + \frac{6Q^2}{r^4} - \frac{21 \alpha Q^4}{10 r^8} - \frac{2\beta}{r^3} \left( 1 - \ln\left(\frac{r}{\beta}\right) \right)\right)\nonumber\\
   &+\frac{2}{r^2}\,\left(\frac{r^2}{\ell_p^2} + \frac{M+\beta/2}{r} - \frac{Q^2}{r^2} + \frac{3\,\alpha\, Q^4}{20\, r^6} - \frac{\beta}{2\,r} \ln\left( \frac{r}{\beta} \right)\right)\left(3 - 3\gamma - \frac{r^2}{\ell_p^2} - \frac{Q^2}{r^2} - \frac{10M}{r} - \frac{27 \alpha Q^4}{20 r^6} + \frac{5 \beta}{r} \ln\left( \frac{r}{\beta} \right) - \frac{2 \beta}{r}\right)\Bigg],\label{pp7}\\
   \omega^2_{\theta}&=-\frac{1}{r^2}\,\left(1-\gamma-\frac{2\,M}{r}+\frac{r^2}{\ell^2_p}+\frac{Q^2}{r^2}-\frac{\alpha\,Q^4}{20\,r^6}+\frac{\beta}{r}\,\mbox{ln}\left(\frac{r}{\beta}\right)\right)\,\left(\frac{r^2}{\ell_p^2} + \frac{M}{r} - \frac{Q^2}{r^2} + \frac{3\,\alpha\, Q^4}{20\, r^6} - \frac{\beta}{2\,r} \ln\left( \frac{r}{\beta} \right) + \frac{\beta}{2\,r} \right)\times\nonumber\\
   &\left(1 - \gamma - \frac{(3M+\beta/2)}{r}+ \frac{2\,Q^2}{r^2} - \frac{\alpha\, Q^4}{5\, r^6} + \frac{3\,\beta}{2\,r} \ln\left( \frac{r}{\beta} \right)\right)^{-1},\label{pp8}\\
   \omega^2_{\phi}&=\frac{1}{r^2}\,\frac{\left(\frac{r^2}{\ell_p^2} + \frac{M}{r} - \frac{Q^2}{r^2} + \frac{3\,\alpha\, Q^4}{20\, r^6} - \frac{\beta}{2\,r} \ln\left( \frac{r}{\beta} \right) + \frac{\beta}{2\,r} \right)}{\sqrt{1 - \gamma - \frac{(3M+\beta/2)}{r}+ \frac{2\,Q^2}{r^2} - \frac{\alpha\, Q^4}{5\, r^6} + \frac{3\,\beta}{2\,r} \ln\left( \frac{r}{\beta} \right)}}.\label{pp9}
\end{align}

\begin{figure}[ht!]
    \centering
    \includegraphics[width=0.3\linewidth]{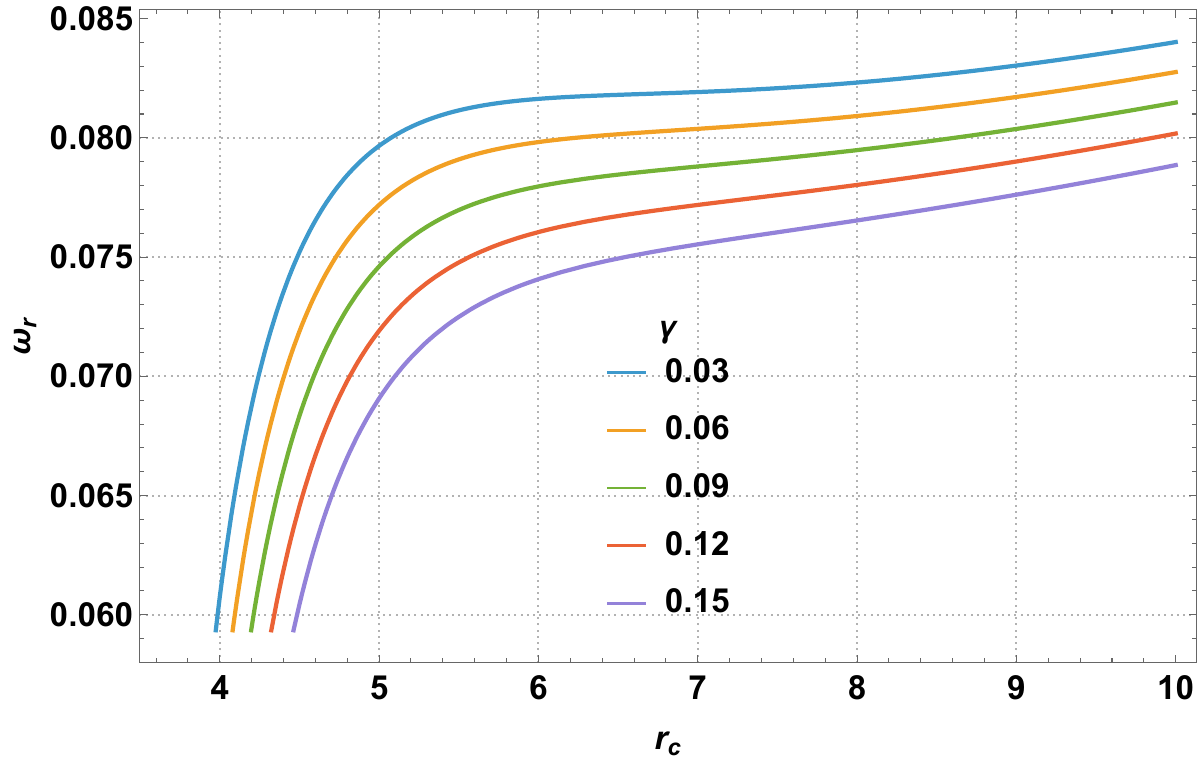}\quad
    \includegraphics[width=0.3\linewidth]{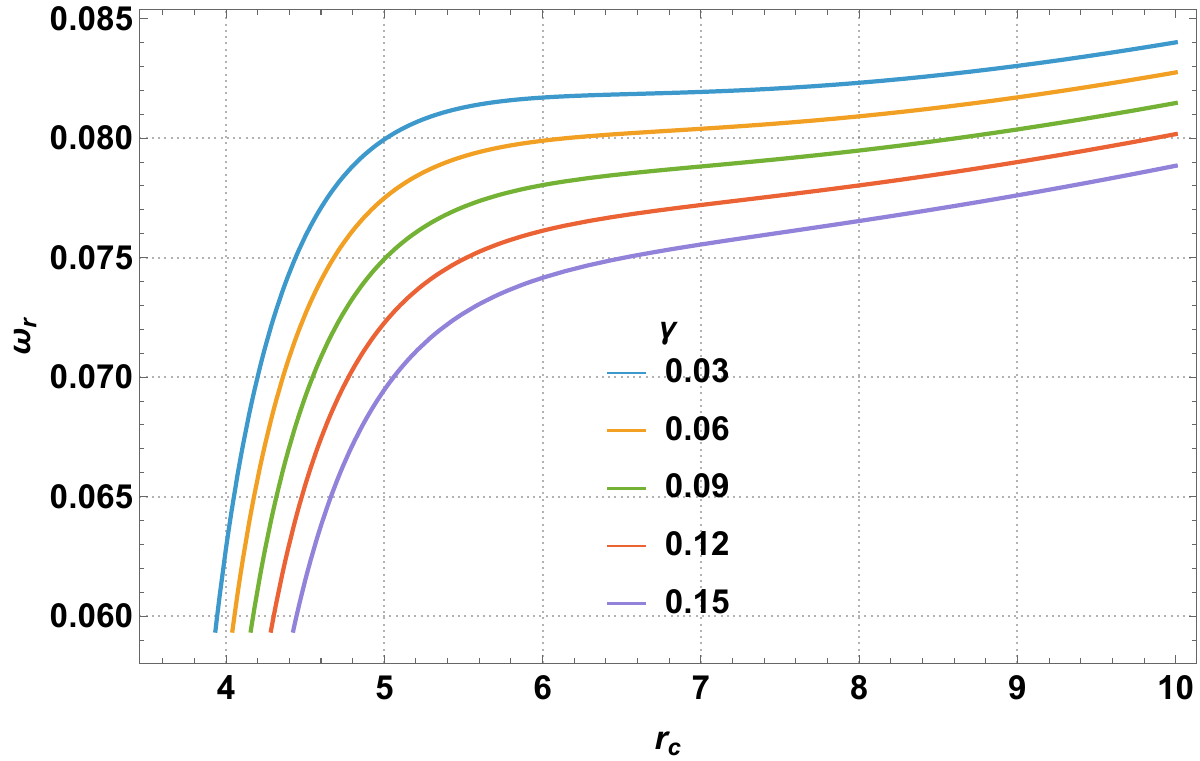}\quad
    \includegraphics[width=0.3\linewidth]{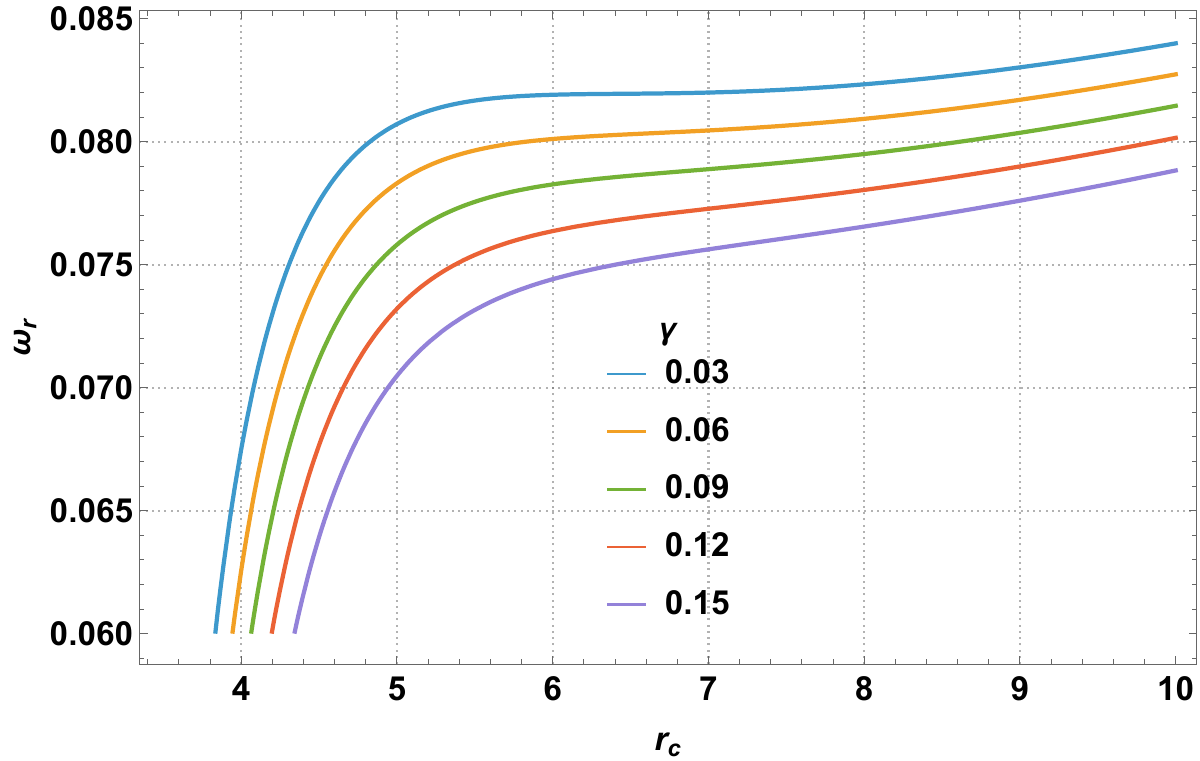}\\
    (a) $Q=0.25$ \hspace{4cm} (b) $Q=0.50$ \hspace{4cm} (c) $Q=0.75$
    \caption{\footnotesize Behavior of the fundamental frequency $\omega_r$ as a function of $r_c$ for different values of $\gamma$.  Here, $M=1,\,\ell_p=25,\,\beta=0.5,\,\alpha=0.25$.}
    \label{fig:frequency-1}
    \includegraphics[width=0.3\linewidth]{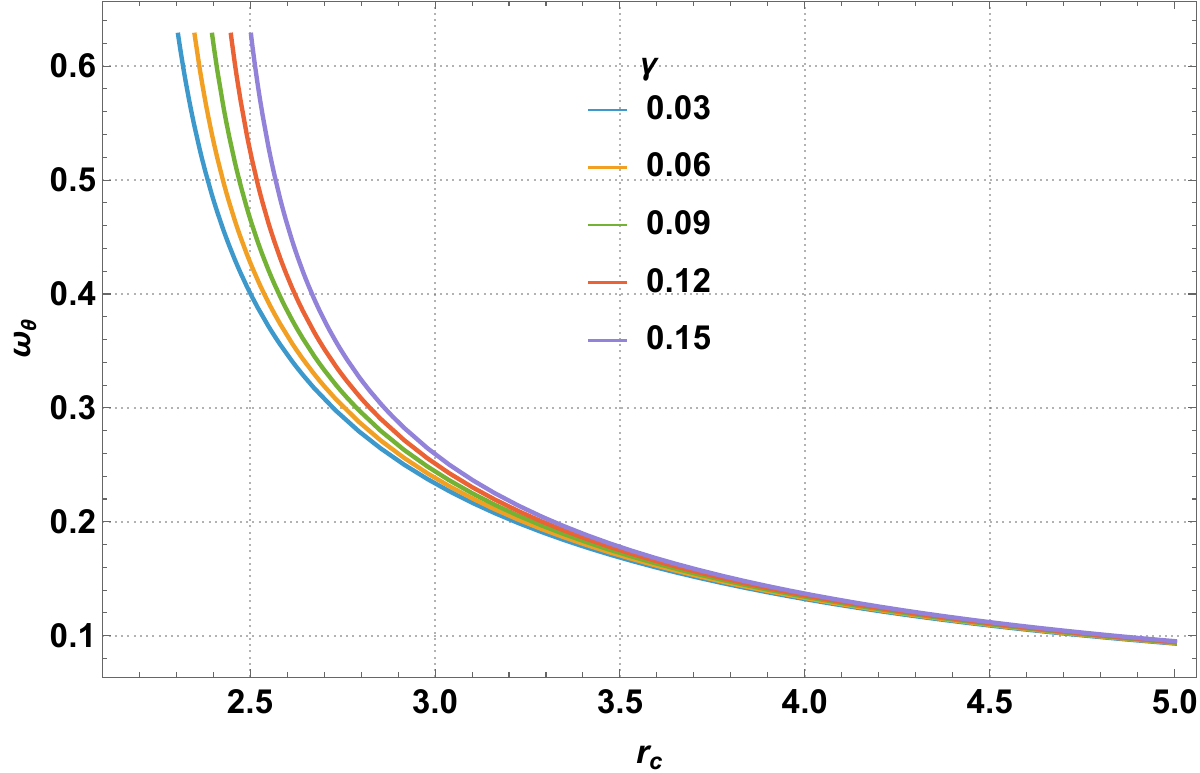}\quad
    \includegraphics[width=0.3\linewidth]{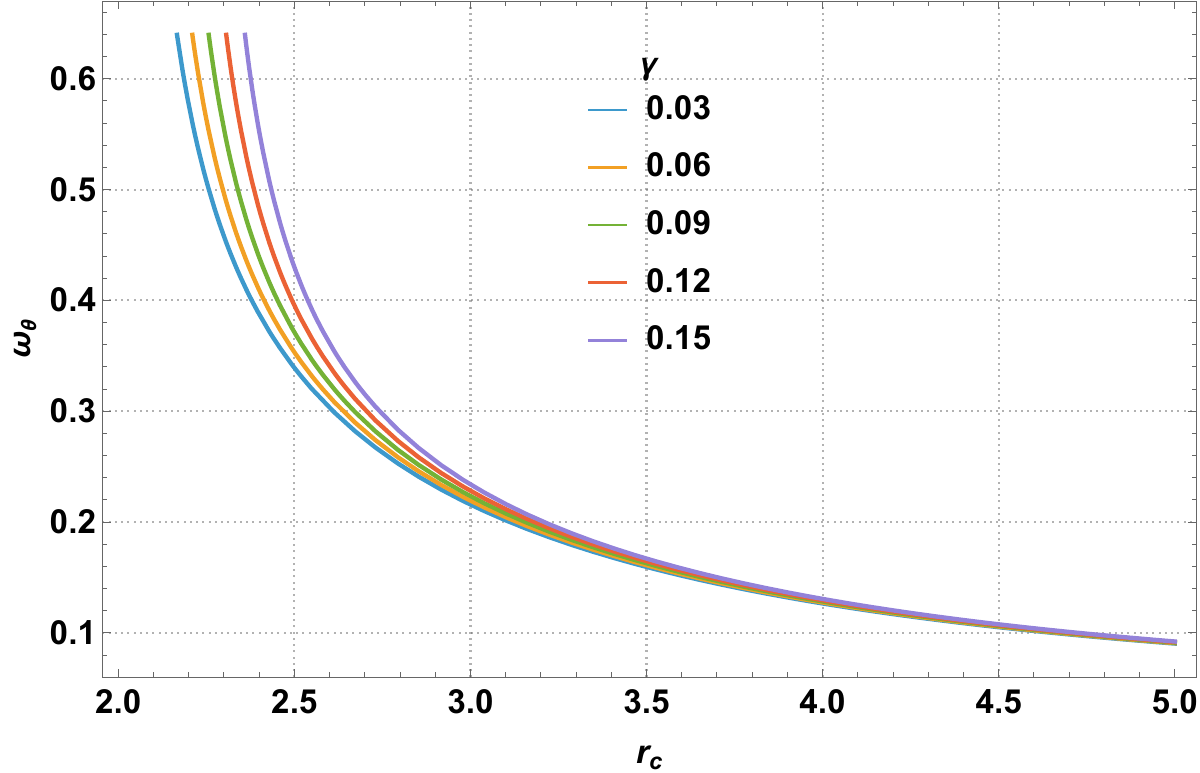}\quad
    \includegraphics[width=0.3\linewidth]{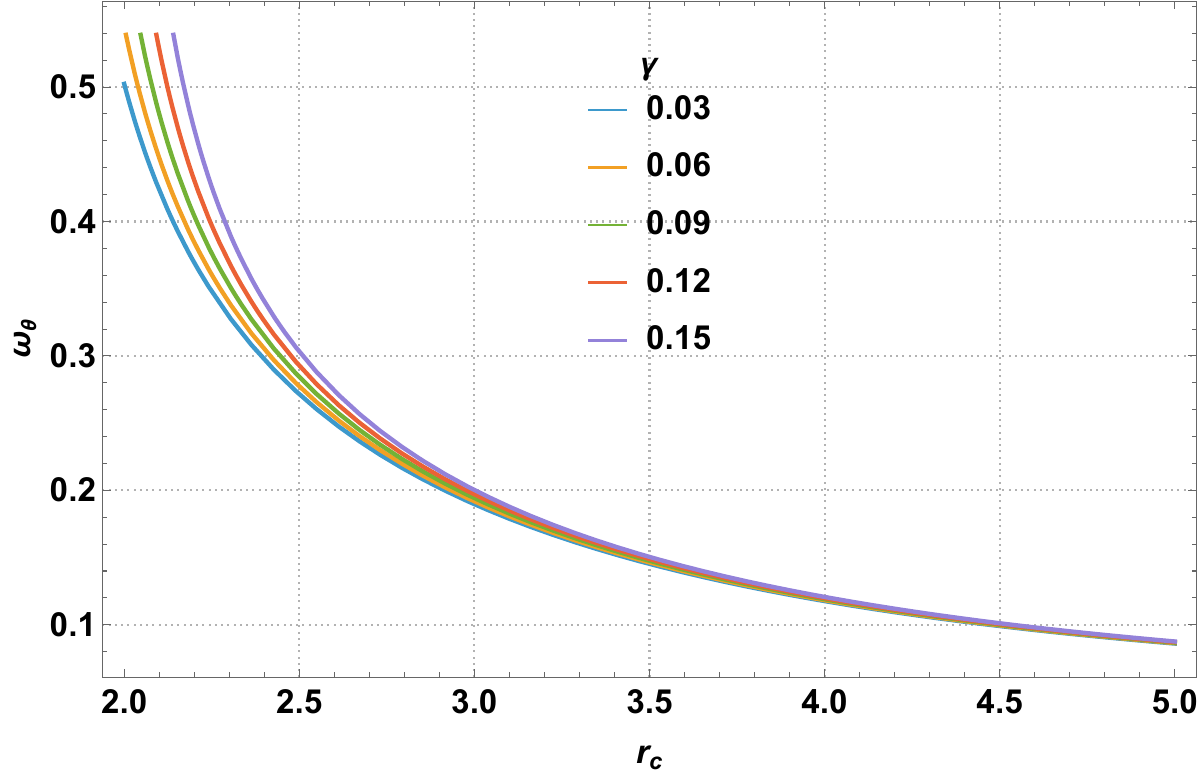}\\
    (a) $Q=0.25$ \hspace{4cm} (b) $Q=0.50$ \hspace{4cm} (c) $Q=0.75$
    \caption{\footnotesize Behavior of the fundamental frequency $\omega_{\theta}$ as a function of $r_c$ for different values of $\gamma$.  Here, $M=1,\,\ell_p=25,\,\beta=0.5,\,\alpha=0.25$.}
    \label{fig:frequency-2}
    \includegraphics[width=0.3\linewidth]{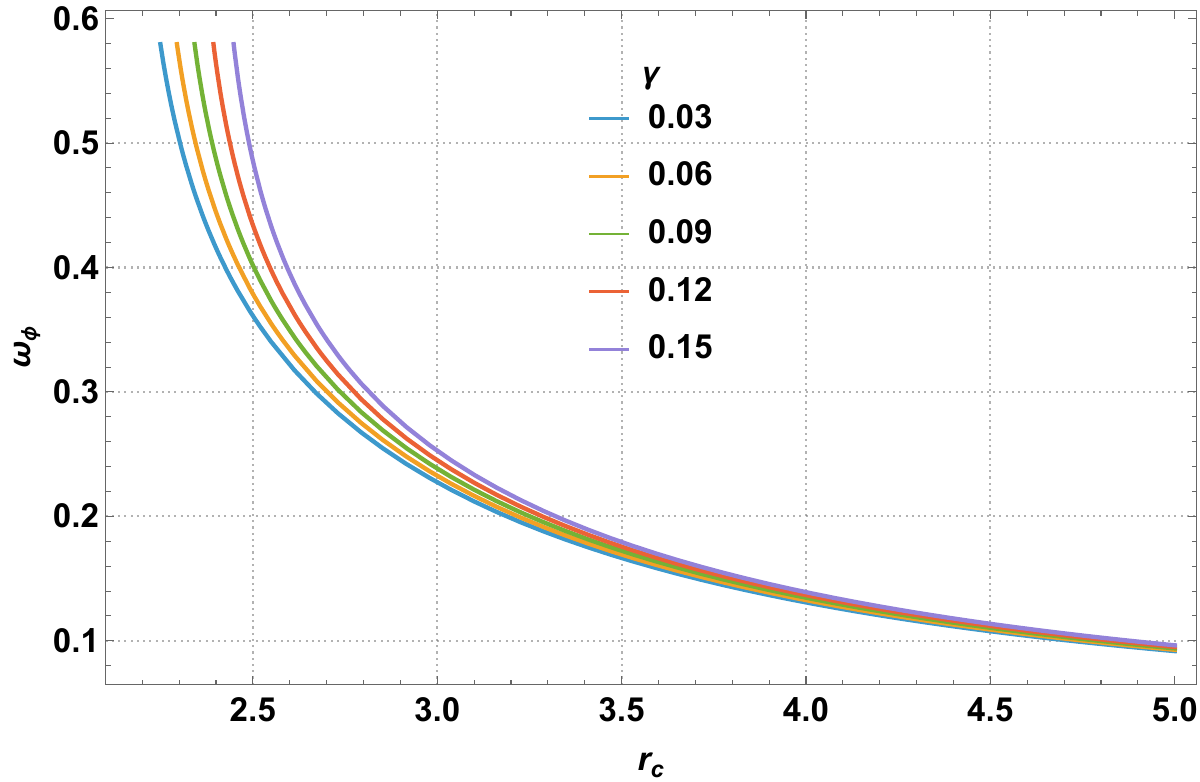}\quad
    \includegraphics[width=0.3\linewidth]{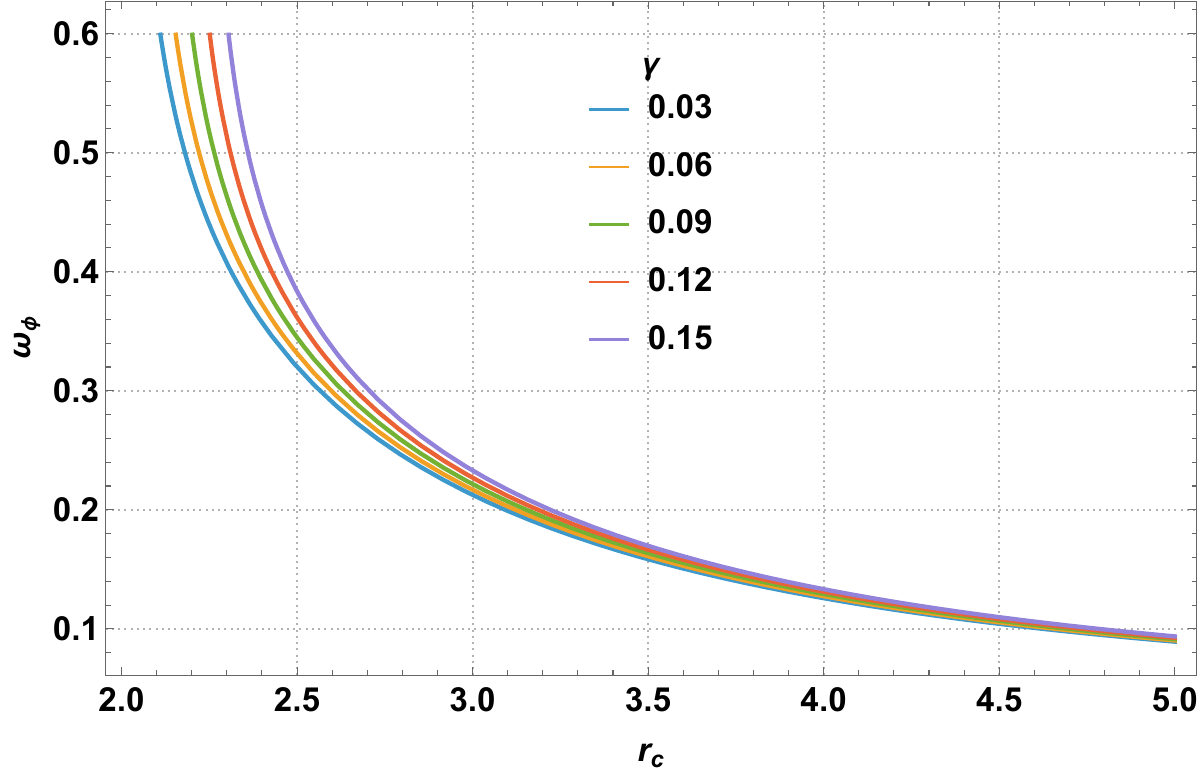}\quad
    \includegraphics[width=0.3\linewidth]{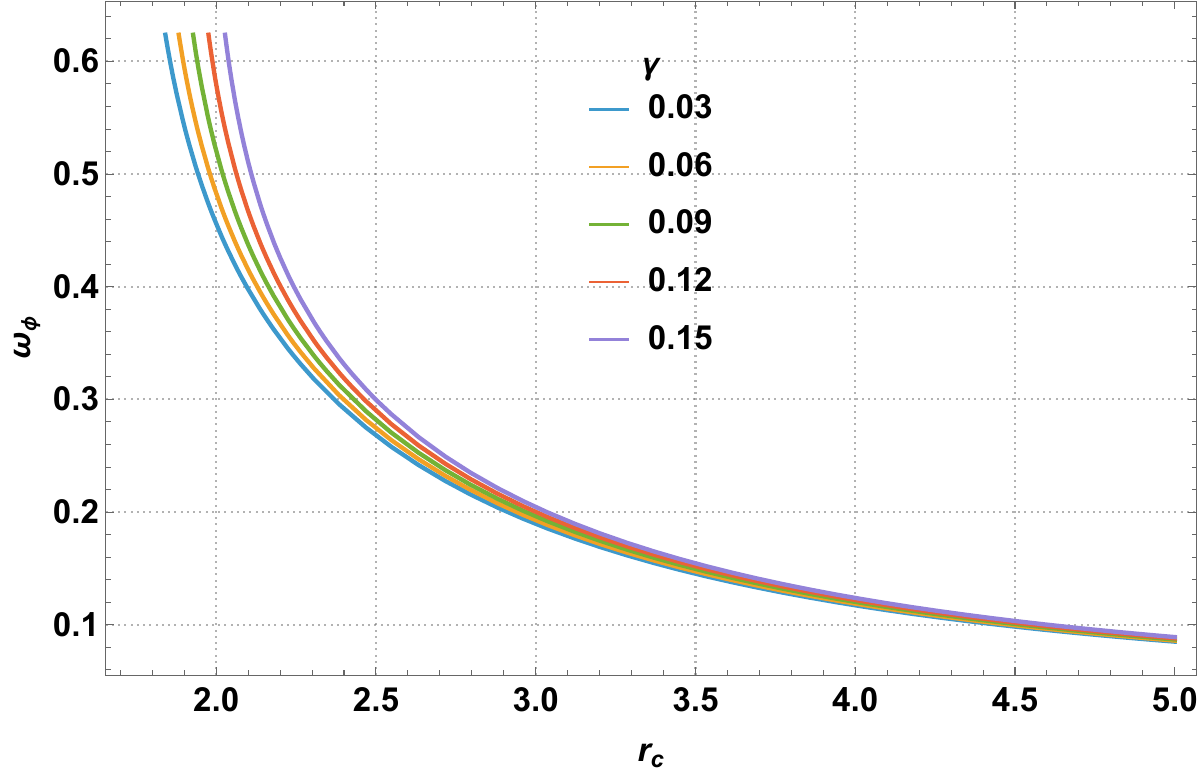}\\
    (a) $Q=0.25$ \hspace{4cm} (b) $Q=0.50$ \hspace{4cm} (c) $Q=0.75$
    \caption{\footnotesize Behavior of the fundamental frequency $\omega_{\phi}$ as a function of $r_c$ for different values of $\gamma$.  Here, $M=1,\,\ell_p=25,\,\beta=0.5,\,\alpha=0.25$.}
    \label{fig:frequency-3}
\end{figure}

Finally, we determine the periastron frequency of a neutral test particle orbiting a non-rotating charged CEH-AdS-BH, with particular attention to small perturbations near the equatorial plane at $\theta = \frac{\pi}{2}$. We perturb the particle slightly from its stable circular orbit to analyze the resulting periastron precession. This perturbation induces oscillations about the equilibrium position, which are characterized by a radial frequency $\Omega_r$. According to the following relationship, the periastron frequency, denoted by $\Omega_p$, is defined as the difference between the orbital (azimuthal) frequency $\Omega_\phi$ and the radial frequency $\Omega_r$, as given by the following relation:
\begin{equation}
  \Omega_p = \Omega_{\phi} - \Omega_r=\left(1 - \gamma - \frac{(3M+\beta/2)}{r}+ \frac{2\,Q^2}{r^2} - \frac{\alpha\, Q^4}{5\, r^6} + \frac{3\,\beta}{2\,r} \ln\left( \frac{r}{\beta} \right)\right)\,\left(\omega^2_{\phi}-\omega_r^2\right).\label{pp10} 
\end{equation}

\section{Scalar Field Perturbations of CEH-AdS-BH Spacetime} \label{sec4}

The investigation of field perturbations around BHs represents a fundamental aspect of theoretical astrophysics, providing crucial insights into BH stability, Hawking radiation properties, and the propagation characteristics of various field types in curved spacetime geometries. In the context of CEH-AdS-BH solutions with exotic matter distributions, scalar field perturbations become particularly intriguing due to the complex interplay between EH NLED corrections, CoS topological effects, and PFDM contributions \cite{isz340}. These exotic components collectively modify the effective potential experienced by propagating scalar fields, leading to distinctive signatures in both the perturbation spectrum and the associated greybody radiation characteristics.

The theoretical framework for analyzing scalar perturbations in modified BH spacetimes has been extensively developed over the past decades, with applications ranging from classical Schwarzschild and Reissner-Nordstr\"{o}m solutions to more exotic configurations involving modified gravity theories \cite{sec4is03}. In our CEH-AdS-BH background, the presence of multiple exotic matter components introduces rich phenomenology that extends far beyond conventional general relativistic treatments. The CoS parameter $\gamma$ modifies the global spacetime structure through angular deficit effects, while the PFDM coupling $\beta$ introduces logarithmic corrections that can significantly alter the potential barrier properties. Simultaneously, EH parameter $\alpha$ contributes higher-order electromagnetic corrections that become particularly relevant in the strong-field regime near the BH horizon.

Understanding these perturbation dynamics becomes essential for several reasons. First, scalar field stability analysis provides direct information about the dynamical stability of the underlying BH solution \cite{sec4is04}. Second, the resulting greybody factors characterize how Hawking radiation deviates from pure blackbody emission, offering potentially observable signatures that could distinguish exotic BH solutions from their conventional counterparts. Third, the perturbation spectrum analysis contributes to our understanding of field propagation in modified gravitational backgrounds, with implications for both theoretical consistency and observational phenomenology.

The dynamics of massless scalar fields in our spacetime are governed by the Klein-Gordon equation, which in covariant form takes the expression:
\begin{equation}
\frac{1}{\sqrt{-g}}\,\partial_{\mu}\left[\left(\sqrt{-g}\,g^{\mu\nu}\,\partial_{\nu}\right)\,\Psi\right]=0,\label{ff1}    
\end{equation}
where $\Psi$ represents the scalar field wave function, $g_{\mu\nu}$ denotes the covariant metric tensor, $g=\det(g_{\mu\nu})$ is the metric determinant, $g^{\mu\nu}$ represents the contravariant metric components, and $\partial_{\mu}$ denotes partial derivatives with respect to the coordinate system.

To analyze the radial wave propagation characteristics, we employ the standard separation of variables technique with the ansatz:
\begin{equation}
   \Psi(t, r,\theta, \phi)=\exp(i\,\omega\,t)\,Y^{m}_{\ell} (\theta,\phi)\,\frac{\psi(r)}{r},\label{ff2}
\end{equation}
where $\omega$ represents the (possibly complex) temporal frequency, $\psi(r)$ describes the radial field component, and $Y^{m}_{\ell}(\theta,\phi)$ denotes the spherical harmonics with multipole indices $\ell$ and $m$.

Through this separation procedure, the Klein-Gordon equation reduces to a one-dimensional Schrödinger-like wave equation:
\begin{equation}
   \frac{\partial^2 \psi(r_*)}{\partial r^2_{*}}+\left(\omega^2-V_\text{scalar}\right)\,\psi(r_*)=0,\label{ff3}
\end{equation}
where the tortoise coordinate transformation is defined by:
\begin{equation}
   dr_*=\frac{dr}{h(r)}.\label{ff4}
\end{equation}

The scalar perturbative potential, which encodes all the geometric and physical effects of our exotic BH solution, takes the form:
\begin{equation}
V_\text{scalar}=\frac{1}{r^2}\,\left(\ell\,(\ell+1)+r\,h'(r)\right)\,h(r),\quad\quad \ell \geq 0.\label{ff5}
\end{equation}

Substituting our metric function from Eq. (\ref{bb2}), this potential becomes:
{\small
\begin{eqnarray}
V_\text{scalar}=\frac{1}{r^2}\,\left[\ell\,(\ell+1)+\frac{2\,M+\beta}{r}-\frac{2\,Q^2}{r^2}+\frac{2\,r^2}{\ell^2_p}+\frac{3\,\alpha\,Q^4}{10\,r^6}-\frac{\beta}{r}\,\ln\left(\frac{r}{\beta}\right)\right]\left(1-\gamma-\frac{2\,M}{r}+\frac{r^2}{\ell^2_p}+\frac{Q^2}{r^2}-\frac{\alpha\,Q^4}{20\,r^6}+\frac{\beta}{r}\,\ln\left(\frac{r}{\beta}\right)\right).\label{ff6}
\end{eqnarray}
}

This expression reveals the intricate structure of the scalar perturbative potential, demonstrating how each exotic matter component contributes distinct modifications. The CoS parameter $\gamma$ appears as a constant shift affecting the overall potential magnitude, while the PFDM coupling $\beta$ introduces both linear and logarithmic terms that significantly alter the potential shape. The EH parameter $\alpha$ contributes higher-order corrections proportional to $Q^4/r^6$, which become increasingly important in the near-horizon regime where strong electromagnetic fields dominate.

\begin{figure}[ht!]
   \centering
   \includegraphics[width=0.32\linewidth]{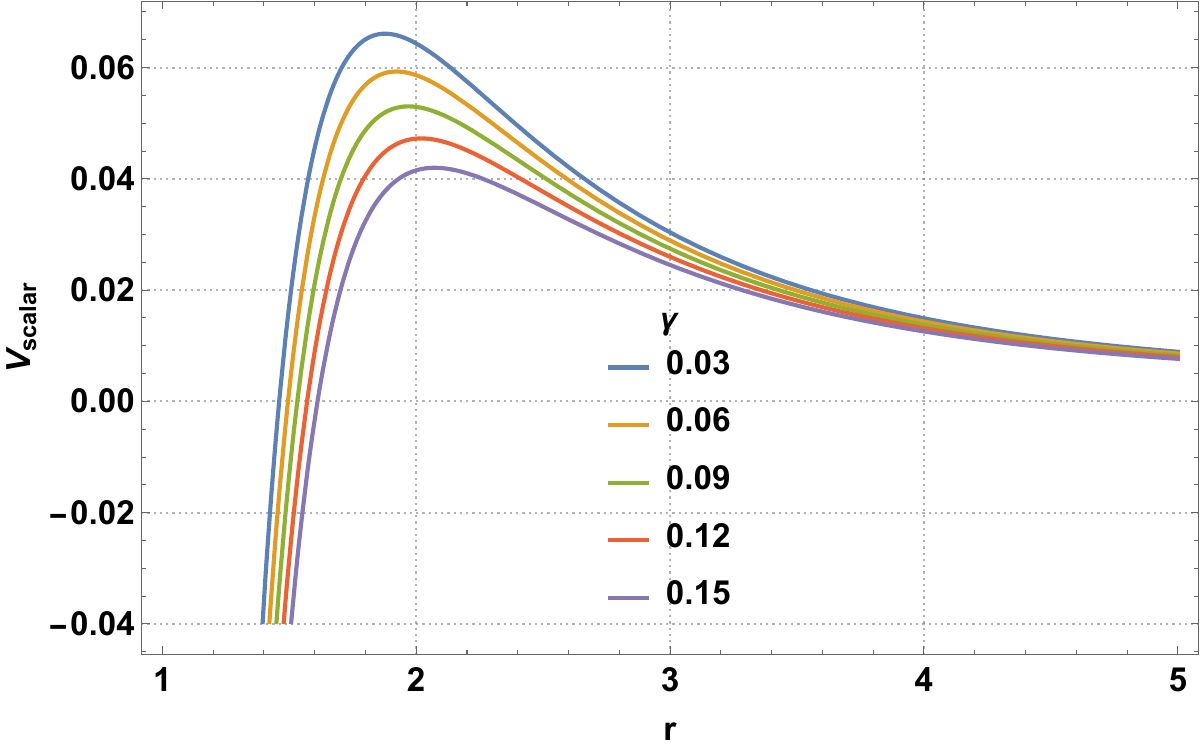}\quad
   \includegraphics[width=0.32\linewidth]{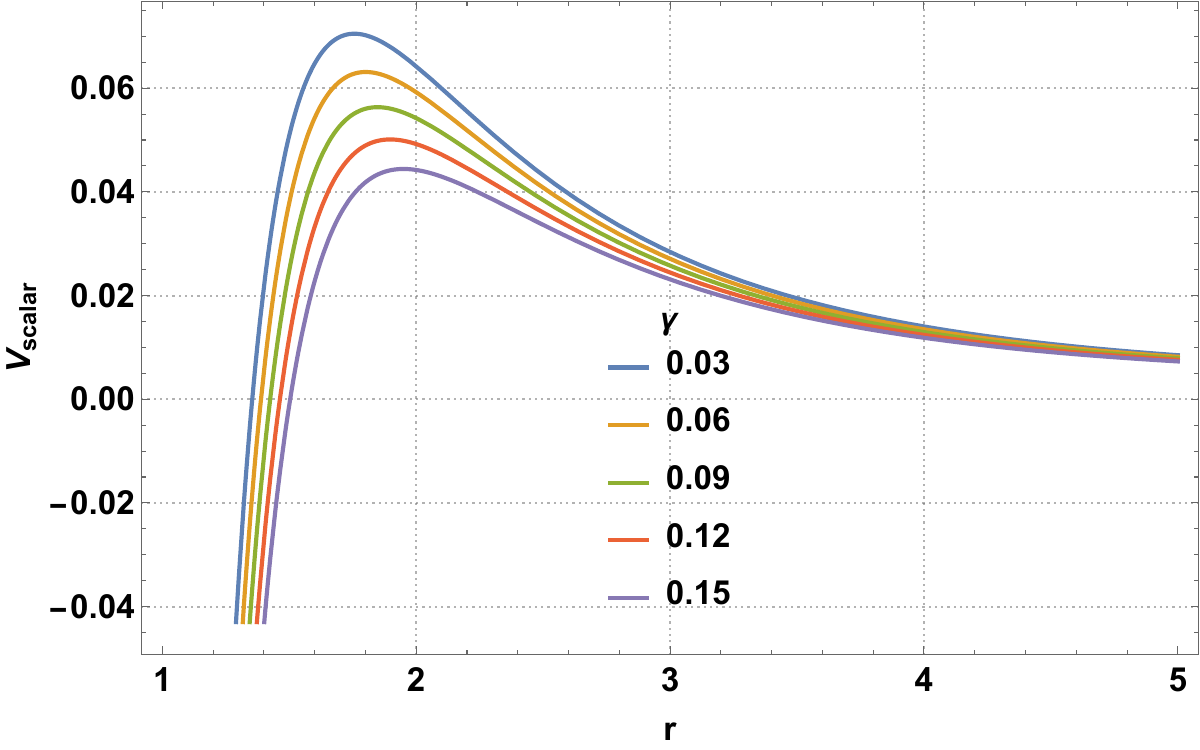}\quad 
   \includegraphics[width=0.32\linewidth]{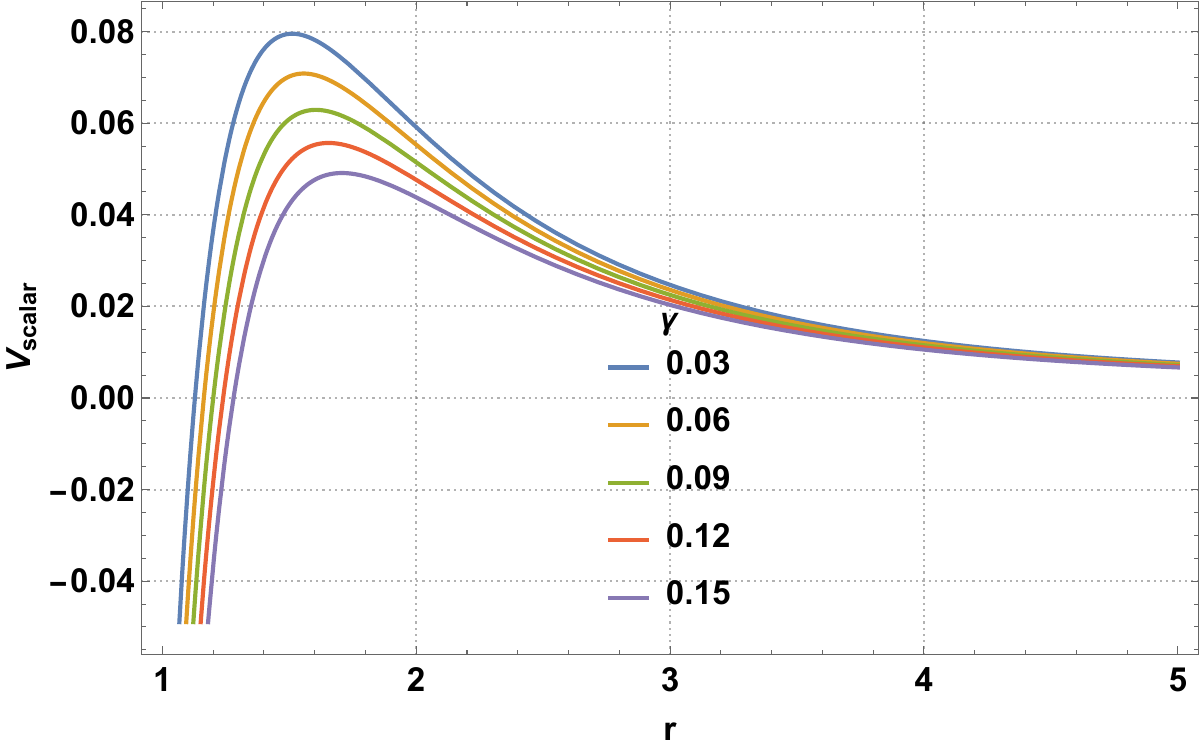}\\
   (a) $Q=0.25$ \hspace{5cm} (b) $Q=0.50$ \hspace{5cm} (c) $Q=0.75$
   \caption{\footnotesize Behavior of the scalar perturbative potential $V_\text{scalar}$ for dominant modes $\ell=0$ by varying CoS parameter $\gamma$. Here, $M=1,\,\ell_p=25,\,\beta=0.5,\,\alpha=0.25$.}
   \label{fig:scalar-potential}
\end{figure}

Figure \ref{fig:scalar-potential} illustrates the radial dependence of the scalar perturbative potential for the monopole mode ($\ell=0$) across different electric charge values. The plots demonstrate several key features: the potential exhibits a characteristic barrier structure that governs wave scattering properties, with the barrier height and shape systematically modified by the CoS parameter $\gamma$. For fixed charge values, increasing $\gamma$ generally reduces the potential magnitude, indicating a weakening of the effective gravitational field experienced by scalar perturbations. This behavior reflects the angular deficit effects introduced by cosmic strings, which effectively reduce the local gravitational strength.

To facilitate analytical progress and reveal the parameter dependence more clearly, we introduce dimensionless variables: $x=r/M$, $y=\beta/M$, $k=M/\ell_p$, $q=Q/M$, and $z=\alpha/M^2$. In these variables, the scalar potential takes the compact form:

{\small
\begin{eqnarray}
M^2\,V_\text{scalar}=\left[\ell\,(\ell+1)+\frac{2+y}{x}-\frac{2\,q^2}{x^2}+2\,k^2\,x^2+\frac{3\,z\,q^4}{10\,x^6}-\frac{y}{x}\,\ln\frac{x}{y}\right]\,\left(1-\gamma-\frac{2}{x}+k^2\,x^2+\frac{q^2}{x^2}-\frac{z\,q^4}{20\,x^6}+\frac{y}{x}\,\ln\frac{x}{y}\right).\label{ff7}
\end{eqnarray}
}

This dimensionless representation clearly reveals the relative importance of different physical effects across various parameter regimes and facilitates numerical analysis of the perturbation spectrum.

\section{Electromagnetic Perturbations in CEH-AdS-BH Spacetime} \label{sec5}

The investigation of EM field perturbations in exotic BH spacetimes represents a crucial complement to scalar field studies, offering unique insights into the electromagnetic response of modified gravitational backgrounds. Unlike scalar perturbations, which probe the geometric curvature properties directly, EM perturbations reveal the intricate coupling between electromagnetic fields and the complex spacetime structure of CEH-AdS-BH solutions with CoS and PFDM \cite{isz34}. The vector nature of electromagnetic fields introduces fundamental differences in the perturbation dynamics, particularly through the absence of monopole modes and the enhanced coupling to spacetime curvature at higher multipole orders.

The theoretical framework for analyzing EM perturbations in curved spacetimes has been extensively developed since the pioneering work of Regge and Wheeler, with subsequent extensions to charged BH backgrounds and modified gravity theories \cite{sec5is03,sec5is04}. In our CEH-AdS-BH system, the presence of multiple exotic matter components creates a rich phenomenological landscape where EH NLED corrections, CoS topological defects, and PFDM distributions collectively influence the electromagnetic wave propagation characteristics. These modifications become particularly pronounced in the strong-field regime, where the interplay between electromagnetic and gravitational effects reaches its most complex manifestation.

The study of EM perturbations serves multiple theoretical and observational purposes. From a fundamental physics perspective, these perturbations provide direct information about the stability properties of the underlying spacetime against electromagnetic field fluctuations \cite{iszn1}. The analysis of quasinormal modes and their associated damping times offers insights into the characteristic timescales governing electromagnetic dissipation processes in exotic BH environments. Furthermore, the EM perturbation spectrum contributes to our understanding of gravitational wave emission mechanisms, particularly in scenarios involving charged matter dynamics around modified BH solutions.

In curved spacetime, EM perturbations are governed by Maxwell's equations in their fully covariant form:
\begin{equation}
\frac{1}{\sqrt{-g}}\left[F_{\alpha \beta }\,g^{\alpha \nu}\,g^{\beta \mu }\sqrt{-g}\,\Psi\right]_{,\mu}=0,\label{em1}
\end{equation}
where $F_{\alpha \beta }=\partial_{\alpha }A_{\beta }-\partial_{\beta}A_{\alpha}$ represents the EM field tensor, $A_{\mu}$ denotes the four-vector potential components, and $\Psi$ characterizes the EM field perturbation amplitude. The metric determinant $g$ and contravariant metric components $g^{\mu\nu}$ encode the full geometric information of our exotic BH spacetime, while the covariant derivative notation $_{,\mu}$ ensures gauge covariance of the field equations.

The separation of variables procedure for EM perturbations in spherically symmetric spacetimes follows established techniques, though the presence of exotic matter components introduces subtle modifications to the standard approach. Following the methodology developed for charged BH studies, the radial wave equation reduces to the canonical Schrödinger-like form:
\begin{equation}
\frac{\partial^2 \psi_\text{em}(r_*)}{\partial r^2_{*}}+\left(\omega^2-V_\text{em}\right)\,\psi_\text{em}(r_*)=0,\label{em2}
\end{equation}
where $r_*$ represents the tortoise coordinate defined by $dr_* = dr/h(r)$, $\omega$ denotes the complex frequency eigenvalue, and $\psi_\text{em}(r_*)$ describes the radial component of the electromagnetic perturbation.

For our CEH-AdS-BH spacetime, the EM perturbative potential exhibits a remarkably structured form that directly reflects the contributions from each exotic matter component:
\begin{equation}
V_\text{em}(r)=\frac{\ell\,(\ell+1)}{r^2}\,h(r)=\left(1-\gamma-\frac{2\,M}{r}+\frac{r^2}{\ell^2_p}+\frac{Q^2}{r^2}-\frac{\alpha\,Q^4}{20\,r^6}+\frac{\beta}{r}\,\ln\left(\frac{r}{\beta}\right)\right)\,\frac{\ell\,(\ell+1)}{r^2},\quad \ell\geq 1.\label{em3}
\end{equation}

This potential structure reveals the fundamental distinction between scalar and EM perturbations: the electromagnetic case requires $\ell\geq 1$ due to the vector nature of EM fields, which inherently lack monopole components. The potential factorizes into two primary contributions: the metric function $h(r)$ encoding all spacetime geometry modifications, and the angular momentum barrier $\ell(\ell+1)/r^2$ characterizing the multipole structure of electromagnetic waves.

The CoS parameter $\gamma$ introduces a uniform multiplicative factor that effectively reduces the overall potential strength, reflecting the angular deficit effects that characterize cosmic string spacetimes. The PFDM coupling $\beta$ contributes through both direct gravitational effects and logarithmic corrections that can substantially modify the potential profile, particularly in intermediate radial regions where these corrections become comparable to the leading-order terms. The EH parameter $\alpha$ introduces highly nonlinear electromagnetic corrections proportional to $Q^4/r^6$, which become increasingly significant in the strong-field regime near the BH horizon where the electromagnetic field strength approaches the QED critical field scale.

\begin{figure}[ht!]
  \centering
  \includegraphics[width=0.32\linewidth]{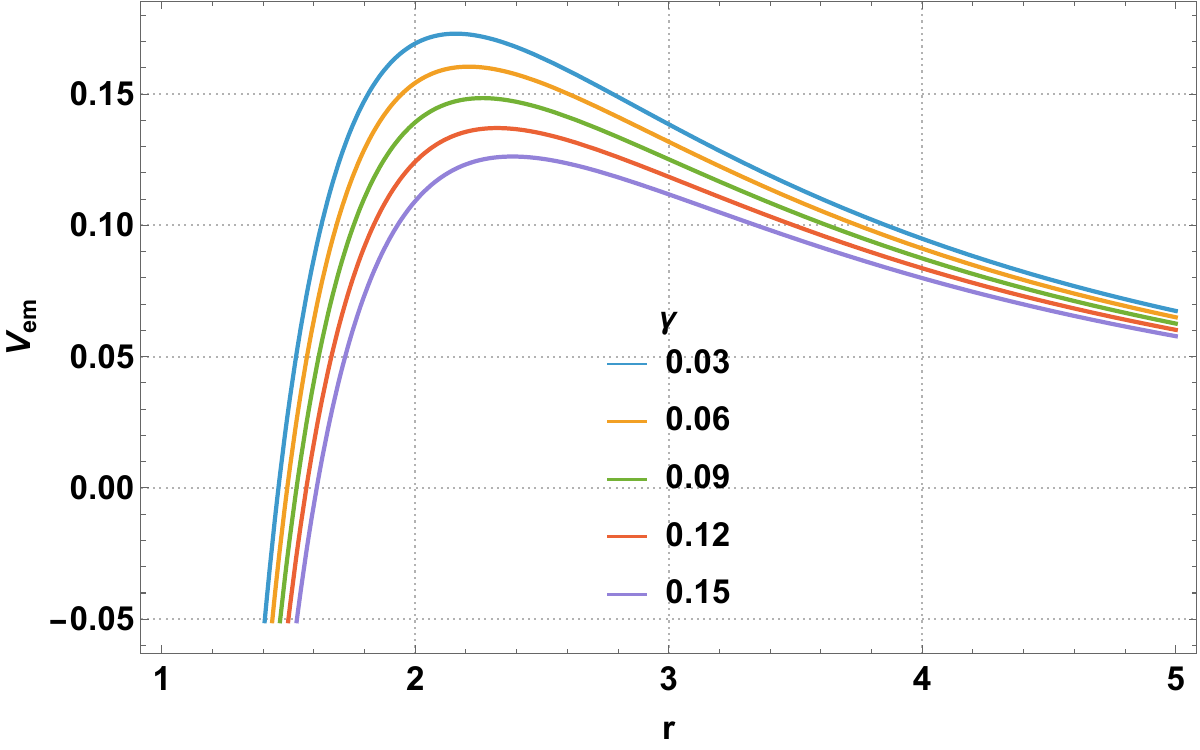}\quad
  \includegraphics[width=0.32\linewidth]{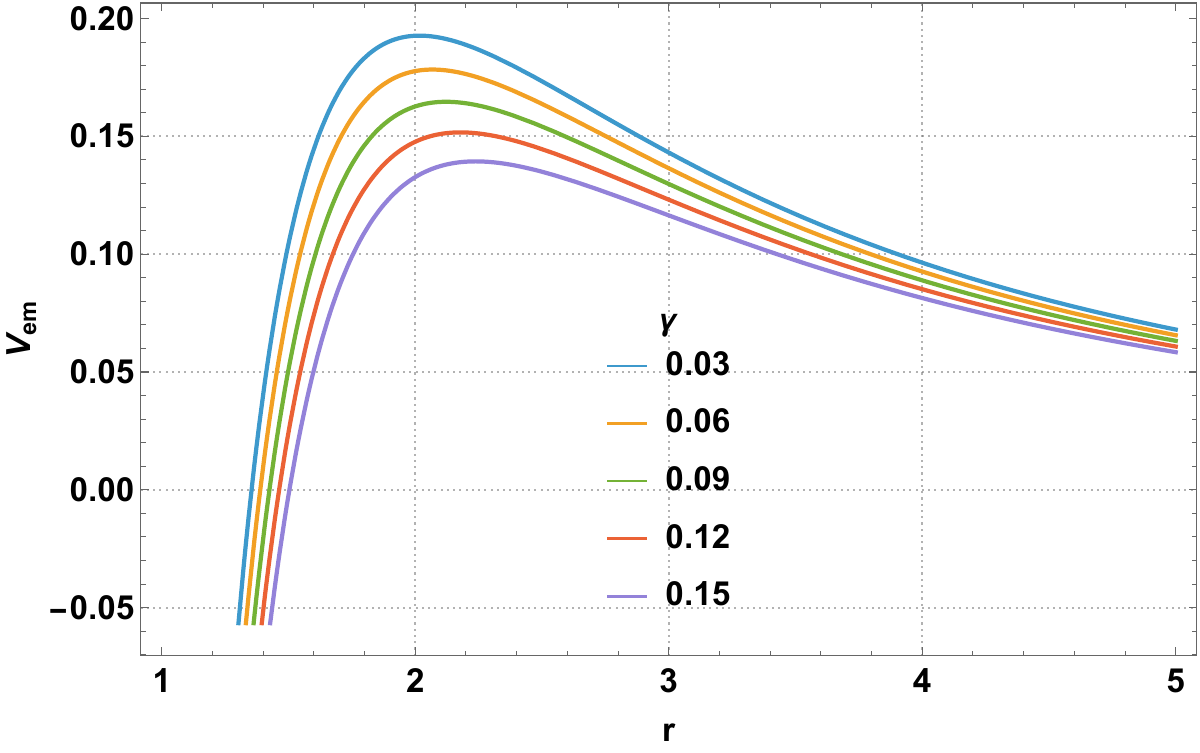}\quad 
  \includegraphics[width=0.32\linewidth]{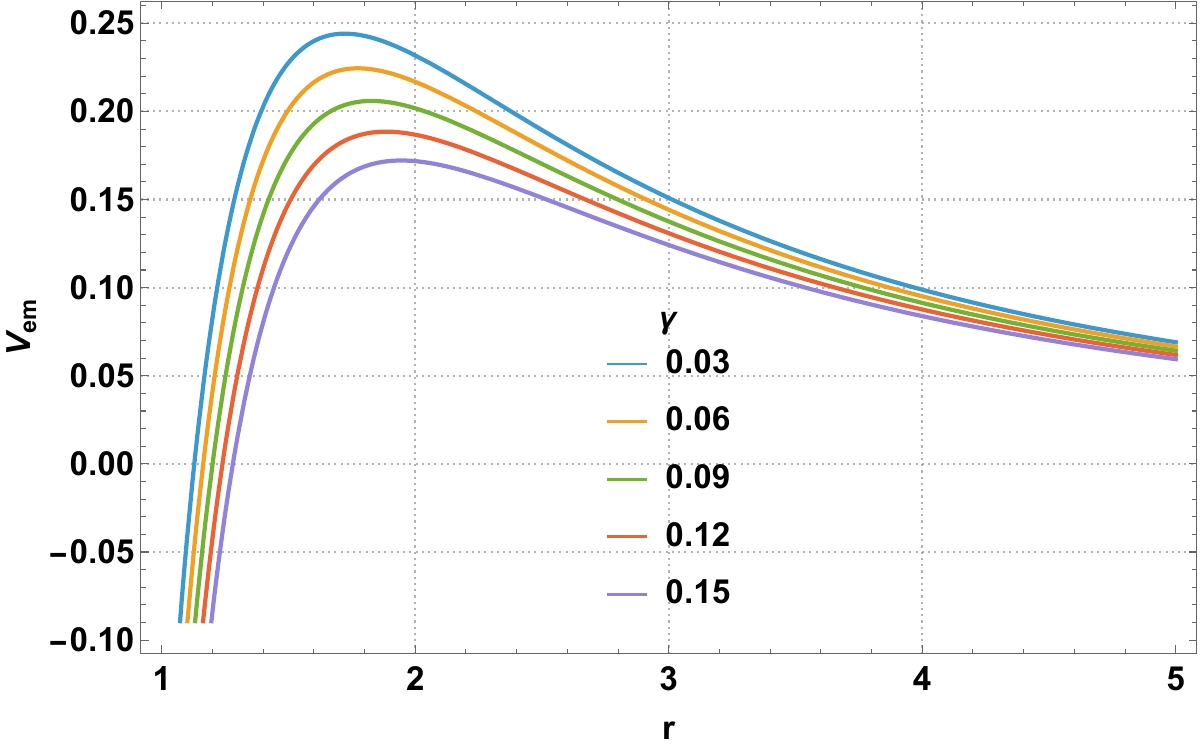}\\
  (a) $Q=0.25$ \hspace{4cm} (b) $Q=0.50$ \hspace{4cm} (c) $Q=0.75$
  \caption{Behavior of the electromagnetic perturbative potential $V_\text{em}$ for dominant modes $\ell=1$ by varying CoS parameter $\gamma$. Here, $M=1,\,\ell_p=25,\,\beta=0.5,\,\alpha=0.25$}
  \label{fig:em-potential}
\end{figure}

Figure \ref{fig:em-potential} illustrates the radial dependence of the EM perturbative potential for the fundamental dipole mode ($\ell=1$) across different electric charge configurations. The systematic variation with the CoS parameter $\gamma$ demonstrates several key features of the electromagnetic wave propagation in our exotic BH background. For low charge values (panel a), the potential exhibits a relatively smooth profile with a characteristic peak that governs wave scattering properties. As the electric charge increases (panels b and c), the potential barrier becomes more pronounced, reflecting the enhanced electromagnetic coupling inherent to the CEH-AdS-BH geometry. The influence of the CoS parameter becomes particularly evident in the systematic downward shift of the potential curves with increasing $\gamma$ values. This behavior directly reflects the angular deficit effects introduced by cosmic strings, which effectively reduce the local gravitational field strength experienced by electromagnetic perturbations. The analysis across different charge values reveals the competing effects of electromagnetic repulsion and gravitational attraction in determining the overall potential structure. The potential barrier structure governs both the bound state spectrum and the scattering phase shifts, which collectively determine the electromagnetic response characteristics of the system. Higher charge configurations exhibit more pronounced potential barriers, leading to enhanced electromagnetic wave reflection and modified transmission coefficients. The long-distance gravitational phenomena are dictated by the AdS curvature scale $\ell_p^{-1}$, and the BH mass $M$ defines the depth of the gravitational potential, possibly revealing observable impacts of exotic BH models \cite{isz32,sec5is07}.

\section{Thermodynamic Properties and Phase Transitions of CEH-AdS-BH with CoS and PFDM} \label{sec6}

The thermodynamic analysis of BHs has emerged as one of the most profound areas of theoretical physics, bridging quantum mechanics, general relativity, and statistical mechanics through the fundamental insights of Hawking and Bekenstein \cite{isz30,sec6is02}. In the context of CEH-AdS-BH solutions with exotic matter distributions, the thermodynamic framework becomes significantly enriched due to the complex interplay between EH NLED corrections, CoS topological effects, and PFDM contributions. These exotic components not only modify the classical thermodynamic quantities but also introduce novel phase transition behaviors and stability criteria that extend far beyond conventional BH thermodynamics \cite{isz28,sec6is04}.

The extended thermodynamic framework for AdS BHs, where the cosmological constant is treated as a thermodynamic pressure, provides a natural setting for investigating phase transitions analogous to those observed in ordinary thermodynamic systems \cite{sec6is05}. This approach, pioneered by Kubiznak and Mann, has revolutionized our understanding of BH thermodynamics by revealing van der Waals-like phase transitions and critical phenomena previously hidden in the conventional thermodynamic treatment \cite{sec6is06}. In our CEH-AdS-BH system, the presence of additional exotic matter components requires further extensions to accommodate the thermodynamic contributions from CoS and PFDM parameters. This leads to a rich phase space structure where multiple competing effects determine the overall system stability and transition characteristics.

The significance of studying thermodynamic phase transitions in modified BH spacetimes extends beyond pure theoretical interest. These investigations provide crucial insights into the fundamental nature of quantum gravity, holographic duality, and the emergence of spacetime from more fundamental degrees of freedom \cite{sec6is07}. The identification of critical points, phase boundaries, and stability regions offers potential observational signatures that could distinguish exotic BH solutions from their classical counterparts through measurements of thermal emission properties and quasi-normal mode spectra.

The thermodynamic investigation begins with the determination of the BH mass through the horizon condition. Setting $h(r_{+})=0$ in our metric function, we obtain:
\begin{equation}
  M=\frac{r_{+}}{2}\left[1-\gamma+\frac{r_{+}^2}{\ell^2_p}+\frac{Q^2}{r_{+}^2}-\frac{\alpha\,Q^4}{20\,r_{+}^6}+\frac{\beta}{r_{+}}\,\ln\left(\frac{r_{+}}{\beta}\right)\right].\label{therm1}
\end{equation}

This mass expression reveals the intricate contributions from each exotic matter component. The CoS parameter $\gamma$ appears as a constant shift that effectively reduces the total mass for given horizon radius, reflecting the angular deficit effects inherent to cosmic string spacetimes. The EH parameter $\alpha$ introduces highly nonlinear electromagnetic corrections that become increasingly important as the charge approaches extremal values, while the PFDM coupling $\beta$ contributes logarithmic terms that can significantly modify the mass-radius relationship.

In the extended phase space approach, the thermodynamic pressure is identified with the AdS curvature radius through:
\begin{equation}
  P=\frac{3}{8\pi\,\ell^2_p}.\label{therm2}
\end{equation}

This identification allows us to rewrite the mass in terms of pressure:
\begin{equation}
  M=\frac{r_{+}}{2}\left[1-\gamma+\frac{8\pi}{3}\,P\,r_{+}^2+\frac{Q^2}{r_{+}^2}-\frac{\alpha\,Q^4}{20\,r_{+}^6}+\frac{\beta}{r_{+}}\,\ln\left(\frac{r_{+}}{\beta}\right)\right].\label{therm3}
\end{equation}

The thermodynamic volume, obtained as the conjugate variable to pressure, takes the familiar form:
\begin{equation}
  V=\left(\frac{\partial M}{\partial P}\right)=\frac{4}{3}\,\pi\,r_{+}^3.\label{therm4}
\end{equation}

Remarkably, despite the complex modifications introduced by exotic matter components, the thermodynamic volume retains its geometric interpretation as proportional to the horizon volume, demonstrating a fundamental robustness of this thermodynamic relationship.

The Hawking temperature, determined from the surface gravity at the event horizon, becomes:
\begin{equation}
  T=\frac{f'(r_{+})}{4\pi}=\frac{1}{4\pi\,r_{+}}\,\left(1 - \gamma+ 8\pi\, P\, r_{+}^2- \frac{Q^2}{r_+^2}+ \frac{\alpha\, Q^4}{4\, r_+^6}+ \frac{\beta}{r_+}\right).\label{therm5}
\end{equation}

This temperature expression exhibits rich parameter dependence that governs the thermal stability and phase transition behavior. The CoS parameter $\gamma$ contributes a constant positive shift to the temperature, while the PFDM parameter $\beta$ introduces an additional $1/r_+$ contribution that becomes dominant in the small horizon limit.

The Hawking-Bekenstein entropy retains its geometric origin:
\begin{equation}
S = \pi\,r_+^2.\label{therm6}
\end{equation}

For completeness, we derive the Gibbs free energy and internal energy:
\begin{equation}
  G=\frac{r_+}{4}\,\left[1 - \gamma- \frac{8\pi}{3} P\, r_+^2+ \frac{3\, Q^2}{r_+^2}- \frac{7\, \alpha\, Q^4}{10\, r_+^6}+ \frac{2\,\beta}{r_+} \ln\left( \frac{r_+}{\beta} \right)- \frac{\beta}{r_+}\right],\label{therm7}
\end{equation}
\begin{equation}
  U=\frac{r_{+}}{2}\left[1-\gamma+\frac{Q^2}{r_{+}^2}-\frac{\alpha\,Q^4}{20\,r_{+}^6}+\frac{\beta}{r_{+}}\,\ln\left(\frac{r_{+}}{\beta}\right)\right].\label{therm7a}
\end{equation}

To facilitate phase space analysis, we express the BH mass in terms of the natural thermodynamic variables:
\begin{equation}
  M(S,P,Q,\gamma,\beta)=\frac{1}{2} \sqrt{\frac{S}{\pi}} \left[1 - \gamma+ \frac{8}{3}\, P\, S+ \frac{\pi\, Q^2}{S}- \frac{\alpha\, \pi^3\, Q^4}{20\, S^3}+ \frac{1}{2}\, \beta\, \sqrt{\frac{\pi}{S}} \ln\left( \frac{S}{\pi\, \beta^2} \right)\right].\label{mass}
\end{equation}

The first law of BH thermodynamics requires modification to account for the additional exotic matter contributions:
\begin{equation}
  dM=T_H\,dS+P\,dV+\Phi_H\,dQ+\mathcal{A}_H\,d\beta+\mathcal{B}_H\,d\gamma,\label{therm8}
\end{equation}
where the intensive thermodynamic variables are defined as:
\begin{align}
  T_H&=\left(\frac{\partial M}{\partial S}\right)_{Q,P,\gamma,\beta},\label{therm9}\\
  V&=\left(\frac{\partial M}{\partial P}\right)_{S,Q,\beta,\gamma}=\frac{4}{3}\,\sqrt{\frac{S^3}{\pi}},\label{therm10}\\
  \mathcal{A}_H&=\left(\frac{\partial M}{\partial \beta}\right)_{S,Q,\gamma,P}=\frac{1}{4}\,\ln\left(\frac{S}{\pi\,\beta^2}\right)-\frac{1}{2},\label{therm11}\\
  \mathcal{B}_H&=\left(\frac{\partial M}{\partial \gamma}\right)_{S,Q,P,\beta}=-\frac{1}{2} \sqrt{\frac{S}{\pi}},\label{therm12}\\
  \Phi&=\left(\frac{\partial M}{\partial Q}\right)_{S,P,\gamma,\beta}=\frac{1}{2} \sqrt{\frac{S}{\pi}}\,\left(\frac{2\pi\,Q}{S}-\frac{\alpha\, \pi^3\, Q^3}{5\, S^3}\right).\label{therm13}
\end{align}

These intensive variables reveal the thermodynamic conjugates to the exotic matter parameters. The quantity $\mathcal{A}_H$ represents the chemical potential associated with PFDM, while $\mathcal{B}_H$ characterizes the thermodynamic response to CoS variations. Both quantities exhibit non-trivial dependencies on the fundamental thermodynamic variables, indicating complex coupling between the exotic matter components and the underlying BH thermodynamics.

In terms of the horizon radius, the temperature becomes:
\begin{equation}
T_{H}=\frac{1}{4\pi\,r_{+}}\left(1-\gamma-\frac{Q^2}{r^2_+}+\frac{Q^4\,\alpha}{4\,r^6_+}+\frac{\beta}{r_+}+8\pi\,P\,r_{+}\right).\label{temp1}
\end{equation}

\begin{figure}[ht!]
  \centering
  \subfloat[$Q=1,\alpha=0.3, \beta =0.4$]{\centering{}\includegraphics[scale=0.55]{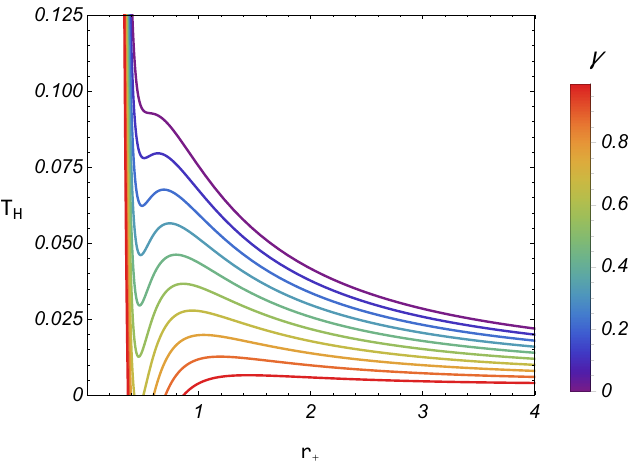}}\quad
  \subfloat[$Q=1,\gamma=0.4, \beta =0.4$]{\centering{}\includegraphics[scale=0.55]{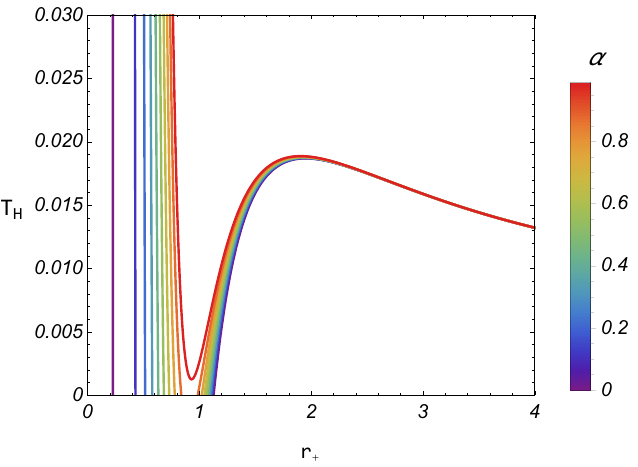}}
 \subfloat[$Q=1,\gamma=0.4, \alpha =0.3$] {\centering{}\includegraphics[scale=0.55]{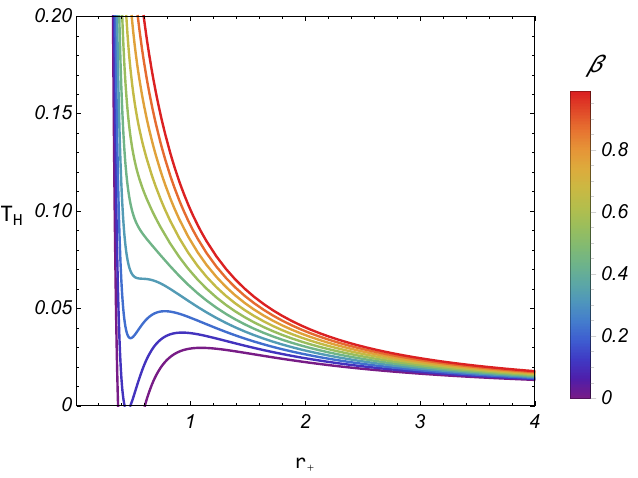}}
  \caption{The Hawking temperature of CEH-AdS-BH with CoS and PFDM showing the influence of the $\alpha$ parameter (left), $\beta$ parameter (middle) and $\gamma$ parameter (right). Here $M=1$ and $\ell_p=25$.}
  \label{temp32}
\end{figure}

Figure \ref{temp32} illustrates the Hawking temperature behavior as a function of horizon radius for various parameter combinations. The temperature minima correspond to thermodynamically unstable configurations, while the regions of positive slope indicate thermodynamically stable phases. The exotic matter parameters shift these critical points, potentially creating new thermodynamic phase structures not present in classical Reissner-Nordström-AdS BHs. This complex behavior demonstrates the intricate interplay between the exotic matter components and their collective influence on the thermal properties. In the left panel, varying the EH parameter $\alpha$ shows how quantum electrodynamic corrections modify the temperature profile, particularly in the strong-field regime. The middle panel demonstrates the influence of PFDM coupling $\beta$, which introduces additional heating effects that become prominent at smaller horizon radii. The right panel reveals how CoS parameter $\gamma$ systematically shifts the temperature curves, reflecting the global geometric modifications introduced by the topological defects.

Thermodynamic stability analysis requires examination of the specific heat capacity, which characterizes the system's response to temperature fluctuations. The specific heat is given by:
\begin{equation}
  C=\frac{dM}{dT_+}=\frac{-16\,\ell_p^2\,r_+^2}{\pi\left(-12\,r_+^8+\ell_p^2\left( -12\,Q^2\,r_+^4+7Q^4\,\alpha+4\,r_+^5(r_++2\,\beta-\gamma\,r_+) \right)  \right)}.\label{heateq}
\end{equation}

\begin{figure}[ht!]
  \centering
  \subfloat[$Q=1,\alpha=0.3, \beta =0.4$]{\centering{}\includegraphics[scale=0.65]{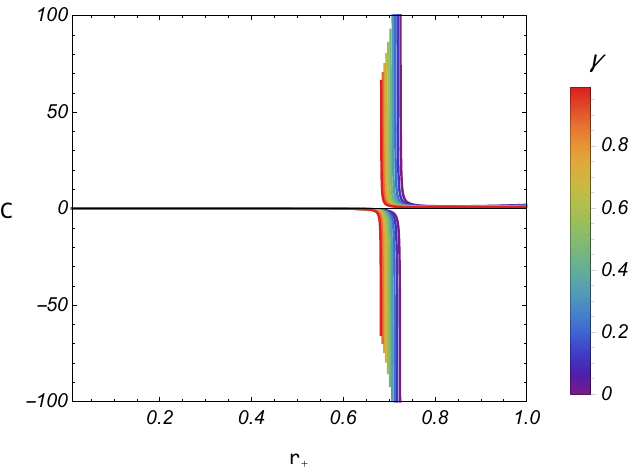}}\qquad
  \subfloat[$Q=1,\alpha=0.3, \beta =0.4$]{\centering{}\includegraphics[scale=0.65]{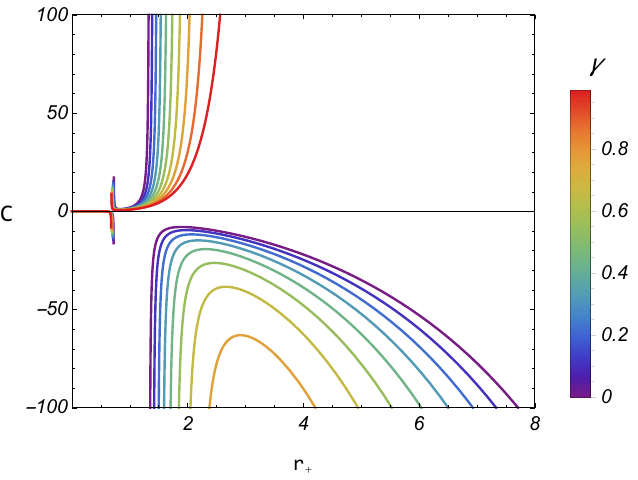}}
  \caption{The specific heat capacity of CEH-AdS-BH with CoS and PFDM showing the influence of the CoS parameter for small values of horizon $r_+$ (left) and large values (right). Here $M=1$ and $\ell_p=25$.}
  \label{heatCapG}
\end{figure}

\begin{figure}[ht!]
  \centering
  \subfloat[$Q=1,\gamma=0.4, \beta =0.4$]{\centering{}\includegraphics[scale=0.65]{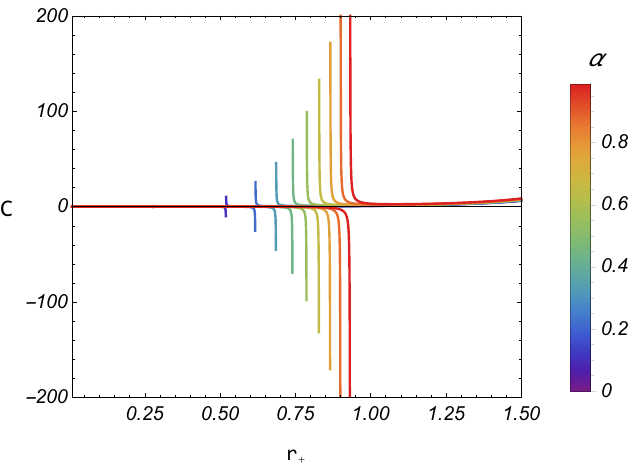}}\qquad
  \subfloat[$Q=1,\gamma=0.4, \beta =0.4$]{\centering{}\includegraphics[scale=0.65]{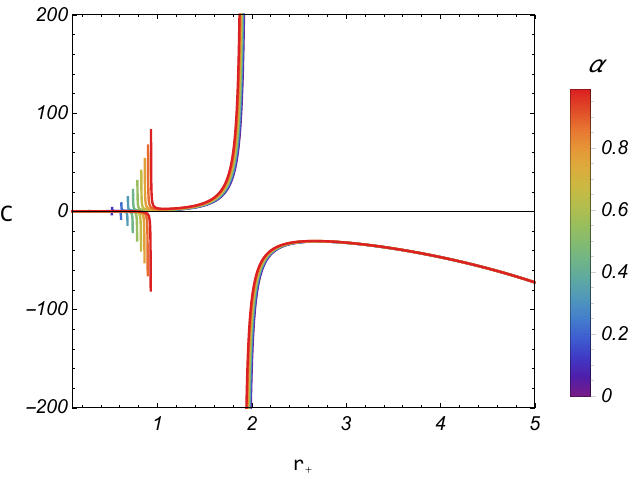}}
  \caption{The specific heat capacity of CEH-AdS-BH with CoS and PFDM showing the influence of the $\alpha$ parameter for small values of horizon $r_+$ (left) and large values (right). Here $M=1$ and $\ell_p=25$.}
  \label{heatCapA}
\end{figure}

\begin{figure}[ht!]
  \centering
  \subfloat[$Q=1,\gamma=0.4, \alpha =0.3$]{\centering{}\includegraphics[scale=0.65]{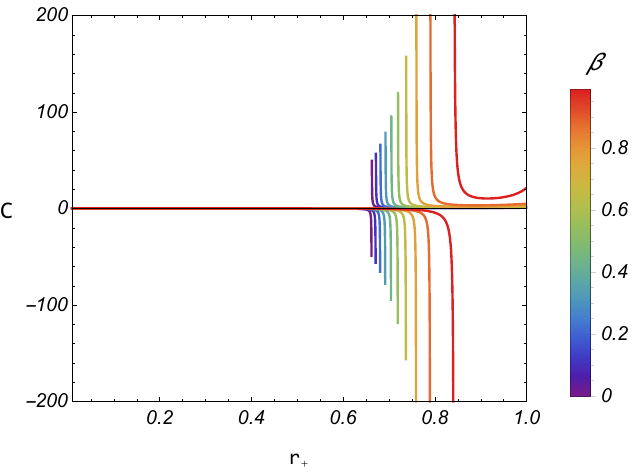}}\qquad
  \subfloat[$Q=1,\gamma=0.4, \alpha =0.3$]{\centering{}\includegraphics[scale=0.65]{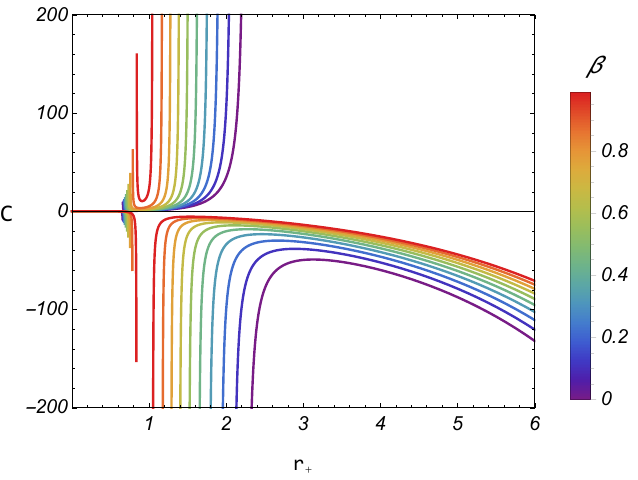}}
  \caption{The specific heat capacity of CEH-AdS-BH with CoS and PFDM showing the influence of the $\beta$ parameter for small values of horizon $r_+$ (left) and large values (right). Here $M=1$ and $\ell_p=25$.}
  \label{heatCapB}
\end{figure}

Figures \ref{heatCapG}, \ref{heatCapA}, and \ref{heatCapB} present comprehensive analyses of the specific heat capacity behavior across different parameter regimes. The plots reveal the divergences in specific heat capacity at $r_{+1}$ and $r_{+2}$ mark second-order phase transition points where the system transitions between thermodynamically stable and unstable regions. Negative specific heat indicates thermal instability where adding energy actually decreases the temperature, characteristic of gravitational systems. Figure \ref{heatCapG} demonstrates how CoS effects modify the stability regions, with larger $\gamma$ values generally expanding the unstable regions. Figure \ref{heatCapA} shows that EH corrections primarily affect the small-radius behavior, where quantum electrodynamic effects become significant. Figure \ref{heatCapB} reveals that PFDM coupling introduces additional complexity in the stability structure, with multiple sign changes possible depending on the parameter values.

Phase transition analysis is completed through examination of the Gibbs free energy:
\begin{equation}
  G=\frac{1}{80}\left( \frac{60Q^2}{r_+}-\frac{20Q^3}{\ell_p^2}-\frac{7Q^4\,\alpha}{r^5_+}-20(\beta+r_+(\gamma-1))+40\,\beta\,\ln\frac{r_+}{|\beta|}  \right).\label{gibs2}
\end{equation}

\begin{figure}[ht!]
  \centering
  \subfloat[$Q=1,\alpha=0.3, \beta =0.4$]{\centering{}\includegraphics[scale=0.55]{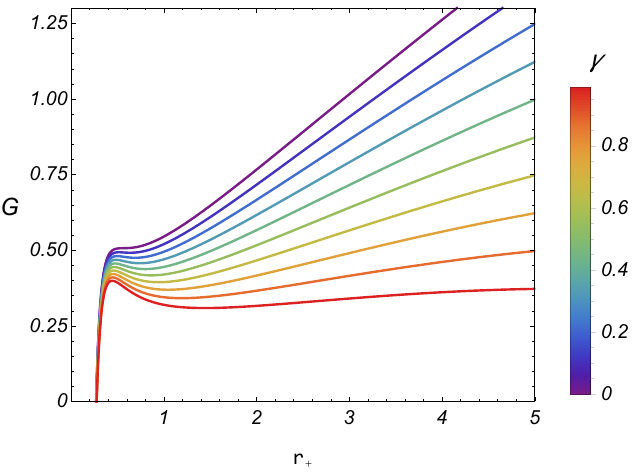}}\quad
  \subfloat[$Q=1,\gamma=0.4, \beta =0.4$]{\centering{}\includegraphics[scale=0.55]{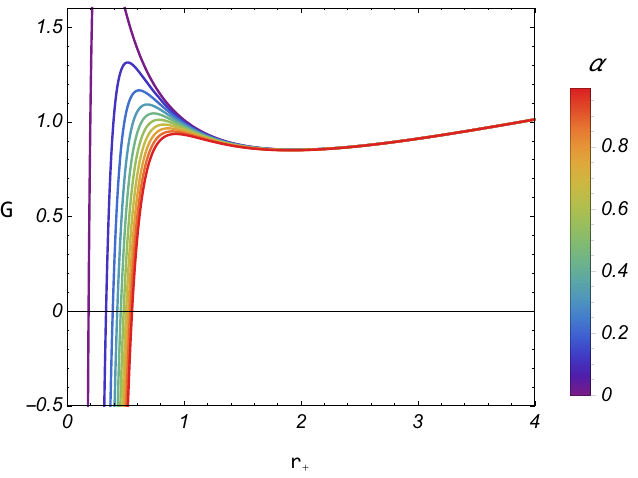}}
 \subfloat[$Q=1,\gamma=0.4, \alpha =0.3$] {\centering{}\includegraphics[scale=0.55]{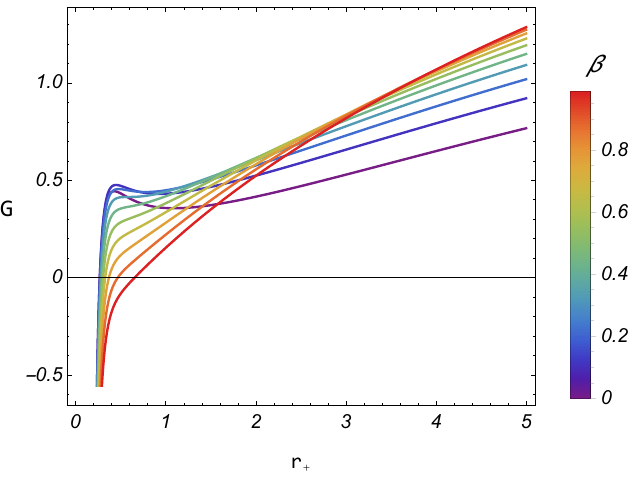}}
  \caption{The Gibbs free energy of CEH-AdS-BH with CoS and PFDM showing the influence of the $\alpha$ parameter (left), $\beta$ parameter (middle) and $\gamma$ parameter (right). Here $M=1$ and $\ell_p=25$.}
  \label{gibbs32}
\end{figure}

Figure \ref{gibbs32} illustrates the Gibbs free energy behavior across different parameter combinations, revealing the characteristic features that govern phase transition dynamics. The free energy profiles demonstrate how the exotic matter components collectively determine the thermodynamic stability and phase structure of the system. The left panel shows that varying the EH parameter $\alpha$ primarily affects the curvature of the free energy profiles, with larger values generally leading to more pronounced minima. The middle panel reveals that PFDM coupling $\beta$ can introduce multiple extrema in the free energy, suggesting the possibility of first-order phase transitions. The right panel demonstrates that CoS parameter $\gamma$ shifts the entire free energy landscape, potentially altering the locations of critical points and phase boundaries.

Our thermodynamic analysis reveals that the exotic matter components introduce rich phase transition behaviors that extend conventional AdS BH thermodynamics into new parameter regimes. These modifications offer potential observational signatures for distinguishing CEH-AdS-BH solutions with CoS and PFDM from their classical counterparts, providing a pathway for experimental verification of these theoretical predictions through precision measurements of thermal emission properties and gravitational wave signatures \cite{isz22}.

\section{Conclusions}\label{sec7}

In this study, we studied physical features of the CEH-AdS-BH spacetimes incorporating both CoS topological defects and PFDM distributions, establishing a comprehensive theoretical framework that extends conventional BH physics into previously unexplored parameter regimes. Our analysis encompassed multiple aspects of modified BH dynamics, from fundamental spacetime structure to complex thermodynamic phase transitions, revealing intricate relationships between exotic matter components and their collective influence on observable phenomena that could provide signatures for distinguishing these systems from classical solutions.

We began our investigation by establishing the foundational spacetime geometry through detailed analysis of the metric function given in Eq.~(\ref{bb2}), which encapsulates the contributions from CoS parameter $\gamma$, EH NLED parameter $\alpha$, and PFDM coupling $\beta$. The horizon structure analysis presented in Table~\ref{ads_eh_horizons_no_naked} demonstrated the diverse range of BH configurations possible within our theoretical framework, from extremal to non-extremal states, and revealed the critical threshold value $\gamma = 0.577216$ beyond which naked singularities emerge. This critical parameter represents a fundamental boundary in the theory, establishing physical constraints on cosmic string effects in BH spacetimes and providing a clear demarcation between viable BH solutions and pathological naked singularity configurations.

The geodesic dynamics investigation revealed fundamental modifications to particle trajectories in the presence of multiple exotic matter components, with each component contributing distinct physical effects that combine in complex nonlinear ways. For photon dynamics, we derived the highly nonlinear trajectory equation given in Eq.~(\ref{cc10}), demonstrating how each parameter contributes distinct nonlinear effects to light propagation through mechanisms ranging from angular deficit modifications to logarithmic dark matter corrections. The effective potential analysis shown in Fig.~\ref{fig:effective-potential} revealed the complex interplay between CoS parameter $\gamma$, EH corrections $\alpha$, and PFDM coupling $\beta$ in determining photon orbital characteristics, with systematic parameter variations producing measurable changes in orbital properties. Our photon sphere analysis, with numerical results presented in Tables~\ref{tab:1}, \ref{tab:2}, and \ref{tab:3}, established systematic trends showing increasing photon sphere radii with CoS parameter values, while the shadow analysis depicted in Fig.~\ref{fig:shadow} demonstrated observationally relevant modifications to BH shadow profiles that could potentially distinguish these exotic solutions from conventional charged BHs through future EHT observations \cite{isz24,isz25}. The shadow size variations of approximately 5-15\% with cosmic string effects provide quantitative predictions that fall within the sensitivity range of current and planned observational campaigns.

For massive test particles, we conducted detailed analyses of circular orbital dynamics and ISCO configurations that revealed competing physical mechanisms governing particle motion in these modified spacetimes. The effective potential profiles shown in Fig.~\ref{fig:effective-potential-neutral} revealed the modified gravitational landscape experienced by neutral particles, where cosmic string effects systematically weaken gravitational binding while electromagnetic charge creates additional repulsive contributions. The ISCO analysis, with results tabulated in Tables~\ref{tab:4} and \ref{tab:5}, established clear parametric dependencies showing outward ISCO shifts of 3-8\% with increasing CoS parameter and inward shifts with higher electric charge values. These orbital modifications have direct implications for accretion disk physics and could manifest as detectable frequency shifts in X-ray timing observations of accreting BH systems.

Our scalar field perturbation analysis established the theoretical foundation for understanding field propagation in CEH-AdS-BH spacetimes, providing crucial insights into stability properties and wave scattering characteristics. The scalar perturbative potential derived in Eq.~(\ref{ff6}) revealed the intricate structure encoding all geometric and physical effects of exotic matter components, while Fig.~\ref{fig:scalar-potential} demonstrated how CoS parameter variations systematically modify the potential barrier properties, with direct implications for wave scattering characteristics and greybody radiation spectra that could serve as observational probes of exotic matter effects. The EM perturbation investigation provided complementary insights into electromagnetic wave propagation characteristics, where EM fields exhibited fundamental differences through their vector nature, requiring $\ell \geq 1$ and lacking monopole components as shown in Eq.~(\ref{em3}). Fig.~\ref{fig:em-potential} illustrated the systematic variations in electromagnetic perturbative potentials with CoS parameter changes, revealing enhanced electromagnetic coupling effects in higher charge configurations that could affect gravitational wave emission and electromagnetic radiation processes around these BHs.

The thermodynamic analysis represented a cornerstone achievement of our investigation, establishing the complete thermodynamic framework for CEH-AdS-BH systems with exotic matter components and revealing novel phase transition behaviors not present in conventional BH thermodynamics. We derived the modified first law of thermodynamics in Eq.~(\ref{therm8}), incorporating additional intensive variables $\mathcal{A}_H$ and $\mathcal{B}_H$ corresponding to PFDM and CoS contributions respectively, which represent new thermodynamic degrees of freedom that could manifest in thermal emission properties and quasi-normal mode spectra. The Hawking temperature analysis revealed complex parameter dependencies, as illustrated in Fig.~\ref{temp32}, showing characteristic features including temperature minima, exponential growth regions, and asymptotic decay behavior reflecting the intricate coupling between exotic matter components and demonstrating how these systems can exhibit thermal behaviors fundamentally different from classical BHs. The thermal stability analysis through specific heat capacity calculations, presented in Figs.~\ref{heatCapG}, \ref{heatCapA}, and \ref{heatCapB}, revealed critical divergence points demarcating thermally stable and unstable regions where the system undergoes second-order phase transitions, with negative specific heat regions indicating gravitational thermal instability characteristic of self-gravitating systems. The phase transition analysis through Gibbs free energy examination, shown in Fig.~\ref{gibbs32}, presented phase structures extending conventional AdS BH thermodynamics into new parameter regimes where multiple competing effects determine the overall system stability and transition characteristics.

Throughout our investigation, we consistently demonstrated that the presence of CoS topological defects, EH NLED corrections, and PFDM distributions collectively create a rich phenomenological landscape that extends far beyond conventional BH physics through synergistic effects that cannot be understood by studying individual exotic components in isolation. The parameter space exploration revealed systematic trends, critical thresholds, and complex nonlinear relationships that govern the overall system behavior, establishing quantitative relationships between exotic matter parameters and observable quantities. These findings establish CEH-AdS-BH solutions with exotic matter as viable theoretical models with potentially observable signatures distinguishable from classical BH solutions, providing specific predictions that could be tested through multiple observational channels. The observational implications of our findings span multiple electromagnetic and gravitational wave detection channels, where the modified photon sphere and shadow characteristics could be probed through high-resolution BH imaging campaigns, while the altered orbital dynamics and fundamental frequencies offer signatures detectable through X-ray timing observations of accreting BH systems \cite{iszx03,iszx04}.

The theoretical framework developed here provides several promising avenues for future research that could bridge fundamental physics with observational astronomy. The extension to rotating CEH-AdS-BH solutions with CoS and PFDM represents a natural next step, requiring advanced mathematical techniques to handle the additional complexity introduced by angular momentum while preserving the essential physics of exotic matter interactions, potentially revealing how these effects couple with BH spin to affect jet formation and electromagnetic emission processes. The exploration of quantum field theory in CEH-AdS-BH backgrounds also offers opportunities to investigate Hawking radiation modifications, vacuum polarization effects, and particle creation processes in the presence of exotic matter components, which could provide insights into quantum gravity phenomenology and the fundamental nature of spacetime. Moreover, investigating the role of CEH-AdS-BH solutions in early universe dynamics, inflation scenarios, and large-scale structure formation represents another frontier for theoretical exploration with potential connections to dark matter phenomenology and modified gravity theories \cite{isz14}, possibly linking BH physics to cosmological observations through the cosmic string and dark matter components that play important roles in both contexts.

Our results demonstrate that exotic matter components can produce measurable deviations from classical BH predictions, with the cosmic string parameter $\gamma = 0.577216$ representing a critical threshold, shadow radius modifications of 5-15\%, ISCO shifts of 3-8\%, and novel thermodynamic phase structures that collectively provide a comprehensive set of observational targets for current and future astronomical facilities. As precision gravitational astronomy continues to advance through improved gravitational wave detectors, enhanced BH imaging capabilities, and more sensitive X-ray timing observations, the distinctive signatures predicted by our analysis may offer new routes for testing fundamental physics in the strong-field regime and potentially discovering evidence for cosmic strings, modified electrodynamics, or exotic dark matter interactions around BHs.

\small

\section*{Acknowledgments}

F.A. acknowledges the Inter University Centre for Astronomy and Astrophysics (IUCAA), Pune, India for granting visiting associateship.  \.{I}.~S. expresses gratitude to T\"{U}B\.{I}TAK, ANKOS, and SCOAP3 for their financial support. He also acknowledges COST Actions CA22113, CA21106, and CA23130 for their contributions to networking.

\section*{Data Availability Statement}

This study is purely theoretical and does not involve the generation or analysis of any datasets. 
Therefore, data sharing is not applicable to this article as no new data were created or analyzed in this study.

\end{document}